\documentclass[prd,aps,
twocolumn,
nofootinbib,
preprintnumbers,
superscriptaddress]{revtex4-2}
\usepackage[T1]{fontenc} 
\usepackage{graphicx}
\graphicspath{{./assets/plots/}}

\usepackage{epsfig}
\usepackage{comment}
\usepackage{xcolor}
\usepackage{rotating}
\usepackage{amssymb}
\usepackage{subfigure}
\usepackage[colorlinks=true, linkcolor=blue, citecolor=blue, urlcolor=blue]{hyperref}
\usepackage{dsfont}
\usepackage{psfrag}
\usepackage{multirow}
\usepackage{amsmath,euscript,array,mathrsfs}
\usepackage{bbold}
\usepackage{epsf}
\usepackage{cleveref}
\usepackage{bbm}
\usepackage[utf8]{inputenc}
\usepackage{placeins}
\usepackage{dcolumn}
\usepackage{tabularx}
\newcolumntype{Y}{>{\centering\arraybackslash}X}

\hypersetup{
    colorlinks=true,
    linkcolor=blue,
    filecolor=magenta,      
    urlcolor=blue,}

\newcommand{\beq}{\begin{equation}}
\newcommand{\eeq}{\end{equation}}
\newcommand{\beqs}{\begin{eqnarray}}
\newcommand{\eeqs}{\end{eqnarray}}
\newcommand{\lsim}{\mathrel{\raisebox{-.6ex}{$\stackrel{\textstyle<}{\sim}$}}}
\newcommand{\gsim}{\mathrel{\raisebox{-.6ex}{$\stackrel{\textstyle>}{\sim}$}}}

\def\hbar{\hspace{0pt}\raisebox{1pt}{$-$} \hspace{-7pt} h}

\newcommand{\be}{\begin{equation}}
\newcommand{\ee}{\end{equation}}

\newcommand{\bea}{\begin{eqnarray}}
\newcommand{\eea}{\end{eqnarray}}

\newcommand{\mel}[3]{\langle#1|#2|#3\rangle}

\def\tr{{\rm tr}}

\def\lbldef#1#2{\expandafter\gdef\csname #1\endcsname {#2}}

\def\href#1#2{#2}

\newcommand{\ber}{\begin{eqnarray}}
\newcommand{\eer}{\end{eqnarray}}

\newcommand{\beqar}{\begin{eqnarray}}

\newcommand{\eeqar}{\end{eqnarray}}

\def\calE{{\mathcal{E}}}

\def\calO{{\mathcal{O}}}

\def\calW{{\mathcal{W}}}

\def\bbt{\mathbb{t}}

\newcommand{\dsl}
  {\kern.06em\hbox{\raise.15ex\hbox{$/$}\kern-.56em\hbox{$\partial$}}}

\newcommand{\eeqarr}{\end{eqnarray}}
\newcommand{\ZZ}{{\rm \kern 0.275em Z \kern -0.92em Z}\;}

  \def\tr{{\hbox{\rm Tr}}}

\def\CC{{\mathchoice
{\rm C\mkern-8mu\vrule height1.45ex depth-.05ex
width.05em\mkern9mu\kern-.05em}
{\rm C\mkern-8mu\vrule height1.45ex depth-.05ex
width.05em\mkern9mu\kern-.05em}
{\rm C\mkern-8mu\vrule height1ex depth-.07ex
width.035em\mkern9mu\kern-.035em}
{\rm C\mkern-8mu\vrule height.65ex depth-.1ex
width.025em\mkern8mu\kern-.025em}}}

\def\RR{{\rm I\kern-1.6pt {\rm R}}}

\def\ZZ{{\rm Z}\kern-3.8pt {\rm Z} \kern2pt}
\def\IB{\relax{\rm I\kern-.18em B}}
\def\ID{\relax{\rm I\kern-.18em D}}
\def\II{\relax{\rm I\kern-.18em I}}
\def\IP{\relax{\rm I\kern-.18em P}}

\newcommand{\bear}{\begin{eqnarray}}
\newcommand{\eear}{\end{eqnarray}}

\def\to{\rightarrow}

\def\tr{{\rm Tr}}
\def\to{\rightarrow}

\def\c{\gamma}

\def\i{\iota}

\def\6{\partial}

\def\bea{\begin{eqnarray}}
\def\eea{\end{eqnarray}}

\def\beqx{\begin{displaymath}}
\def\eeqx{\end{displaymath}}

\newcommand{\bmat}{\left(\begin{array}}
\newcommand{\emat}{\end{array}\right)}

\newcommand{\Edit}[1]{#1}

\def\c{\chi}

\def\i{\iota}

\def\bo{{\raise-.3ex\hbox{\large$\Box$}}}               
\def\face{{\raise.2ex\hbox{$\displaystyle \bigodot$}\mskip-2.2mu \llap {$\ddot \smile$}}}
\def\>{\rangle}                                      
\def\<{\langle}                                      

\def\leftrightarrowfill{$\mathsurround=0pt \mathord\leftarrow \mkern-6mu
        \cleaders\hbox{$\mkern-2mu \mathord- \mkern-2mu$}\hfill
        \mkern-6mu \mathord\rightarrow$}        
\def\dvec#1{\vbox{\ialign{##\crcr
        \leftrightarrowfill\crcr\noalign{\kern-1pt\nointerlineskip}
        $\hfil\displaystyle{#1}\hfil$\crcr}}}           
\def\tr{{\rm tr \,}}                                    

\def\-{\hphantom{-}}

\def\dee{\mathrm{d}}

\newcommand \captionchisquarespfour {3.06}
\newcommand \captionchisquarespeight {3.02}
\newcommand \captionchisquarelargeN {3.29}

\begin{document}
\title{On the spectrum of mesons in quenched $Sp(2N)$ gauge theories}

\preprint{CTPU-PTC-23-50}

\author{Ed Bennett}
\email{e.j.bennett@swansea.ac.uk}
\affiliation{Swansea Academy of Advanced Computing, Swansea University, Fabian Way, SA1 8EN, Swansea, Wales, UK}
\author{Jack Holligan}
\email{holligan@msu.edu}
\affiliation{Biomedical and Physical Sciences Building, Michigan State University, East Lansing, Michigan, USA, 48824}
\author{Deog Ki Hong}
\email{dkhong@pusan.ac.kr}
\affiliation{Department of Physics, Pusan National University, Busan 46241, Korea}
\affiliation{Institute for Extreme Physics, Pusan National University, Busan 46241, Korea}
\author{Jong-Wan Lee}
\email{j.w.lee@ibs.re.kr}
\affiliation{Department of Physics, Pusan National University, Busan 46241, Korea}
\affiliation{Institute for Extreme Physics, Pusan National University, Busan 46241, Korea}
\affiliation{Particle Theory and Cosmology Group, Center for Theoretical Physics of the Universe,
Institute for Basic Science (IBS), Daejeon, 34126, Korea}
\author{C.-J.~David~Lin}
\email{dlin@nycu.edu.tw}
\affiliation{Institute of Physics, National Yang Ming Chiao Tung University, 1001 Ta-Hsueh Road, Hsinchu 30010, Taiwan}
\affiliation{Center for High Energy Physics, Chung-Yuan Christian University,
Chung-Li 32023, Taiwan}
\affiliation{Centre for Theoretical and Computational Physics, National Yang Ming Chiao Tung University, 1001 Ta-Hsueh Road, Hsinchu 30010, Taiwan}
\author{Biagio Lucini}
\email{b.lucini@swansea.ac.uk}
\affiliation{Department of Mathematics, Faculty  of Science and Engineering,
Swansea University, Fabian Way, SA1 8EN Swansea, Wales, UK}
\affiliation{Swansea Academy of Advanced Computing, Swansea University,
Fabian Way, SA1 8EN, Swansea, Wales, UK}
\author{Maurizio Piai}
\email{m.piai@swansea.ac.uk}
\affiliation{Department of Physics, Faculty  of Science and Engineering,
Swansea University,
Singleton Park, SA2 8PP, Swansea, Wales, UK}
\author{Davide Vadacchino}
\email{davide.vadacchino@plymouth.ac.uk}
\affiliation{Centre for Mathematical Sciences, University of Plymouth, Plymouth, PL4 8AA, United Kingdom}


\begin{abstract}
   We report  the findings of our extensive study of the spectra of flavoured mesons in lattice gauge theories with 
   symplectic gauge group and fermion matter content treated in the quenched approximation. 
   For the $Sp(4)$, $Sp(6)$, and  $Sp(8)$ gauge groups, 
   the (Dirac) fermions transform in either the fundamental, or the 2-index, antisymmetric or symmetric, representations.
   This study sets the stage for future precision calculations with dynamical fermions 
   in the low mass region of lattice parameter space.
  Our results have potential  phenomenological applications ranging from composite Higgs models, to  
  top (partial) compositeness, to dark matter models with composite, strong-coupling 
  dynamical origin. 
   Having adopted the Wilson flow as a scale-setting procedure, we apply Wilson chiral perturbation theory to extract the continuum and massless limits for the observables of interest. The resulting measurements are used to perform a simplified extrapolation to the large-$N$ limit, hence drawing a preliminary connection with gauge theories with unitary groups. We conclude with a brief discussion of the  Weinberg sum rules.
   
\end{abstract}

\maketitle

\tableofcontents

\section{Introduction}

Strongly coupled gauge theories that live in four space-time dimensions, have gauge group $Sp(2N)$ 
(for $N\in \mathbb{Z}^+$), and are coupled to fermion matter fields, 
 have a plethora of applications in proposals of new physics that 
 extend the Standard Model (SM) of particle physics.
 They can provide the  microscopic origin of Composite Higgs Models (CHMs)~\cite{Kaplan:1983fs,
Georgi:1984af,Dugan:1984hq},\footnote{An overview of the field can be found in the review papers in Refs.~\cite{
Panico:2015jxa,Witzel:2019jbe,
Cacciapaglia:2020kgq},  the tables in Refs.~\cite{
Ferretti:2013kya,Ferretti:2016upr,Cacciapaglia:2019bqz},
the incomplete selection of useful papers in Refs.~\cite{
Katz:2005au,Barbieri:2007bh,
Lodone:2008yy,Gripaios:2009pe,Mrazek:2011iu,Marzocca:2012zn,Barnard:2013zea,
Grojean:2013qca,Cacciapaglia:2014uja,
Ferretti:2014qta,Arbey:2015exa,Cacciapaglia:2015eqa,Vecchi:2015fma,Ma:2015gra,
Feruglio:2016zvt,DeGrand:2016pgq,Fichet:2016xvs,
Galloway:2016fuo,Agugliaro:2016clv,Belyaev:2016ftv,Csaki:2017cep,Chala:2017sjk,Golterman:2017vdj,
Csaki:2017jby,Serra:2017poj,Alanne:2017rrs,Alanne:2017ymh,Sannino:2017utc,Alanne:2018wtp,Bizot:2018tds,
Cai:2018tet,Agugliaro:2018vsu,BuarqueFranzosi:2018eaj,Cacciapaglia:2018avr,Gertov:2019yqo,Ayyar:2019exp,
Cacciapaglia:2019ixa,Appelquist:2020bqj,BuarqueFranzosi:2019eee,
Cacciapaglia:2019dsq,Cacciapaglia:2020vyf,Cai:2020njb,
Dong:2020eqy,Cacciapaglia:2021uqh,Banerjee:2022izw,
Appelquist:2022qgl}
and in
Refs.~\cite{
Contino:2003ve,Agashe:2004rs,Agashe:2005dk,Agashe:2006at,
Contino:2006qr,Falkowski:2008fz,Contino:2010rs,Contino:2011np,
Caracciolo:2012je,Erdmenger:2020lvq,Erdmenger:2020flu,
Elander:2020nyd,Elander:2021bmt,Elander:2021kxk,Elander:2023aow,Erdmenger:2023hkl}.}
and have been exploited to explain the origin of the 
large mass of the top quark,
via the implementation of top partial compositeness (TPC)~\cite{Kaplan:1991dc},\footnote{The reader may find it useful to 
refer to the more recent, critical discussions in Refs.~\cite{Grossman:1999ra,Gherghetta:2000qt,Chacko:2012sy}.}
and can be used to explain for the origin of dark matter, through the Strongly Interacting Massive Particle (SIMP) paradigm~\cite{Hochberg:2014dra,Hochberg:2014kqa,Hochberg:2015vrg},\footnote{An incomplete list of relevant papers includes also Refs.~\cite{Hansen:2015yaa,Bernal:2015xba, Bernal:2017mqb,Berlin:2018tvf,
Bernal:2019uqr,Tsai:2020vpi,Kondo:2022lgg}.} and they might even affect
the detectable gravitational wave (GW) stochastic background~\cite{Witten:1984rs,Kamionkowski:1993fg,Allen:1996vm,Schwaller:2015tja, Croon:2018erz,Christensen:2018iqi}, if responsible for a phase transition in the early universe~\cite{Huang:2020crf,Halverson:2020xpg,Kang:2021epo,Reichert:2021cvs,Reichert:2022naa,Pasechnik:2023hwv}.\footnote{A number of present and future experiments might detect such effects, see for example Refs.~\cite{Seto:2001qf,
Kawamura:2006up,Crowder:2005nr,Corbin:2005ny,Harry:2006fi,
Hild:2010id,Yagi:2011wg,Sathyaprakash:2012jk,Thrane:2013oya,
Caprini:2015zlo,
LISA:2017pwj,
LIGOScientific:2016wof,Isoyama:2018rjb,Baker:2019nia,
Brdar:2018num,Reitze:2019iox,Caprini:2019egz,
Maggiore:2019uih}.}

In all these applications, the strong-coupling regime of the theory
plays a central role, and hence one must develop and apply non-perturbative instruments in order to 
extract quantitative information about the dynamics of the theories of interest.
The natural framework for such endeavour is lattice gauge theory.
Until recently, the literature on $Sp(2N)$ lattice gauge theories was
rather limited~\cite{Holland:2003kg}. The discovery of the Higgs boson~\cite{ATLAS:2012yve,CMS:2012qbp}
has triggered a new wave of interest in extensions of the SM with strongly coupled origin,
which has motivated the start of an extensive programme of exploration of $Sp(2N)$ gauge theories on
 the lattice~\cite{Bennett:2017kga,
Lee:2018ztv,Bennett:2019jzz,
Bennett:2019cxd,Bennett:2020hqd,
Bennett:2020qtj,Lucini:2021xke,
Bennett:2021mbw,Bennett:2022yfa,
Bennett:2022gdz,Bennett:2022ftz,
Hsiao:2022gju,Hsiao:2022kxf,
Maas:2021gbf,Zierler:2021cfa,
Kulkarni:2022bvh,Bennett:2023wjw,Bennett:2023rsl,
Bennett:2023gbe,Mason:2023ixv,Forzano:2023duk}, 
as candidates for the dynamical origin of CHMs.\footnote{In parallel, 
extensive work on 
lattice gauge theories with  $SU(2)$~\cite{Hietanen:2014xca,Detmold:2014kba,
Arthur:2016dir,Arthur:2016ozw,Pica:2016zst,Lee:2017uvl,Drach:2017btk,Drach:2020wux,Drach:2021uhl, 
Drach:2021uhl} and $SU(4)$~\cite{Ayyar:2017qdf,Ayyar:2018zuk,Ayyar:2018ppa,
 Ayyar:2018glg,Cossu:2019hse,Lupo:2021nzv,Hasenfratz:2023sqa} gauge groups
 relevant to CHMs  has been performed.
 Lattice results on the $SU(3)$ theory with $N_f=8$ fundamental fermions transforming in the fundamental representation~\cite{
LatKMI:2014xoh,Appelquist:2016viq,LatKMI:2016xxi,
Gasbarro:2017fmi,LatticeStrongDynamics:2018hun,
LatticeStrongDynamicsLSD:2021gmp,Hasenfratz:2022qan, LSD:2023uzj,LatticeStrongDynamics:2023bqp}
 have also been used in the CHM context~\cite{Appelquist:2020bqj,Appelquist:2022qgl}---see also related earlier work in Refs.~\cite{Vecchi:2015fma,Ma:2015gra,BuarqueFranzosi:2018eaj}.}

This paper reports on  new results
obtained by considering $Sp(2N)$ lattice gauge theories, with $N=2,\, 3,\, 4$, coupled to fermion matter fields treated in the 
quenched approximation. The effects due to  fermions 
are not included in the Monte Carlo algorithm generating the available ensembles of configurations, 
but only in the formulation of the operators used
to probe the underlying Yang-Mills dynamics.

There are three main, compelling motivations to perform an extended study of these theories
with such  approximation.
 Firstly, at least in the CHM and SIMP contexts, the regions of parameter space of interest for phenomenological applications
 are often characterised by  moderately heavy fermions  and sizable amounts of explicit symmetry breaking. 
This is needed, among other reasons, by model building consideration. In a realistic, complete model, one must
ensure that the masses of towers of new composite states  be large enough to have escaped direct detection so far. 
Furthermore, some of the new composite states
 must decay only via weak interactions, introduced by couplings to the SM fields. This can be achieved by making the particles heavy enough to forbid kinematically some direct decay within the strong-coupling sector. 
If these conditions are met, the quenched approximation may already be precise enough to produce 
 useful estimates of the relevant spectroscopy parameters (masses and decay constants), with comparatively low investment of 
 computing time and resources. In addition, the quenched approximation captures at once large classes of models, that differ only by the number of fermions, while the study of dynamical fermions requires dedicated Monte Carlo calculations
 for each choice of matter field content.

 The second motivation is of a technical nature, 
 and is closely related to  the final comment we made in the previous paragraph, that
already  highlights both flexibility and applicability of quenched calculations.  
 Whatever the model of interest, a systematic and rigorous dynamical study demands to benchmark it against a simpler, well understood reference example. Doing so allows to control possible systematic  effects and to prevent 
 unwanted misinterpretation of the results. It also provides a way to  gauge the size of the dynamical effects due to fermions.
 This is particularly important, somewhat paradoxically, when studying theories for which one expects large
 effects to arise due to the fermion dynamics. For example, this is the case when one is looking for quasi-conformal behaviour 
 (and large anomalous dimensions)
 in the long distance physics of theories that are expected to be close to the boundary of the conformal window.
 
The third motivation for this study is that the quenched approximation, for fermions in the fundamental representation, provides
a natural connection to other approaches to non-perturbative physics, in particular those relying on the large-$N$ limit  
and holography~\cite{Maldacena:1997re,Gubser:1998bc,Witten:1998qj,Aharony:1999ti}.
We will not further discuss this point in the paper, but it is remarkable, for example, that the recent explosion of interest 
in gauge-gravity dualities has provided instruments that are particularly well suited to study the quenched, large-$N$ limit of 
non-Abelian gauge theories.
It is worthy of notice that the large-$N$ limit of $Sp(2N)$ theories 
is expected to yield the same results, 
in a common sector of the physical spectrum, as the 
 large-$N_c$ limit for $SU(N_c)$ theories, 
 for which the literature on lattice 
 numerical studies is more developed---see for instance 
 Refs.~\cite{Lucini:2012gg,GarciaPerez:2020gnf,Hernandez:2020tbc,DeGrand:2023hzz}.

We study $Sp(2N)$ gauge theories with $N \geq 2$ that are asymptotically free.
For the quenched fermion matter fields, we restrict attention to the three smallest possible representations: the fundamental $(\rm f)$, and the 2-index antisymmetric $(\rm as)$, and symmetric---adjoint---$(\rm s)$ representations.
For example, the symmetry-breaking pattern of the 
$Sp(2N)$ theory with $N_{(\rm f)}=2$ fundamental (Dirac) fermions is described by the $SU(4)/Sp(4)$ coset
relevant to minimal CHMs. With the addition of $N_{(\rm as)}=3$ (Dirac) fermions in the antisymmetric representation
this theory also provides a potential microscopic realisation of top partial compositeness~\cite{Barnard:2013zea}.
But it is worth noting that the $N_{(\rm f)}=0$ and $N_{(\rm as)}=3$ theory is also a potential completion for a CHM~\cite{Cacciapaglia:2019ixa}.

The paper is organised as follows.
In Sect.~\ref{Sec:lattice}, we define the theory of interest and explain the lattice technology we deploy for this study.
We present our main results for the spectra of mesons
 in Sect.~\ref{Sec:results}, organising the material by gauge group and by representation.
 We briefly discuss the Weinberg sum rules, in Sect.~\ref{Sec:Weinberg}.
 In Sect.~\ref{Sec:conclusions}, we summarise the main lessons we learned, and outline future research opportunities.
The paper is supplemented by an extensive Appendix, that shows the technical details 
characterising the intermediate numerical results that we analysed to arrive at our main results.

\section{Lattice Theory and Observables}
\label{Sec:lattice}
In this section, we define the lattice theories of interest and the ensembles we generated, 
as well as the observables we computed and analysed. We present our strategy for handling finite volume and
finite spacing effects, and our scale-setting procedure. In doing so, we reorganise and consistently integrate material presented elsewhere, in particular in Refs.~\cite{Bennett:2017kga,Bennett:2019cxd,Bennett:2020qtj,Bennett:2022ftz},
but we also expand  this material, and specialise it to the case of interest, as appropriate.
We then describe the continuum and massless limit extrapolations, that rely on Wilson chiral perturbation theory (W$\chi$PT)~\cite{
Sheikholeslami:1985ij,Rupak:2002sm} (we found it useful also to read Ref.~\cite{Sharpe:1998xm}, as well as some of the literature on improvement~\cite{Symanzik:1983dc,Luscher:1996sc}).

\subsection{Action and ensembles}

\begin{table}[t]
    \caption{Lattice ensembles analysed in the Yang-Mills $Sp(2N)$ gauge theories of interest. 
    We report the value of $N$, of the lattice coupling, $\beta$, of the spatial, $N_s$, and temporal, $N_t$,
     extension of the lattice, as well as the gradient flow scale, $w_0$, expressed in units of the lattice spacing, $a$.
    \label{tab:latticeDetails}\\}
    \centering
    \begin{tabular}{|c|c|c|c|}
\hline
$N$ & $\beta$ & $N_s^3 \times N_t$ & $w_0 / a$ \\
\hline
\hline
\multirow{5}*{2} & 7.62 & $24^3 \times 48$ & $1.31800(95)$ \\
\cline{2-4} & 7.7 & $48^3 \times 60$ & $1.45284(40)$ \\
\cline{2-4} & 7.85 & $48^3 \times 60$ & $1.76364(63)$ \\
\cline{2-4} & 8.0 & $48^3 \times 60$ & $2.10735(99)$ \\
\cline{2-4} & 8.2 & $48^3 \times 60$ & $2.6188(23)$
\\
\hline
\hline
\multirow{5}*{3} & 15.6 & $24^3 \times 48$ & $1.29831(67)$ \\
\cline{2-4} & 16.1 & $24^3 \times 48$ & $1.8000(17)$ \\
\cline{2-4} & 16.5 & $48^3 \times 96$ & $2.24078(99)$ \\
\cline{2-4} & 16.7 & $48^3 \times 96$ & $2.5040(11)$ \\
\cline{2-4} & 17.1 & $48^3 \times 96$ & $3.0768(24)$
\\
\hline
\hline
\multirow{5}*{4} & 26.5 & $24^3 \times 48$ & $1.34724(48)$ \\
\cline{2-4} & 26.7 & $48^3 \times 96$ & $1.44654(17)$ \\
\cline{2-4} & 26.8 & $48^3 \times 96$ & $1.50491(17)$ \\
\cline{2-4} & 27.0 & $48^3 \times 96$ & $1.62325(25)$ \\
\cline{2-4} & 27.3 & $60^3 \times 120$ & $1.80187(25)$
\\
\hline
\end{tabular}

\end{table}

The calculation of the mass spectrum of mesons in the quenched approximation is carried out on  configurations
sampled using the standard Wilson action for gauge group $Sp(2N)$:
\begin{equation}\label{eq.WilsonAction}
    S_W\equiv \beta\sum_{x}\sum_{\mu<\nu}\left(1-\frac{1}{2N}\Re\,\tr P_{\mu\nu}(x)\right)\,,
\end{equation}
where $\beta\equiv4N/g^2$, $g$ is the coupling strength, $\Re$ denotes the real part and $\tr$ denotes the trace of the gauge matrix. 
The plaquette, $P_{\mu\nu}(x)$,  is defined on the smallest closed path in the $(\mu,\nu)$ plane with origin at lattice site $x$. A gauge link in the $\mu$ direction, originating at $x$, is denoted by the group element $U_{\mu}(x)$, hence the plaquette is
\begin{equation}\label{eq.Plaquette}
    P_{\mu\nu}(x)=U_{\mu}(x)U_{\nu}(x+a\hat{\mu})U^{\dagger}_{\mu}(x+a\hat{\nu})U^{\dagger}_{\nu}(x)\,,
\end{equation}
with $\hat{\mu}$, $\hat{\nu}$ denoting the unit vectors in the $\mu$, $\nu$ directions, respectively,  and $a$ the lattice spacing. We study observables for $Sp(2N)$ with
 $N=2$, $3$, and $4$. (Later on we will present a simplified extrapolation towards asymptotically large values of $N$.) 

The properties characterising our ensembles are detailed in \Cref{tab:latticeDetails},
in which for each ensemble we specify $N$,  the coupling $\beta$, the extents of the spatial, $N_s$, and temporal, $N_t$, directions
of the lattice, and the gradient flow scale $w_0/a$---described in more detail in \Cref{sec:scaleSetting}. Each meson measurement is performed from 200 lattice configurations.
We use the same lattice inputs for $Sp(4)$ as  in Ref.~\cite{Bennett:2019cxd}, 
to allow for direct comparison, while the choices of $\beta$ are a representative subgroup of those employed in
Ref.~\cite{Bennett:2020qtj}---see also Ref.~\cite{Bennett:2022ftz}---but we are using much larger volumes,
in order to reduce finite volume effects, as discussed in Appendix~\ref{Sec:FV}. A single update in the Markov chain consists of one application of the heatbath algorithm to each lattice link~\cite{KENNEDY1985393} and four applications of the overrelaxation algorithm~\cite{PhysRevD.29.306,PhysRevD.23.2901}. This is defined as a single ``sweep''. We perform an initial 600 sweeps to thermalise the lattice and thereafter apply 12 sweeps between each of the 200 configurations,
 to reduce autocorrelation. We checked that none of the ensembles used for this analysis show significant evidence of topological freezing.

\subsection{Mesons}

\begin{table}[t]
    \caption{Interpolating operators, $\mathcal{O}_M$, appearing in the correlation functions computed for this publication. For each operator, we indicate their name, label, Dirac algebra structure,  spin, $J$, and parity, $P$, of the associated mesons. We find it convenient to also associate each operator with the meson sourced by the analogous QCD operator. 
    \label{tab:operators}\\ }
    \centering
    \begin{tabular}{|c|c|c|c|c|}
    \hline
    Channel & Label & $\mathcal{O}_M$ & $J^P$ & QCD meson \\
    \hline\hline
    Pseudoscalar & PS & $\overline{\psi}\gamma_5\psi$ & $0^-$ & $\pi$ \\
    \hline
    Scalar & S & $\overline{\psi}\psi$ & $0^+$ & $a_0$ \\
    \hline
    Vector & V & $\overline{\psi}\gamma_{\mu}\psi$ & $1^-$ & $\rho$ \\
    \hline
    Axial-vector & AV & $\overline{\psi}\gamma_5\gamma_{\mu}\psi$ & $1^+$ & $a_1$ \\
    \hline
    Tensor & T & $\overline{\psi}\gamma_0\gamma_{\mu}\psi$ & $1^-$ & $\rho$ \\
    \hline
    Axial-tensor & AT & $\overline{\psi}\gamma_5\gamma_0\gamma_{\mu}\psi$ & $1^+$ & $b_1$ \\
    \hline
    \end{tabular}
\end{table}

The observable quantities of interest for this paper are the flavour non-singlet meson masses and the related decay constants. They are measured by examining the large-time behaviour of two-point 
correlation functions involving interpolating operators sourcing the mesons, which we denote as $\mathcal{O}_M$. We list the 
interesting  operators and their properties in \Cref{tab:operators}.

Masses and decay constants of mesons made of fermions transforming in the fundamental representation in a channel labelled as $M$ are denoted as $m_M$ and $f_M$, respectively. Because we study mesons comprised of fermions  transforming in three distinct representations of the group, in order to distinguish them we change the aspect of the label. While retaining upper case labels for 
the $(\rm f)$ fermions, mesons made of $(\rm as)$ fermions have lower case labels, and  calligraphic lettters are used for the labels of mesons made of $(\rm s)$ fermions.  For example, the pseudoscalar masses in the $(\rm f)$, $(\rm as)$, and $(\rm s)$ cases are
denoted by $m_{\rm PS}$, $m_{\rm ps}$, and $m_{\mathcal{PS}}$, respectively. In \Cref{tab:operators} and in rest of this subsection, we denote a general channel by an uppercase, e.g. as $M$, to lighten the notation, the replacements needed for the other two cases being clear from the context.

\begin{table}[t]
\caption[$Sp(4)$ chiral extrapolations]{Massless and continuum extrapolation of the decay constants, $\hat{f}$, and masses, $\hat{m}$, expressed in units of the gradient flow scale, $w_0$, for quenched mesons in the $Sp(4)$ theory, for the three representations considered in this study. The uncertainties reported are determined by the extrapolation, starting from the statistical uncertainties of the individual measurements. Reduced chi-squared values, $\chi^2/N_{\rm d.o.f.}$, that are greater than $3.0$ are highlighted in red
(e.g., \captionchisquarespfour). All $Sp(4)$ measurements with (f) or (as) mesons have been performed on newly generated configurations, and agree, within errors, with those reported in Ref.~\cite{Bennett:2019cxd}. The massless and continuum extrapolations for  $\hat{f}_{\rm AV}^2$ and $\hat{f}_{\rm av}^2$, which are displayed by the plots in Appendix \ref{sec:CMExtrapolations}, are affected by large systematics due to numerical noise.\\
\label{tab:Sp4ChiralExtrap}}
\centering
\begin{tabular}{|c|c|c|c|}
\hline
\multicolumn{4}{|c|}{$Sp(4)$}\\
\hline
Representation & Channel & Chiral limit & $\chi^2/{\rm d.o.f.}$\\
\hline
\multirow{8}{*}{Fundamental} & $\hat{f}_{\rm PS}$ & 0.0818(11) & 1.30\\
\cline{2-4}
 & $\hat{f}_{\rm V}$ & 0.1603(29) & 1.60\\
\cline{2-4}
 & $\hat{f}_{\rm AV}$ & 0.215(12) & 1.10\\
\cline{2-4}
 & $\hat{m}_{\rm V}$ & 0.6022(49) & 1.63\\
\cline{2-4}
 & $\hat{m}_{\rm AV}$ & 1.087(43) & 1.04\\
\cline{2-4}
 & $\hat{m}_{\rm S}$ & 1.052(38) & 2.97\\
\cline{2-4}
 & $\hat{m}_{\rm T}$ & 0.5991(81) & 1.47\\
\cline{2-4}
 & $\hat{m}_{\rm AT}$ & 1.145(48) & 2.21\\
\hline
\hline
\multirow{8}{*}{Antisymmetric} & $\hat{f}_{\rm ps}$ & 0.1084(12) & 1.06\\
\cline{2-4}
 & $\hat{f}_{\rm v}$ & 0.1917(66) & 1.37\\
\cline{2-4}
 & $\hat{f}_{\rm av}$ & 0.254(17) & 1.67\\
\cline{2-4}
 & $\hat{m}_{\rm v}$ & 0.7459(90) & 1.21\\
\cline{2-4}
 & $\hat{m}_{\rm av}$ & 1.270(63) & 1.33\\
\cline{2-4}
 & $\hat{m}_{\rm s}$ & 1.129(65) & 1.71\\
\cline{2-4}
 & $\hat{m}_{\rm t}$ & 0.774(14) & 1.95\\
\cline{2-4}
 & $\hat{m}_{\rm at}$ & 1.408(75) & 1.97\\
\hline
\hline
\multirow{8}{*}{Symmetric} & $\hat{f}_{\mathcal{PS}}$ & 0.1535(19) & 2.42\\
\cline{2-4}
 & $\hat{f}_{\mathcal{V}}$ & 0.276(12) & 1.56\\
\cline{2-4}
 & $\hat{f}_{\mathcal{AV}}$ & 0.406(20) & 2.47\\
\cline{2-4}
 & $\hat{m}_{\mathcal{V}}$ & 0.881(11) & 1.29\\
\cline{2-4}
 & $\hat{m}_{\mathcal{AV}}$ & 1.460(71) & 2.07\\
\cline{2-4}
 & $\hat{m}_{\mathcal{S}}$ & 1.284(54) & \textcolor{red}{3.06}\\
\cline{2-4}
 & $\hat{m}_{\mathcal{T}}$ & 0.902(16) & \textcolor{red}{3.07}\\
\cline{2-4}
 & $\hat{m}_{\mathcal{AT}}$ & 2.077(99) & \textcolor{red}{3.38}\\
\hline
\end{tabular}

\end{table}

\begin{table}[t]
\caption[$Sp(6)$ chiral extrapolations]{Massless and continuum extrapolation of the decay constants, $\hat{f}$, and masses, $\hat{m}$, expressed in units of the gradient flow scale, $w_0$, for quenched mesons in the $Sp(6)$ theory, for the three representations considered in this study. The uncertainties reported are determined by the extrapolation, starting from the statistical uncertainties of the individual measurements. 
\label{tab:Sp6ChiralExtrap}\\ }
\centering
\begin{tabular}{|c|c|c|c|}
\hline
\multicolumn{4}{|c|}{$Sp(6)$}\\
\hline
Representation & Channel & Chiral limit & $\chi^2/{\rm d.o.f.}$\\
\hline
\multirow{8}{*}{Fundamental} & $\hat{f}_{\rm PS}$ & 0.1073(20) & 1.56\\
\cline{2-4}
 & $\hat{f}_{\rm V}$ & 0.1922(67) & 1.84\\
\cline{2-4}
 & $\hat{f}_{\rm AV}$ & 0.235(17) & 1.78\\
\cline{2-4}
 & $\hat{m}_{\rm V}$ & 0.5890(86) & 1.71\\
\cline{2-4}
 & $\hat{m}_{\rm AV}$ & 1.062(55) & 1.59\\
\cline{2-4}
 & $\hat{m}_{\rm S}$ & 0.846(64) & 1.38\\
\cline{2-4}
 & $\hat{m}_{\rm T}$ & 0.610(13) & 1.11\\
\cline{2-4}
 & $\hat{m}_{\rm AT}$ & 1.090(68) & 2.06\\
\hline
\hline
\multirow{8}{*}{Antisymmetric} & $\hat{f}_{\rm ps}$ & 0.1940(32) & 2.89\\
\cline{2-4}
 & $\hat{f}_{\rm v}$ & 0.353(18) & 1.90\\
\cline{2-4}
 & $\hat{f}_{\rm av}$ & 0.267(24) & 1.26\\
\cline{2-4}
 & $\hat{m}_{\rm v}$ & 0.782(12) & 1.47\\
\cline{2-4}
 & $\hat{m}_{\rm av}$ & 1.026(83) & 1.13\\
\cline{2-4}
 & $\hat{m}_{\rm s}$ & 0.897(68) & 1.00\\
\cline{2-4}
 & $\hat{m}_{\rm t}$ & 0.779(19) & 1.24\\
\cline{2-4}
 & $\hat{m}_{\rm at}$ & 1.468(94) & 1.67\\
\hline
\hline
\multirow{8}{*}{Symmetric} & $\hat{f}_{\mathcal{PS}}$ & 0.2142(51) & 2.72\\
\cline{2-4}
 & $\hat{f}_{\mathcal{V}}$ & 0.476(19) & 2.28\\
\cline{2-4}
 & $\hat{f}_{\mathcal{AV}}$ & 0.426(39) & 1.97\\
\cline{2-4}
 & $\hat{m}_{\mathcal{V}}$ & 0.912(10) & 1.56\\
\cline{2-4}
 & $\hat{m}_{\mathcal{AV}}$ & 1.027(93) & 1.42\\
\cline{2-4}
 & $\hat{m}_{\mathcal{S}}$ & 0.673(62) & 0.59\\
\cline{2-4}
 & $\hat{m}_{\mathcal{T}}$ & 0.893(19) & 2.08\\
\cline{2-4}
 & $\hat{m}_{\mathcal{AT}}$ & 1.61(14) & 1.07\\
\hline
\end{tabular}

\end{table}

\begin{table}[t]
\caption[$Sp(8)$ chiral extrapolations]{Massless and continuum extrapolation of the decay constants, $\hat{f}$, and masses, $\hat{m}$, expressed in units of the gradient flow scale, $w_0$, for quenched mesons in the $Sp(8)$ theory, for the three representations considered in this study. The uncertainties reported are determined by the extrapolation, starting from the statistical uncertainties of the individual measurements. Reduced chi-squared values, $\chi^2/N_{\rm d.o.f.}$, that are greater than $3.0$ are highlighted in red (e.g., \captionchisquarespeight).
\label{tab:Sp8ChiralExtrap}\\ }
\centering
\begin{tabular}{|c|c|c|c|}
\hline
\multicolumn{4}{|c|}{$Sp(8)$}\\
\hline
Representation & Channel & Chiral limit & $\chi^2/{\rm d.o.f.}$\\
\hline
\multirow{8}{*}{Fundamental} & $\hat{f}_{\rm PS}$ & 0.1117(20) & 0.48\\
\cline{2-4}
 & $\hat{f}_{\rm V}$ & 0.1921(80) & 0.68\\
\cline{2-4}
 & $\hat{f}_{\rm AV}$ & 0.245(19) & 1.69\\
\cline{2-4}
 & $\hat{m}_{\rm V}$ & 0.5782(66) & 0.74\\
\cline{2-4}
 & $\hat{m}_{\rm AV}$ & 0.993(56) & 1.38\\
\cline{2-4}
 & $\hat{m}_{\rm S}$ & 0.856(52) & 0.72\\
\cline{2-4}
 & $\hat{m}_{\rm T}$ & 0.573(11) & 1.23\\
\cline{2-4}
 & $\hat{m}_{\rm AT}$ & 0.963(65) & 0.82\\
\hline
\hline
\multirow{8}{*}{Antisymmetric} & $\hat{f}_{\rm ps}$ & 0.2152(43) & 1.85\\
\cline{2-4}
 & $\hat{f}_{\rm v}$ & 0.434(19) & 1.20\\
\cline{2-4}
 & $\hat{f}_{\rm av}$ & $\cdots$ & $\cdots$ \\
\cline{2-4}
 & $\hat{m}_{\rm v}$ & 0.7955(63) & 1.23\\
\cline{2-4}
 & $\hat{m}_{\rm av}$ & $\cdots$ & $\cdots$ \\
\cline{2-4}
 & $\hat{m}_{\rm s}$ & $\cdots$ & $\cdots$ \\
\cline{2-4}
 & $\hat{m}_{\rm t}$ & 0.8085(88) & 0.94\\
\cline{2-4}
 & $\hat{m}_{\rm at}$ & $\cdots$ & $\cdots$ \\
\hline
\hline
\multirow{8}{*}{Symmetric} & $\hat{f}_{\mathcal{PS}}$ & 0.2380(64) & 1.80\\
\cline{2-4}
 & $\hat{f}_{\mathcal{V}}$ & 0.677(15) & 1.63\\
\cline{2-4}
 & $\hat{f}_{\mathcal{AV}}$ & $\cdots$ & $\cdots$ \\
\cline{2-4}
 & $\hat{m}_{\mathcal{V}}$ & 0.9513(53) & \textcolor{red}{3.02}\\
\cline{2-4}
 & $\hat{m}_{\mathcal{AV}}$ & $\cdots$ & $\cdots$ \\
\cline{2-4}
 & $\hat{m}_{\mathcal{S}}$ & $\cdots$ & $\cdots$ \\
\cline{2-4}
 & $\hat{m}_{\mathcal{T}}$ & 0.9608(72) & 1.81\\
\cline{2-4}
 & $\hat{m}_{\mathcal{AT}}$ & $\cdots$ & $\cdots$ \\
\hline
\end{tabular}

\end{table}

\begin{table}[t]
\caption[$Sp(\infty)$ chiral extrapolations]{Massless and continuum extrapolation of the decay constants, $\hat{f}$, and masses, $\hat{m}$, expressed in units of the gradient flow scale, $w_0$, for quenched mesons, extrapolated to the $Sp(\infty)$ theory, for the three representations considered in this study. The uncertainties reported are determined by the extrapolation, starting from the statistical uncertainties of the individual measurements. Reduced chi-squared values, $\chi^2/N_{\rm d.o.f.}$, that are greater than $3.0$ are highlighted in red (e.g., \captionchisquarelargeN). When the large-$N$ extrapolation has been performed with only two data points, this has been done by simple error propagation, solving for the coefficients of a linear extrapolation.  Only statistical uncertainties have been included.
A few extrapolations for the heaviest states made of fermions in large representations result 
 in negative values of mass squared, due to the existence of large systematic errors affecting these few observables. 
The large-$N$ extrapolation for (as) and (s) fermions do not agree, except for the pseudoscalar decay constant, as discussed in the main body of the paper.
\label{tab:largeNChiralExtrap}\\}
\centering
\begin{tabular}{|c|c|c|c|}
\hline
\multicolumn{4}{|c|}{$Sp(\infty)$}\\
\hline
Representation & Channel & Chiral limit & $\chi^2/{\rm d.o.f.}$\\
\hline
\multirow{8}{*}{Fundamental} & $\hat{f}^2_{\rm PS}/N_c$ & 0.01913(77) & \textcolor{red}{3.29}\\
\cline{2-4}
 & $\hat{f}^2_{\rm V}/N_c$ & 0.0523(50) & 1.31\\
\cline{2-4}
 & $\hat{f}^2_{\rm AV}/N_c$ & 0.073(16) & 0.00\\
\cline{2-4}
 & $\hat{m}^2_{\rm V}$ & 0.307(15) & 0.08\\
\cline{2-4}
 & $\hat{m}^2_{\rm AV}$ & 0.84(22) & 0.29\\
\cline{2-4}
 & $\hat{m}^2_{\rm S}$ & 0.28(18) & 1.25\\
\cline{2-4}
 & $\hat{m}^2_{\rm T}$ & 0.315(25) & \textcolor{red}{3.36}\\
\cline{2-4}
 & $\hat{m}^2_{\rm AT}$ & 0.62(26) & 0.58\\
\hline
\hline
\multirow{8}{*}{Antisymmetric} & $\hat{f}^2_{\rm ps}/N_c^2$ & 0.0851(26) & 2.68\\
\cline{2-4}
 & $\hat{f}^2_{\rm v}/N_c^2$ & 0.323(26) & 0.60\\
\cline{2-4}
 & $\hat{f}^2_{\rm av}/N_c^2$ & 0.085(43) & $\cdots$\\
\cline{2-4}
 & $\hat{m}^2_{\rm v}$ & 0.710(24) & 0.04\\
\cline{2-4}
 & $\hat{m}^2_{\rm av}$ & -0.07(60) & $\cdots$\\
\cline{2-4}
 & $\hat{m}^2_{\rm s}$ & -0.14(47) & $\cdots$\\
\cline{2-4}
 & $\hat{m}^2_{\rm t}$ & 0.705(36) & 0.83\\
\cline{2-4}
 & $\hat{m}^2_{\rm at}$ & 2.50(93) & $\cdots$\\
\hline
\hline
\multirow{8}{*}{Symmetric} & $\hat{f}^2_{\mathcal{PS}}/N_c^2$ & 0.0901(45) & 0.01\\
\cline{2-4}
 & $\hat{f}^2_{\mathcal{V}}/N_c^2$ & 0.730(35) & \textcolor{red}{20.18}\\
\cline{2-4}
 & $\hat{f}^2_{\mathcal{AV}}/N_c^2$ & 0.22(11) & $\cdots$\\
\cline{2-4}
 & $\hat{m}^2_{\mathcal{V}}$ & 1.033(28) & 1.93\\
\cline{2-4}
 & $\hat{m}^2_{\mathcal{AV}}$ & -1.10(71) & $\cdots$\\
\cline{2-4}
 & $\hat{m}^2_{\mathcal{S}}$ & -1.94(37) & $\cdots$\\
\cline{2-4}
 & $\hat{m}^2_{\mathcal{T}}$ & 1.033(40) & \textcolor{red}{5.77}\\
\cline{2-4}
 & $\hat{m}^2_{\mathcal{AT}}$ & -0.8(.6) & $\cdots$\\
\hline
\end{tabular}

\end{table}

The zero-momentum 2-point correlation function of operators $\calO_M$ and $\calO_{M'}$ is defined as
\begin{equation}
    C_{M,M'}(t)\equiv \sum_{\Vec{x}}\mel{0}{\calO_M(\Vec{x},t)\calO^{\dagger}_{M'}(\Vec{0},0)}{0}\,.
\end{equation}
We set $M=M'$, and examine the large-time behaviour of the correlation function,
that we approximate as follows:
\begin{eqnarray}
  C_{M,M}&&(t\to\infty)  \simeq\\
   &&\frac{|\mel{0}{\calO_M}{M}|^2}{2m_M}\left(e^{-m_Mt}+e^{-m_M(T-t)}\right)\,,\nonumber
\end{eqnarray}
having ignored contamination from states other than the lightest one.
In the case of S, T and AT channels, we measure only the mass of the ground state composite particle. In the other three cases (PS, V and AV), we extract also the decay constant, besides the mass. To do so, for $V$ and $AV$ channels 
we exploit the fact that the matrix elements obey the following relations:
\begin{eqnarray}
    \mel{0}{\calO^{\Edit{\mu}}_{\rm V}}{\rm V}&=&f_{\rm V}m_{\rm V}\epsilon^{\mu}\\
    \mel{0}{\calO^{\Edit{\mu}}_{\rm AV}}{\rm AV}&=&f_{\rm AV}m_{\rm AV}\epsilon^{\mu}
\end{eqnarray}
where $\epsilon^{\mu}$ is the polarisation vector, normalised so that $\epsilon^{\mu}\epsilon_{\mu}=1$.

For the decay constant of the pseudoscalar mesons, we use one additional correlation function:
\begin{equation}
    C_{\rm AV,PS}(t)=\sum_{\Vec{x}}\mel{0}{\calO_{\rm AV}(\Vec{x},t)\calO^{\dagger}_{\rm PS}(\Vec{0},0)}{0}\,.
\end{equation}
Its large-time behaviour is expected to be described as
\begin{eqnarray}
    C_{\rm AV,PS}&&(t\to\infty) \simeq\\
&&\frac{f_{\rm PS}\mel{0}{\calO_{PS}}{PS}^*}{2m_{PS}}\left(e^{-m_{PS}t}-e^{-m_{\rm PS}(T-t)}\right)\,.\nonumber
\end{eqnarray}
The normalisations are chosen so that the corresponding 
decay constant in QCD is $f_{\pi}\simeq 93$ MeV.

From the large-time behaviour of all these correlation functions, we can hence measure nine observables, for each of the three types of fermions, and for each of the three gauge groups.

\begin{figure}[t]
    \centering
    \includegraphics[width=0.45\textwidth]{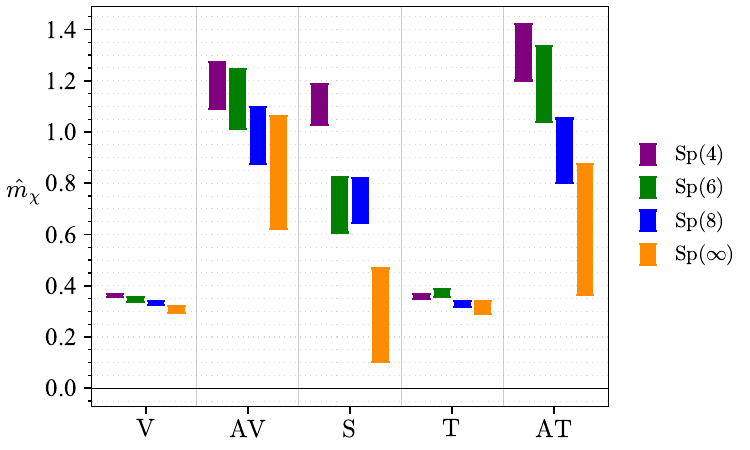}
    \vskip0.1cm
    \includegraphics[width=0.45\textwidth]{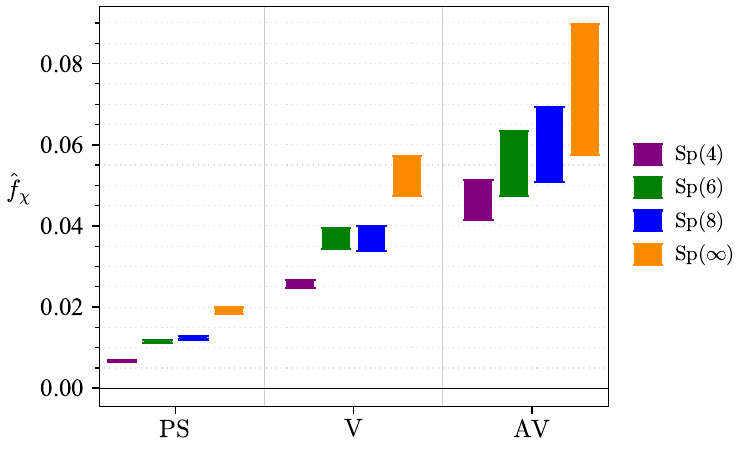}
    \caption[Fundamental]{Masses (top) and decay constants (bottom) squared of mesons in the $Sp(N_c)$ theory with
    (quenched)  matter consisting of  fermions transforming in the fundamental representation, $(\rm f)$, extrapolated to the massless and continuum limits, expressed in units of the gradient flow scale, $w_0$,
    computed for $N_c = 4,\, 6,\, 8$, and 
 further extrapolated to  $N_c\rightarrow \infty$.}
    \label{fig:Fundamental}
\end{figure}

\begin{figure}[t]
    \centering
        \includegraphics[width=0.5\textwidth]{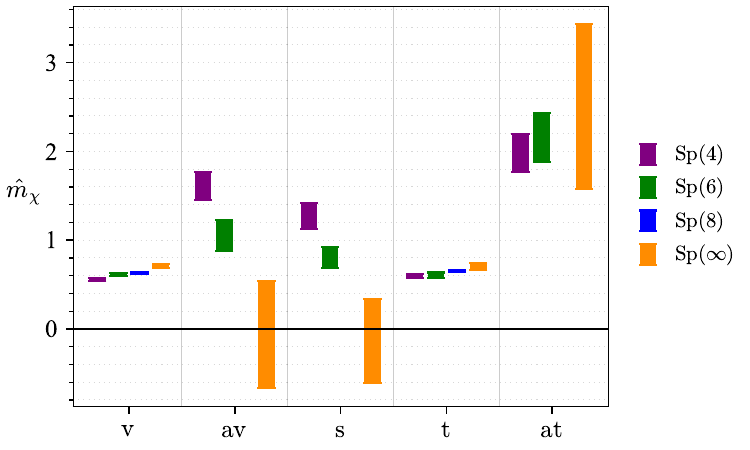}
         \vskip0.1cm
    \includegraphics[width=0.5\textwidth]{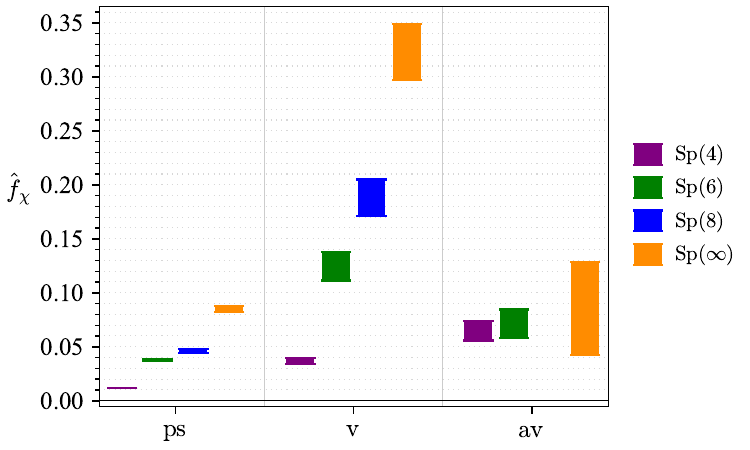}
    \caption[Antisymmetric]{
    Masses (top) and decay constants (bottom) squared of mesons in the $Sp(N_c)$ theory with
    (quenched)  matter consisting of  fermions transforming in the 2-index antisymmetric representation, $(\rm as)$, extrapolated to the massless and continuum limits, expressed in units of the gradient flow scale, $w_0$,
    computed for $N_c = 4,\, 6,\, 8$, and 
 further extrapolated to  $N_c\rightarrow \infty$.
}
    \label{fig:Antisymmetric}
\end{figure}

\begin{figure}[t]
    \centering
    \includegraphics[width=0.5\textwidth]{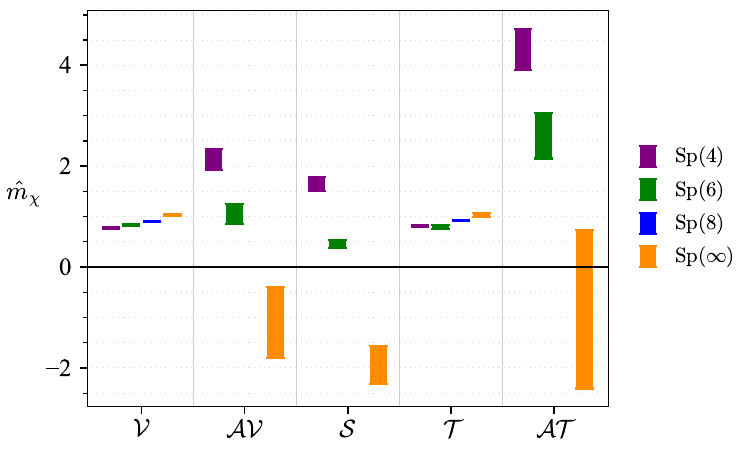}
          \vskip0.1cm   
        \includegraphics[width=0.5\textwidth]{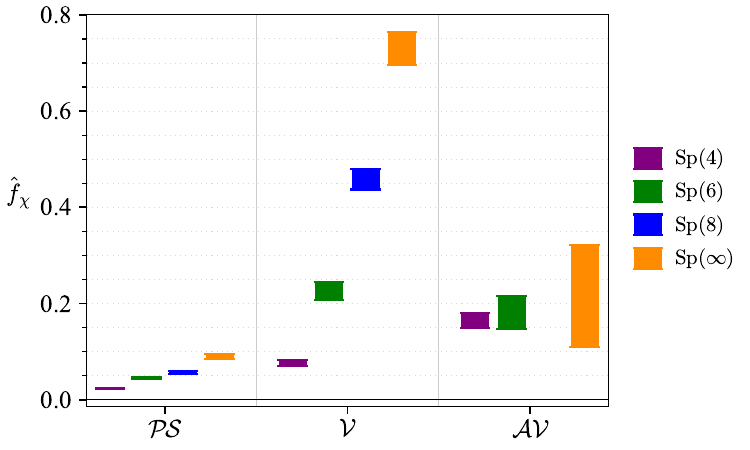}
    \caption[Symmetric]{    Masses (top) and decay constants (bottom) squared of mesons in the $Sp(N_c)$ theory with
    (quenched)  matter consisting of  fermions transforming in the 2-index symmetric representation, $(\rm s)$, extrapolated to the massless and continuum limits, expressed in units of the gradient flow scale, $w_0$,
    computed for $N_c = 4,\, 6,\, 8$, and 
 further extrapolated to  $N_c\rightarrow \infty$.
}
    \label{fig:Symmetric}
\end{figure}

\subsection{Scale setting}\label{sec:scaleSetting}

We  adopt a scale-setting procedure that is especially suited to studies
of novel strongly coupled models, and is  based on the gradient flow and its
lattice implementation, the Wilson flow~\cite{Luscher:2010iy,Luscher:2013vga}.
We follow the same process outlined in Ref.~\cite{Bennett:2022ftz}, in the context of the $Sp(2N)$ 
lattice programme, and report here only 
basic definitions necessary to fix the notation in the following.

The gradient flow for gauge fields, $B_{\mu}(\bbt, x)$, is defined by solving  in five space-time dimensions
 the differential equation
\begin{equation}
    \frac{\mathrm{d} B_{\mu}(\bbt,x)}{\mathrm{d}\bbt}=
    D_{\nu}F_{\mu\nu}(\bbt,x)\,,~~~B_{\mu}(0,x)=A_{\mu}(x)\,,
\end{equation}
where $\bbt$ is known as flow-time, $D_{\mu}\equiv \partial_{\mu}+[B_{\mu},\cdot]$
and $F_{\mu\nu}(\bbt,\,x)\equiv [D_\mu,\,D_\nu]$.
The flow defined by the above equation drives the configuration 
$A_\mu(x)$ at $\bbt=0$ of the gauge fields towards 
a stationary point of the continuum Yang-Mills action. It is possible 
to show that, at leading order in the gauge coupling, it implements a Gaussian 
smoothening of the field over a region of mean-square radius $\sqrt{8\bbt}$. A
renormalized coupling, $\alpha$, can then be defined at this scale as follows
\begin{equation}
\alpha(\mu^{-1}=\sqrt{8\bbt}) \equiv k_\alpha \bbt^2 \langle E(\bbt) \rangle \equiv
k_\alpha \mathcal{E}(\bbt)\,,
\end{equation}
where $E(\bbt)=\tfrac{1}{4} F_{\mu\nu}(\bbt)F^{\mu\nu}(\bbt)$, and
$k_\alpha$ is a numerical coefficient that can be extracted from perturbation theory~\cite{Luscher:2011bx}.
A reference scale $\bbt_0$ can be defined implicitly 
as follows:
\begin{equation}
\calE(\bbt_0) = 
\left.\bbt^2\langle E(\bbt)\rangle\right|_{\bbt=\bbt_0} = \calE_0\,,
\end{equation}
with a conventional choice of $\calE_0$. Alternatively,
one can define the related quantity 
\begin{equation}
\mathcal{W}(\bbt) = \bbt\frac{\rm d}{\rm d \bbt} \calE(\bbt)\,,
\end{equation}
and the scale $w_0$ as
\begin{equation}
\calW(\bbt=w_0^2) =\calW_0\,,
\end{equation}
for an appropriate, conventional choice of $\calW_0$~\cite{BMW:2012hcm}. 

On the lattice, the \emph{Wilson} flow $V_\mu(\bbt)$ is defined 
by solving the differential system:
\begin{eqnarray}
    \frac{\dee V_{\mu}(\bbt,x)}{\rm d \bbt}&=&
    -g_0^2(\partial_{x,\mu}S^{\rm flow}[V_{\mu}])V(\bbt,x)\,,\\
    V_{\mu}(0,x)&=&U_{\mu}(x)\,,
\end{eqnarray}
where $S^{\rm flow}[V_{\mu}]$ is the Wilson action. The configurations
$U_\mu(x)$ in a given ensemble are used as initial conditions for the system,  and the 
flow is obtained by numerical integration. The lattice 
observables are then computed with the resulting, finite flow-time, smoothened configurations. 
In order to compute the Wilson flow scale, $w_0$, on the lattice, we adopted the
clover-leaf discretization for $E(\bbt)$. We follow the strategy
described in detail in Ref.~\cite{Bennett:2022ftz} in order to fix 
 reference values for $\calW_0$ for different choices of $N$.  We summarise 
  in Table~\ref{tab:latticeDetails} the resulting value of 
  $1/\hat{a}\equiv w_0 / a$ thus obtained.
  In the following, we adopt thehatted notation to present dimensional
 quantities in units of the gradient flow scale, i.e. $\hat{m}=m w_0$, for a generic mass $m$.

\subsection{Continuum and massless extrapolation}

As we discussed in the introduction to this paper, the quenched approximation is expected to yield reasonably good estimates for observable quantities when the fermion contribution to the dynamics is small.
This is the case for moderately large fermion masses, but also when the number of $(f)$-type fermions is small while the number of colours is large. 
We extrapolate our numerical lattice data to the continuum and massless limit simultaneously.
As we look at comparatively large groups, such as $Sp(8)$, it is also interesting to perform extrapolations to the large-$N$ limit as well.
Yet, before proceeding to describe our analysis, we alert the reader to use  some caution when using the results of such extrapolations in phenomenological applications, in view of the systematic uncertainty intrinsic in the quenched approximation.

Having set the scale using the Wilson flow, we 
follow a procedure inspired by W$\chi$PT prescription~\cite{Sheikholeslami:1985ij,Rupak:2002sm}, truncated at the next-to-leading 
order---see also Refs.~\cite{Bennett:2019jzz,Bennett:2019cxd} for earlier implementations of this strategy in $Sp(2N)$ theories.
  The same formal expression holds for  masses and decay constants:
\begin{eqnarray}\label{eq.ContChiExtrapolation}
    \hat{m}^{2,\rm NLO}_{\rm M}&=&\hat{m}^{2,\c}_{\rm M}(1+L^0_{m,\rm M}\hat{m}^2_{\rm PS})+W^0_{m,\rm M}\hat{a}\,,\\
    \hat{f}^{2,\rm NLO}_{\rm M}&=&\hat{f}^{2,\c}_{\rm M}(1+L^0_{f,\rm M}\hat{m}^2_{\rm PS})+W^0_{f,\rm M}\hat{a}\,,
\end{eqnarray}
where the superscript  $\c$  denotes a quantity in the massless limit, with $1/\hat{a}\equiv w_0/a$, and $\hat{m}_{\rm PS}$  the mass of 
 pseudoscalar meson in units of the gradient flow scale (in the appropriate representation of $Sp(2N)$).
The coefficients appearing on the right-hand side of these relations are extracted by fitting numerical results obtained with 
different values of  lattice coupling, $\beta$, and  fermion masses.

\section{Summary of Results}
\label{Sec:results}

\begin{figure}[t]
    \centering
    \begin{tabular}{c}
        \includegraphics[width=75mm]{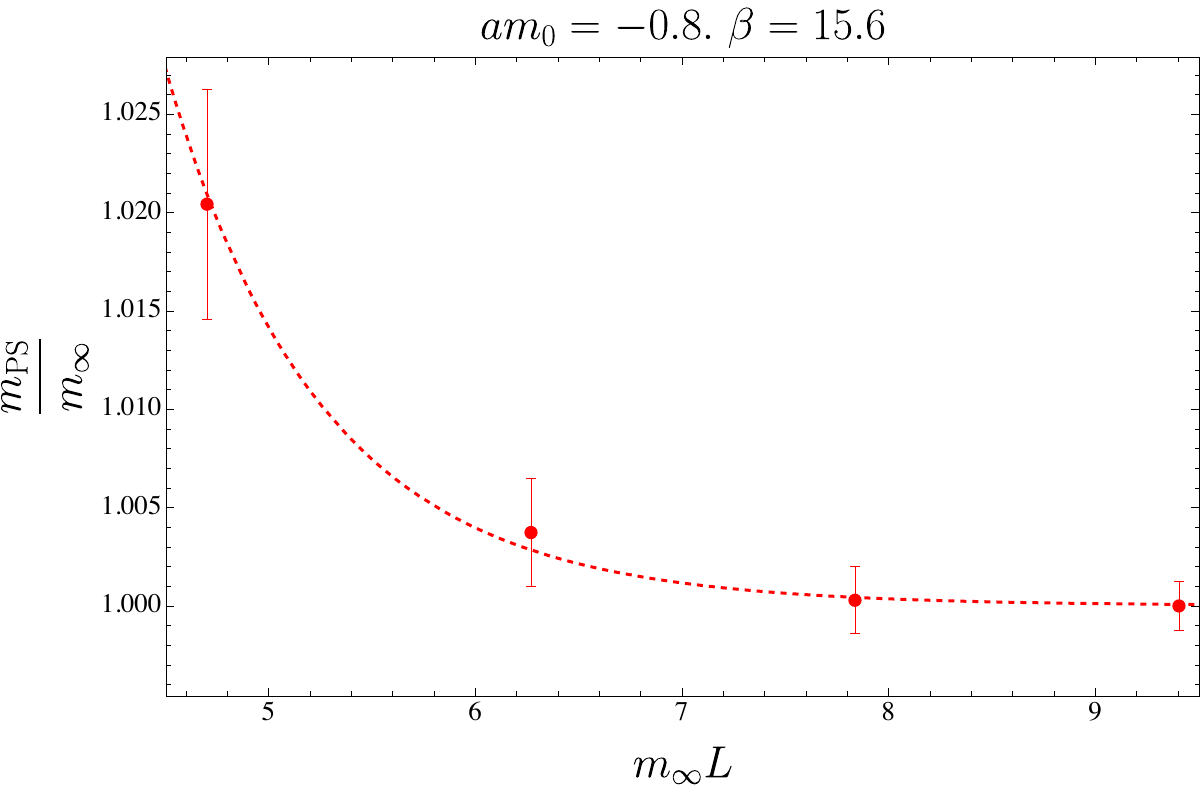}\\ \includegraphics[width=75mm]{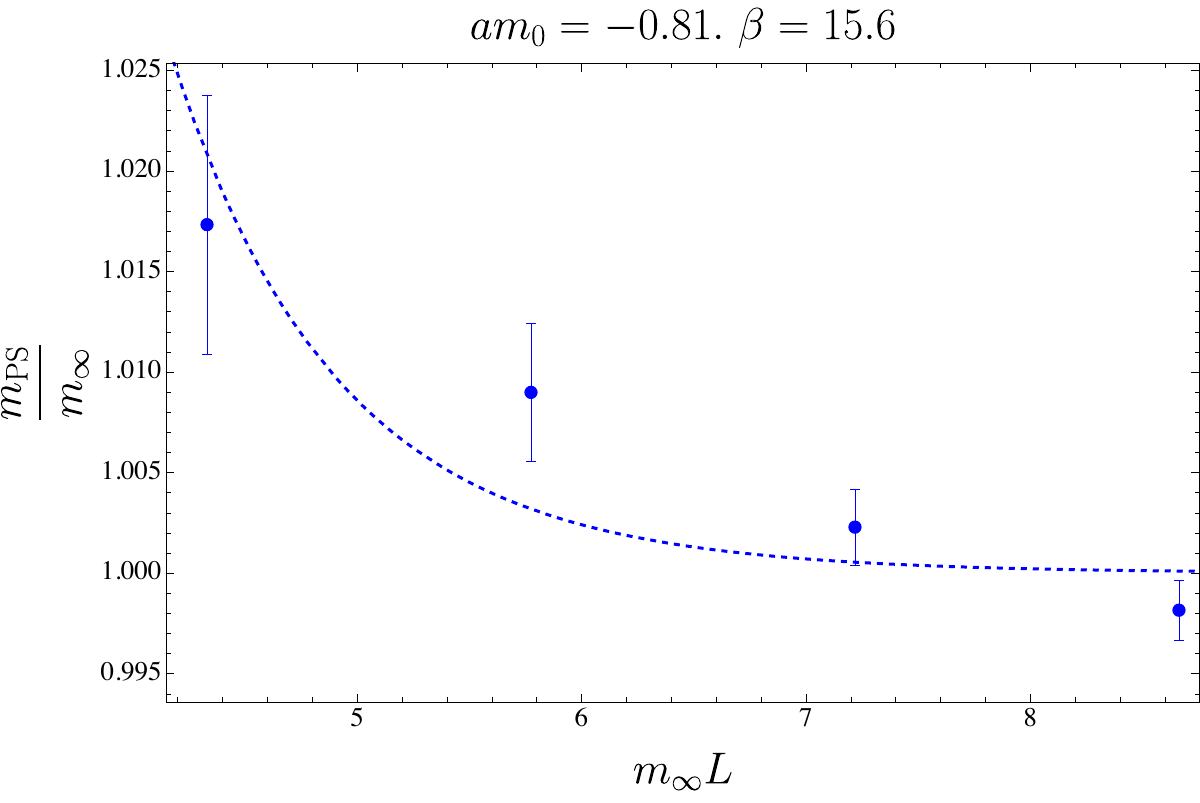}
    \end{tabular}
    \caption{
    Masses  of $\rm PS$ mesons made of $(\rm f)$ fermions in the $Sp(6)$ theory, plotted as a function of $m_{\rm PS} L$, 
    where $L=N_s a$ is the extent of the spatial lattice direction, for two representative choices of fermion mass. 
    We normalise the mass to its infinite volume extrapolation ($m_{\rm PS}/m_{\infty}$). 
    The dashed line is the  prediction based on the infinite volume formula: $1+A\frac{e^{-m_{\infty}L}}{(m_{\infty}L)^{3/2}}$,
    where $A$ and $m_{\infty}$ are fitting parameters. 
\label{Fig:FV0}
}
\label{fig:sp6FFiniteSize}
\end{figure}

In this section, we display, summarise, and critically discuss our final results, extrapolated to the massless and continuum limits. 
Details about the intermediate results can be found in the Appendix,
and in the data release~\cite{datarelease}. Given the correlator, $C(t)$, we can extract the effective mass accounting for both forward- and backward-propagations in Euclidean time, $t$, defining the effective mass as
\begin{equation}\label{eq.mEff}
    m_{\rm eff}(t)={\rm arccosh}\left[\frac{C(t+a)+C(t-a)}{2C(t)}\right].
\end{equation}
We include in the analysis  only
numerical results obtained from ensembles for which we found unambiguous evidence of a clear plateau 
in the effective-mass plot. 

We restricted attention to cases in which finite-volume effects
 are smaller than the statistical uncertainties---see Appendix~\ref{Sec:FV} and Fig.~\ref{Fig:FV0}.
Our results for the continuum, massless extrapolations
are listed in Tables~\ref{tab:Sp4ChiralExtrap}, \ref{tab:Sp6ChiralExtrap}, and~\ref{tab:Sp8ChiralExtrap}.
The masses and decay constants of mesons made of (quenched) fermions of type
$(\rm f)$, $(\rm as)$, and $(\rm s)$, respectively, are displayed in 
Figs.~\ref{fig:Fundamental},~\ref{fig:Antisymmetric}~and~\ref{fig:Symmetric}. 
All these plots show the $1\sigma$-equivalent best-fit ranges,
obtained by bootstrapping the statistical error through the maximum likelihood process based upon W$\chi$PT. 
We report the mass of the lowest excitation in the V, AV, S, T, and AT channels,
and the decay constants  of the PS, V, and AV lightest states, omitting  few cases in which the measurements are inconclusive.

Before discussing the individual results, we highlight the presence of five major sources of systematic effects in this study.
First and foremost, the calculations use quenched fermions, hence part of the dynamics is not included faithfully in the Monte Carlo process generating the ensembles.
One expects the results to be reasonably accurate in the limit in which the number of fermions is small, or their mass is large. 
Available measurements for the $Sp(4)$ theory with $N_{(\rm f)}=2$ fermions transforming in the fundamental representation 
suggest that the discrepancy might not exceed the level of $10\%\div 25\%$, but this conclusion depends on the observable of interest~\cite{Bennett:2019jzz}.
In order to achieve a better precision, particularly to include large number of fermions---for example, by approaching the lower end of the conformal window---dedicated calculations with dynamical fermions are needed.

In the numerical calculations the fermion mass is  large enough
to kinematically prevent the lightest V meson decay to PS pairs.
On theoretical grounds, we know that the quenched approximation may 
lead to unitarity problems in the low mass region, and hence we avoid it.
Empirically,   we also found that finite-volume effects become severe when we use light  masses in the fermion propagators,
hence we restricted attention to choices for which  $m_{\rm PS}/m_{\rm V}\gsim 0.6$.
The reader should hence exercise some caution in using the results
of next-to-leading order W$\chi$PT to extrapolate to massless and continuum limits.

A third limitation is given by the fact that some of the meson masses are comparatively large, when expressed in lattice units.
We retained in the analysis only ensembles and choices of the fermion masses for which ${m}_{\rm PS} a \ll 1$,
but the masses of the $\rm AV$, $\rm AT$, and $\rm S$ states are often far larger.
This lattice artefact manifests itself as a deterioration of the signal in the effective mass plots,
particularly  in the S and AT channels for the (as) and (s) fermions.

One  way to ameliorate the aforementioned difficulty would be to perform the study on finer lattices, hence raising the 
intrinsic cutoff of the  theory and reducing finite-spacing effects. 
To do so would require the adoption of larger values of the lattice coupling, 
$\beta$. Unfortunately, by doing so autocorrelation grows, thermalisation takes longer, and the 
calculations would become  too expensive to justify within the quenched approximation. Furthermore, this might lead to topological freezing, especially for large groups,  $Sp(6)$ and $Sp(8)$. 

A simple way of visualizing the size
 of finite-spacing effects is to display the measurements of masses and decay constants
of the mesons at finite $\beta$, together with their extrapolations obtained with W$\chi$PT.
We report in the Appendix a catalogue of such plots. The extrapolations 
for the mass of the $\rm V$ and $\rm T$ states are affected by rather large finite-spacing effects.
For the purpose of this paper, of benchmarking the space of $Sp(2N)$ theories
 coupled to matter fermion fields, this is adequate.
Future precision  studies with dynamical fermions will require a more radical approach, possibly involving improving the action.

Finally, we conducted a quite extensive study of the size of finite-volume effects---see Appendix~\ref{Sec:FV},
as well as the example in Fig.~\ref{fig:sp6FFiniteSize}.
Given the comparative simplicity of the dynamics implemented in the ensemble generation, 
we could generate many ensembles, by varying the volume up to
to $\tilde V =60^3 \times 120 \times a^4$, hence ensuring
that this source of systematic effects can be completely ignored.
Interestingly, we found that for finite volume effects to be smaller than  statistical uncertainties
we must use volumes for which $m_{\rm PS} L \gsim 8$ for $Sp(6)$ (as shown in Fig.~\ref{Fig:FV0}), or even $m_{\rm PS} L \gsim 11$ for $Sp(8)$. 
This finding highlights the need to perform dedicated studies of finite volume effects 
in  calculations with 
 dynamical fermions, as
such strong requirements might prove computationally challenging to meet\footnote{Note that finite volume effects can be more severe in the quenched approximation, see Refs.~\cite{Bernard:1995ez, Lin:2002aj}.}.

  \begin{table}[t]
\caption{Numerical results for the sum rules $s_0$, $s_1$, and $s_2$, as defined in the main text,
which include only the lightest bound states.
All results for $Sp(2N)$ use extrapolations  to the massless and continuum limits of the
 (quenched) theories discussed in the body of the paper. The $SU(3)$ case
 is included for comparison, and \Edit{uses numerical values from Ref.~\cite{Erlich:2005qh} and references therein},
 for finite mass of the two  $(\rm f)$ fermions. The uncertainties are computed with simple error propagation, ignoring correlations.
 The $Sp(8)$ case is incomplete, as some measurements are missing, as explained in the main text.\\ 
} 
\label{Tab:s}
\centering
\begin{tabular}{|c|c|c|c|}
\hline
Theory & $s_0$ & $s_1$ & $s_2$ \\
\hline
\hline
$Sp(4)$, (f) & $0.397(75)$  & $-1.07(21)$ & $-4.88(82)$ \\
$Sp(4)$, (as) & $0.33(10)$  & $-1.08(28)$ & $-4.09(94)$ \\
$Sp(4)$, (s) & $0.26(18)$  & $-1.48(31)$ & $-4.95(99)$ \\
\hline
$Sp(6)$, (f) & $0.72(15)$  & $-0.81(25)$ & $-3.88(94)$ \\
$Sp(6)$, (as) & $1.70(35)$  & $0.12(14)$ & $0.01(26)$ \\
$Sp(6)$, (s) & $1.25(63)$  & $-0.00(17)$ & $-0.02(28)$ \\
\hline
$Sp(8)$, (f) & $0.62(19)$  & $-0.96(30)$ & $-3.8(1.0)$ \\
$Sp(8)$, (as) & $\cdots$ & $\cdots$ & $\cdots$ \\
$Sp(8)$, (s) & $\cdots$ & $\cdots$ & $\cdots$ \\
\hline
$Sp(\infty)$, (f) & $1.04(44)$ $N_c$ & $-0.77(35)$ & $-2.8(1.4)$ \\
$Sp(\infty)$, (as) & $0.2(1.5) \times 10^{2}$ $N_c^2$ & $0.47(14)$ & $1.02(22)$ \\
$Sp(\infty)$, (s) & $11.3(2.1)$ $N_c^2$ & $0.58(15)$ & $1.31(26)$ \\
\hline\hline
\parbox[t]{2.7cm}{$SU(3)$, (f) \\ ($m_\pi=139.6\textnormal{ MeV}$)} & $0.298(55)$  & $-0.35(18)$ & $-1.48(44)$ \\
\hline
\end{tabular}

\end{table}

Having discussed the main sources of systematic uncertainty, we can now proceed to comment on our results
for the physical observables, starting from 
 the case of matter transforming  in the fundamental representation.
The top panel of Fig.~\ref{fig:Fundamental} shows that the lightest state is a $\rm V$ meson, corresponding to the $\rho$ meson 
in QCD. The degeneracy between $\rm V$ and $\rm T$ channels agrees with current algebra, 
within the uncertainties,
 for all $Sp(N_c=2N)$ theories considered here.
 All  states in $\rm AV$, $\rm AT$ and $\rm S$ channels are heavier, and affected by sizable errors.
 Their masses, expressed in units of $w_0$,  tend consistently to decrease with $N_c$,
 but appear to converge to a finite result.
 Conversely, the decay constants squared (bottom panel of Fig.~\ref{fig:Fundamental}) grow proportionally to $N_c$, 
 as expected from large-$N_c$ arguments. Even after taking into account  their leading-order $N_c$ behavior, we find residual dependence on $N_c$, as discussed in the following subsection.
 Figs.~\ref{fig:Antisymmetric} and~\ref{fig:Symmetric} display the same information, but for
 mesons made of $(as)$ and $(a)$ fermions. Again, the vector and tensor states are the lightest, and degenerate,
 as expected. The decay constant for mesons made of matter transforming in the 2-index representations
 scale with $N_c^2$.

 \subsection{Towards large $N$}
  \label{Sec:largeN}

 \Cref{fig:Fundamental,fig:Antisymmetric,fig:Symmetric}
   display also the result of extrapolating the numerical results to the large-$N_c$ limit.
 This is performed by assuming that all the squares of the meson masses exhibit the following behavior: 
 \beqs
 \hat{m}_M^2(N_c)&=& \hat{m}_M^2(\infty)+\frac{\Delta \hat{m}_M^2(\infty)}{N_c}\,.
 \eeqs
In the case of the square of the decay constants,
we assume the following relations to hold:
\beqs
  \frac{\hat{f}_M^2(N_c)}{N_c}&=& \frac{\hat{f}_M^2(\infty)}{N_c}+\frac{\Delta \hat{f}_M^2(\infty)}{N^2_c}\,,\\
   \frac{\hat{f}_m^2(N_c)}{N_c^2}&=& \frac{\hat{f}_m^2(\infty)}{N_c^2}+\frac{\Delta \hat{f}_m^2(\infty)}{N_c^3}\,,\\
  \frac{\hat{f}_{\cal M}^2(N_c)}{N_c^2}&=& \frac{\hat{f}_{\cal M}^2(\infty)}{N_c^2}+\frac{\Delta \hat{f}_{\cal M}^2(\infty)}{N^3_c}\,,
 \eeqs
 for mesons constituted of (f), (as), and (s) fermions, respectively.
 As (in most cases) three independent measurements are available, obtained for $Sp(4)$, $Sp(6)$, and $Sp(8)$,
 we  apply a maximum likelihood analysis to extract the two unknown coefficients, 
 and perform the $N_c\rightarrow +\infty$ extrapolations.
 
 In the case of fermions transforming on the antisymmetric and symmetric representations,  mesons and decay constants tend to be larger 
than in the fundamental case, but are affected by bigger uncertainties. We can still verify that the lightest states in the $\rm V$ and $\rm T$ channel are degenerate, as expected,
but in several examples we are not able to measure the mass and decay constant for the $Sp(8)$ case, 
as shown in Table~\ref{tab:Sp8ChiralExtrap}.
 In such occurrences, the large-$N_c$ limit is obtained by simple \Edit{extrapolation} from the two available data points---see Table~\ref{tab:largeNChiralExtrap}.

We do not find agreement in the large-$N$ extrapolations of the properties of mesons made of (as) and (s) fermions, 
with the noticeable exception of the decay constant of the pseudoscalar state.
This fact, combined with the large value of some $\chi^2/N_{\rm d.o.f}$, and  with the fact that for many observables we could
not use $Sp(8)$ results, suggests that the large-$N$ extrapolations for the mesons made of (s) fermions are affected by large systematic uncertainties, and should not be used in phenomenological studies. 
We decided to report these results, despite their poor quality, to illustrate  the fact 
that, in order to study the large-$N$ limit of this type of mesons, a more refined numerical strategy will be needed.
We remind the reader that our main objective in this paper is to benchmark what is achievable within this large class of theories,
hence even such negative result is of some value.
Similar conservative arguments may apply also to the $Sp(8)$ theory with (as) fermions, while for $Sp(4)$ and $Sp(6)$ the measurements performed with (as) fermions yield reasonable results, and
the values of $\chi^2/N_{\rm d.o.f}$ are acceptable.

\begin{figure*}[t]
    \centering
    \begin{tabular}{cc}
        \includegraphics[width=0.4\textwidth]{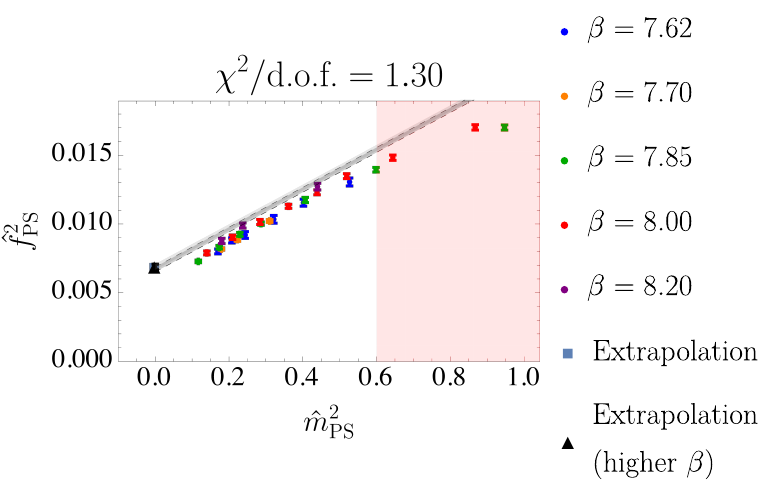} & \includegraphics[width=0.4\textwidth]{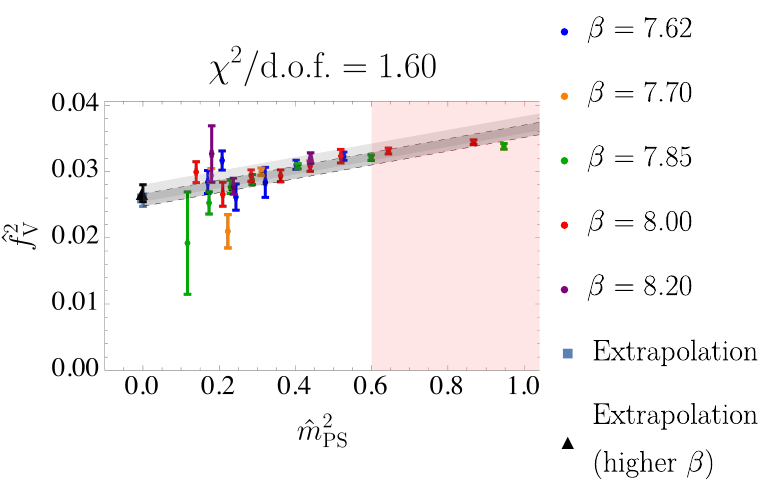}\\
        \multicolumn{2}{c}{\includegraphics[width=0.4\textwidth]{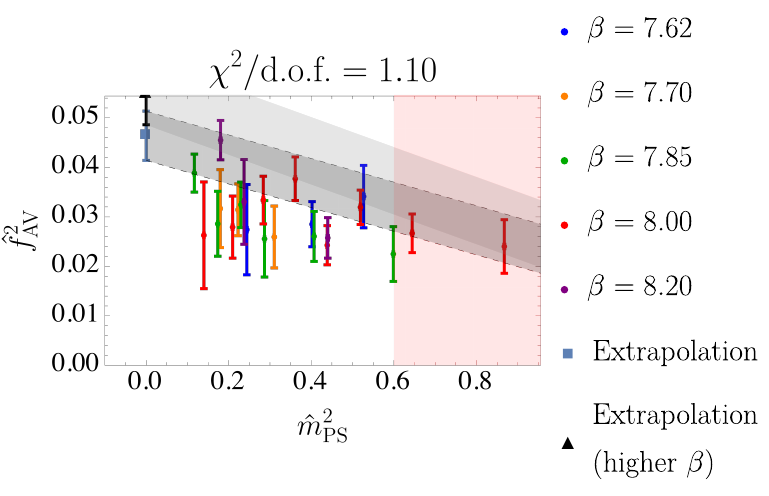}}
    \end{tabular}
    \caption[$Sp(4)$ chiral decay constants for fundamental fermions]{Decay constants squared in the $\rm PS$, $\rm V$, and $\rm AV$ channels comprised of fermions in the fundamental representation of $Sp(4)$. The reduced chi-squared value is printed at the top of each plot. Data points in the pink shaded region are not included in the curve-fitting procedure. The grey band represents the continuum and massless extrapolation, with the blue square being the observable and the vertical width corresponding to the statistical error. In instances where a reliable extrapolation cannot be made, no grey band is shown. All quantities are expressed in units of the gradient flow scale, $w_0$. The extrapolation with the smallest $\beta$ value removed is shown as a lighter grey band and a black triangle in cases where data were available at the smallest $\beta$ value.
\label{fig:sp4Fchiraldecay}}
\end{figure*}

\begin{figure*}[t]
    \centering
    \begin{tabular}{cc}
        \includegraphics[width=0.4\textwidth]{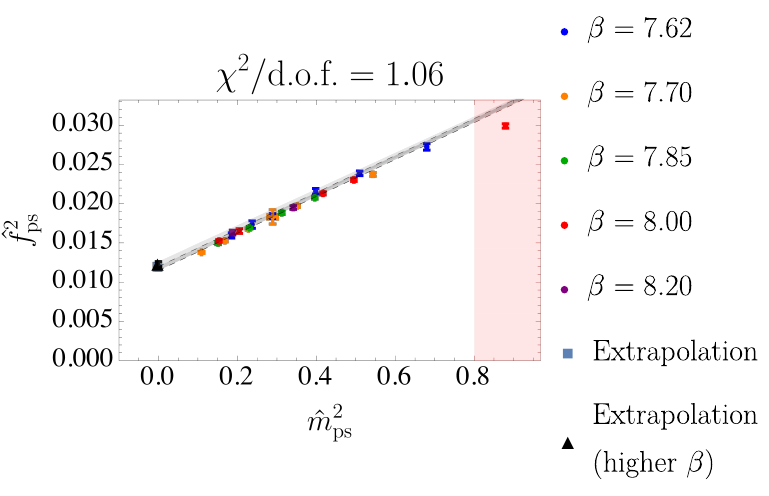} & \includegraphics[width=0.4\textwidth]{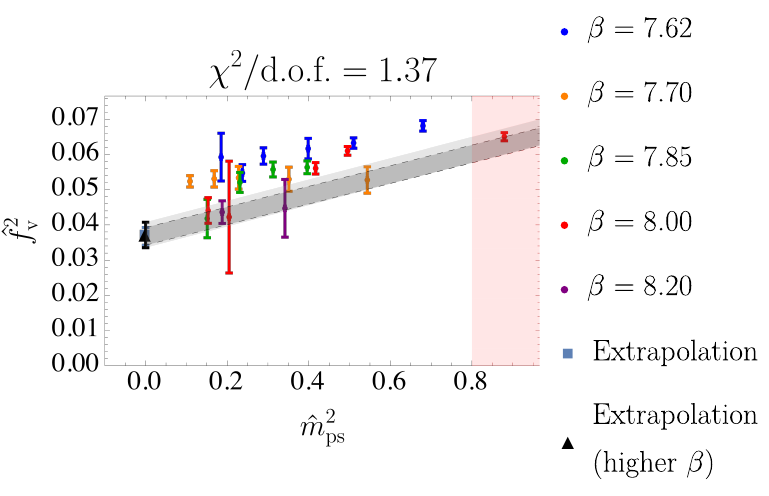}\\
        \multicolumn{2}{c}{\includegraphics[width=0.4\textwidth]{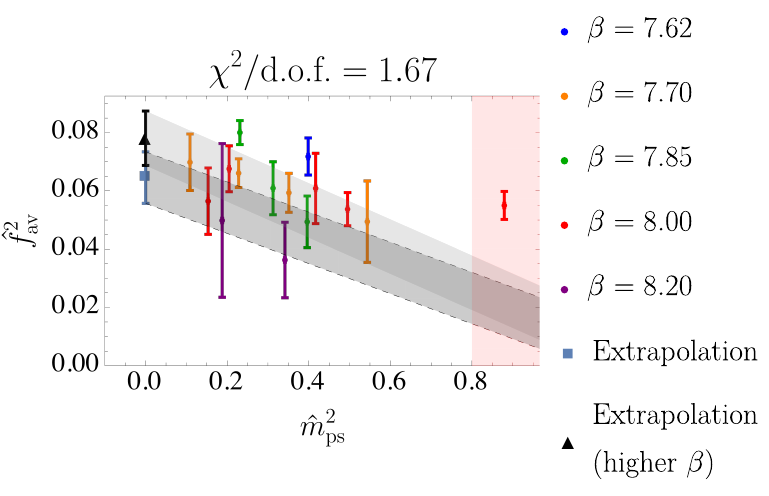}}
    \end{tabular}
    \caption[$Sp(4)$ chiral decay constants for antisymmetric fermions]{Decay constants squared in the $\rm ps$, $\rm v$, and $\rm av$ channels comprised of fermions in the antisymmetric representation of $Sp(4)$. The reduced chi-squared value is printed at the top of each plot. Data points in the pink shaded region are not included in the curve-fitting procedure. The grey band represents the continuum and massless extrapolation with the blue square being the observable  and the vertical width corresponding to the statistical error. In instances where a reliable extrapolation cannot be made, no grey band is shown. All quantities are expressed in units of the gradient flow scale, $w_0$. The extrapolation with the smallest $\beta$ value removed is shown as a lighter grey band and a black triangle in cases where data were available at the smallest $\beta$ value.
\label{fig:sp4ASchiraldecay}}
\end{figure*}

\begin{figure*}[t]
    \centering
    \begin{tabular}{cc}
        \includegraphics[width=0.4\textwidth]{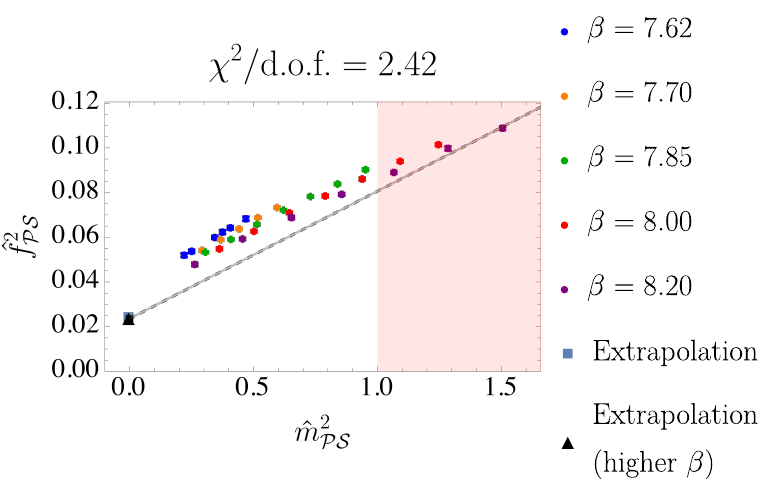} & \includegraphics[width=0.4\textwidth]{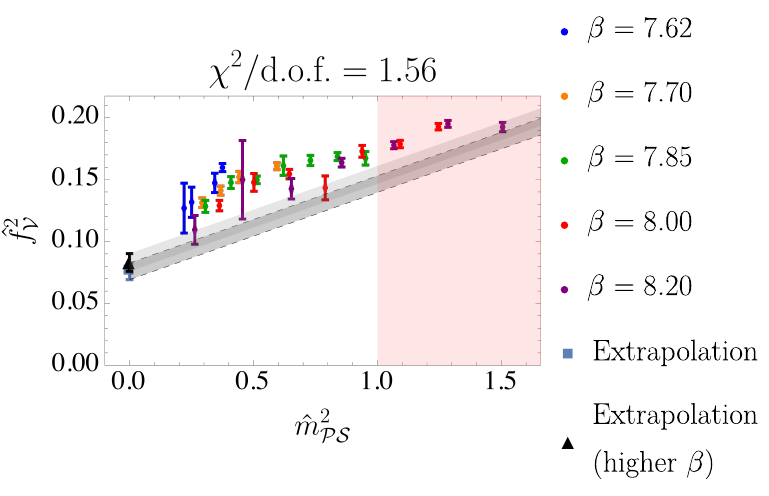}\\
        \multicolumn{2}{c}{\includegraphics[width=0.4\textwidth]{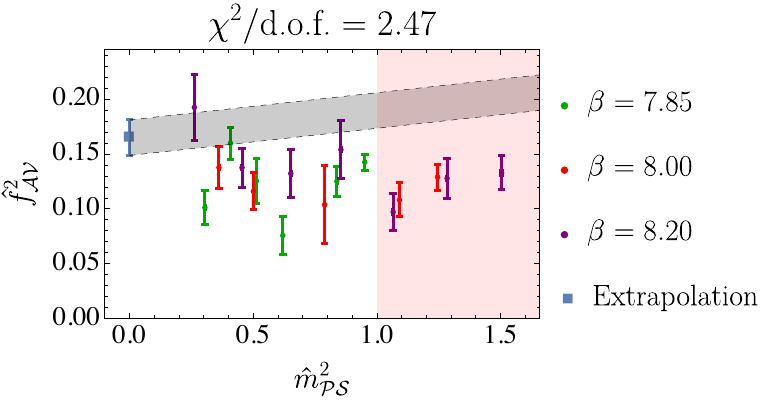}}
    \end{tabular}
    \caption[$Sp(4)$ chiral decay constants for symmetric fermions]{Decay constants squared in the ${\cal PS}$, $\cal V$, and ${\cal AV}$ channels comprised of fermions in the symmetric representation of $Sp(4)$. The reduced chi-squared value is printed at the top of each plot. Data points in the pink shaded region are not included in the curve-fitting procedure. The grey band represents the continuum and massless extrapolation with the blue square being the observable and the vertical width corresponding to the statistical error. In instances where a reliable extrapolation cannot be made, no grey band is shown. All quantities are expressed in units of the gradient flow scale, $w_0$. The extrapolation with the smallest $\beta$ value removed is shown as a lighter grey band and a black triangle in cases where data were available at the smallest $\beta$ value.
\label{fig:sp4Schiraldecay}}
\end{figure*}

\begin{figure*}[t]
    \centering
    \begin{tabular}{cc}
    \includegraphics[width=0.4\textwidth]{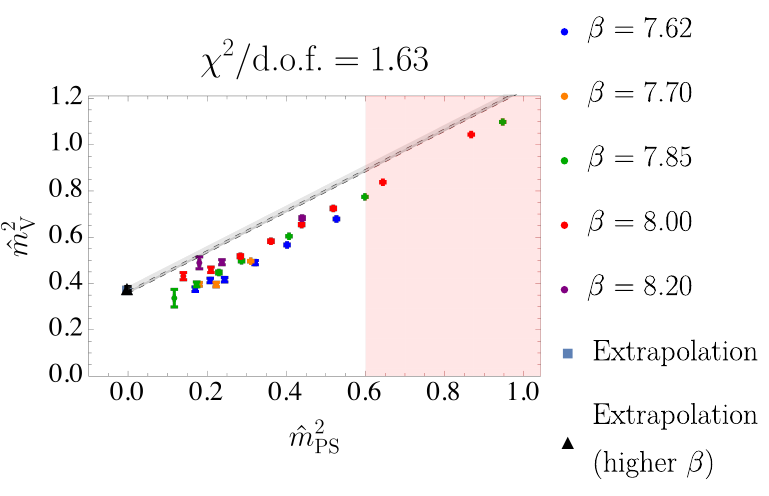} & \includegraphics[width=0.4\textwidth]{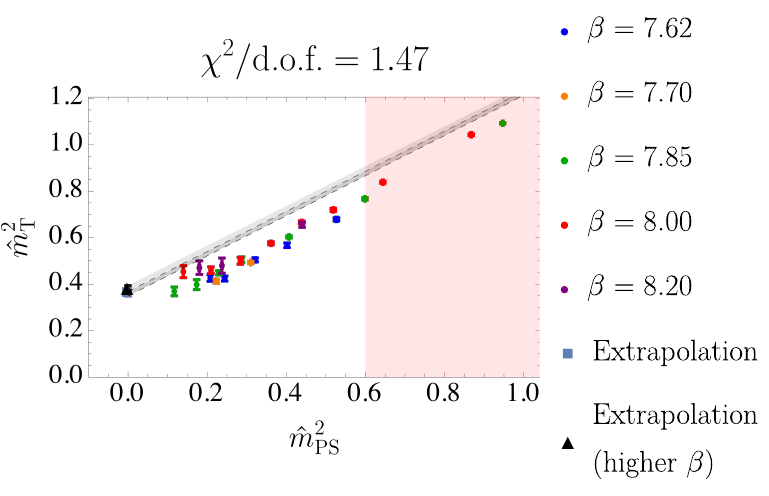}\\
    \includegraphics[width=0.4\textwidth]{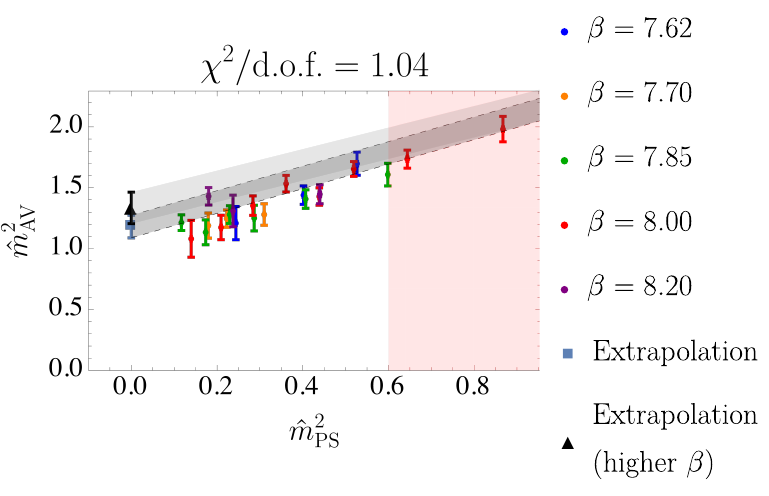} & \includegraphics[width=0.4\textwidth]{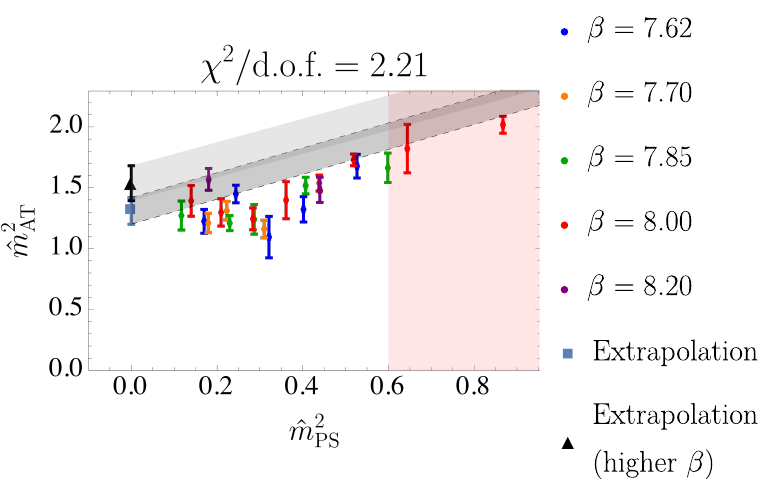}\\
    \multicolumn{2}{c}{\includegraphics[width=0.4\textwidth]{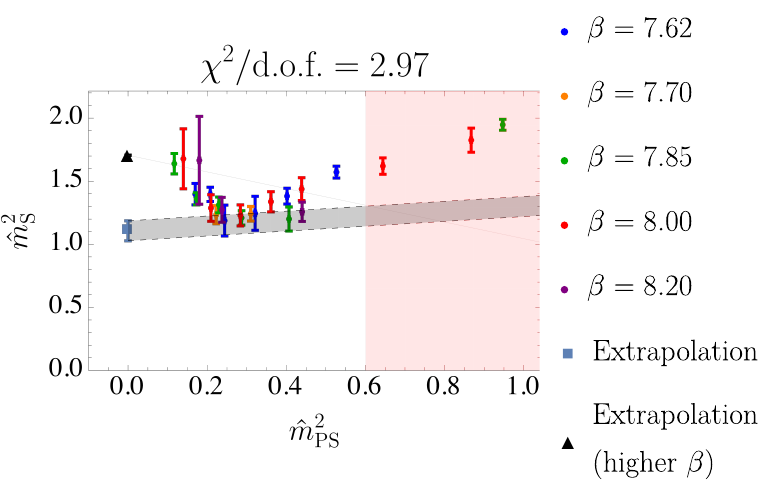}}
    \end{tabular}
    \caption[$Sp(4)$ chiral masses for fundamental fermions]{Masses squared in the $\rm V$, $\rm T$,  $\rm AV$,  $\rm AT$, and $ \rm S$ channels comprised of fermions in the fundamental representation of $Sp(4)$. The reduced chi-squared value is printed at the top of each plot. Data points in the pink shaded region are not included in the curve-fitting procedure. The grey band represents the continuum and massless extrapolation with the blue square being the observable  and the vertical width corresponding to the statistical error. In instances where a reliable extrapolation cannot be made, no grey band is shown. All quantities are expressed in units of the gradient flow scale, $w_0$. The extrapolation with the smallest $\beta$ value removed is shown as a lighter grey band and a black triangle in cases where data were available at the smallest $\beta$ value.
\label{fig:sp4Fchiralmass}}
\end{figure*}

\begin{figure*}[t]
    \centering
    \begin{tabular}{cc}
    \includegraphics[width=0.4\textwidth]{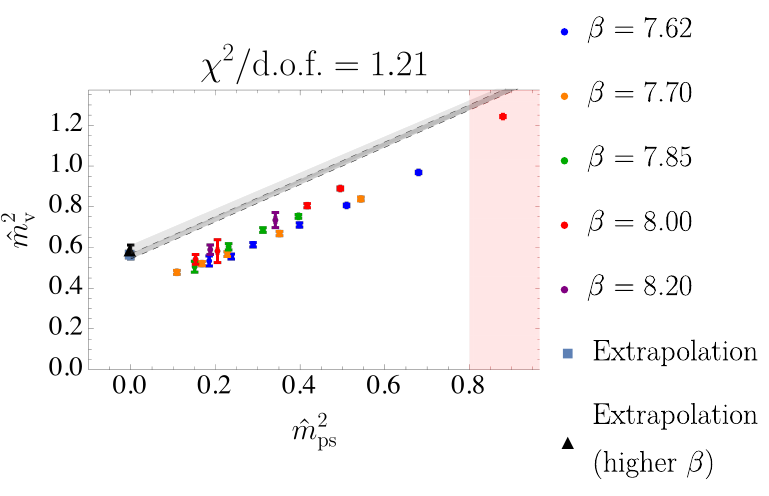} & \includegraphics[width=0.4\textwidth]{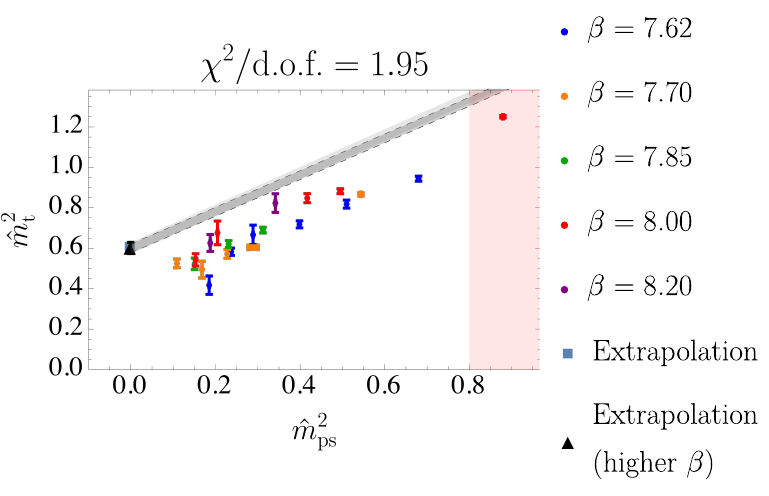}\\
    \includegraphics[width=0.4\textwidth]{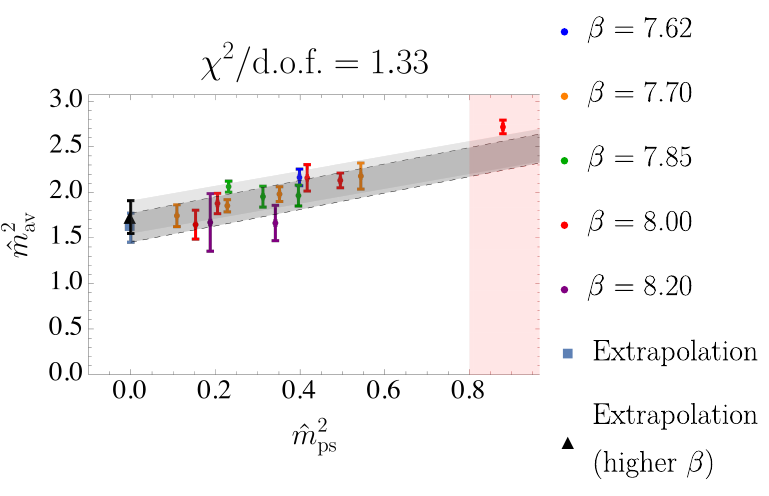} & \includegraphics[width=0.4\textwidth]{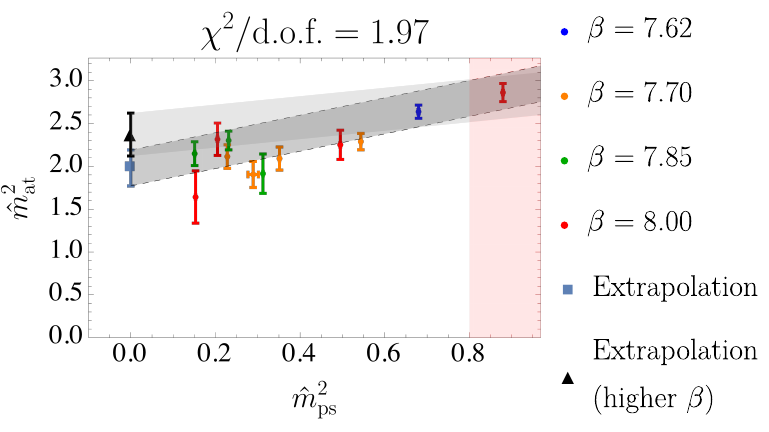}\\
    \multicolumn{2}{c}{\includegraphics[width=0.4\textwidth]{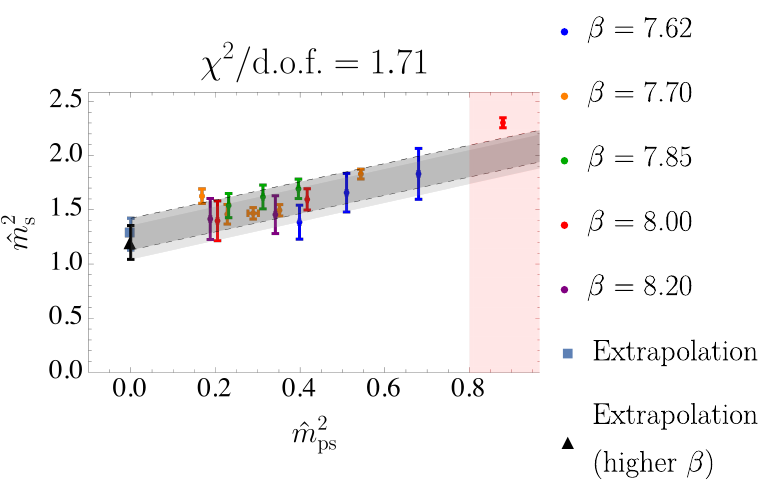}}
    \end{tabular}
    \caption[$Sp(4)$ chiral masses for antisymmetric fermions]{Masses squared in the $\rm v$, $\rm t$,  $\rm av$,  $\rm at$, and $ \rm s$  channels comprised of fermions in the antisymmetric representation of $Sp(4)$. The reduced chi-squared value is printed at the top of each plot. Data points in the pink shaded region are not included in the curve-fitting procedure. The grey band represents the continuum and massless extrapolation with the blue square being the observable  and the vertical width corresponding to the statistical error. In instances where a reliable extrapolation cannot be made, no grey band is shown. All quantities are expressed in units of the gradient flow scale, $w_0$. The extrapolation with the smallest $\beta$ value removed is shown as a lighter grey band and a black triangle in cases where data were available at the smallest $\beta$ value.
\label{fig:sp4ASchiralmass}}
\end{figure*}

\begin{figure*}[t]
    \centering
    \begin{tabular}{cc}
    \includegraphics[width=0.4\textwidth]{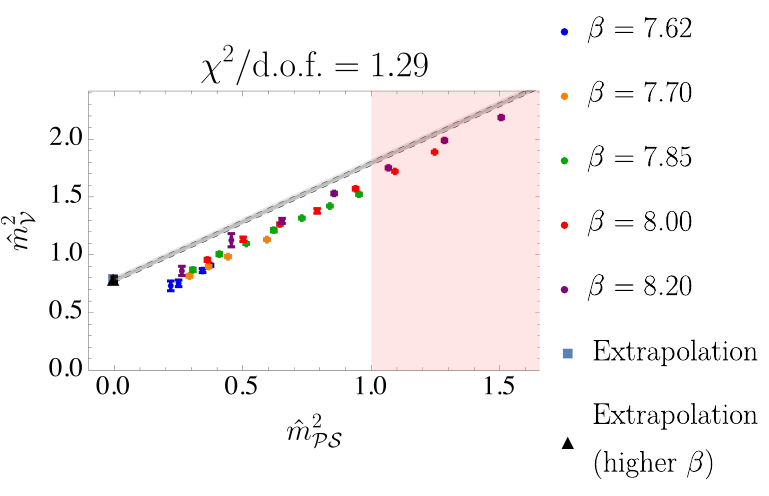} & \includegraphics[width=0.4\textwidth]{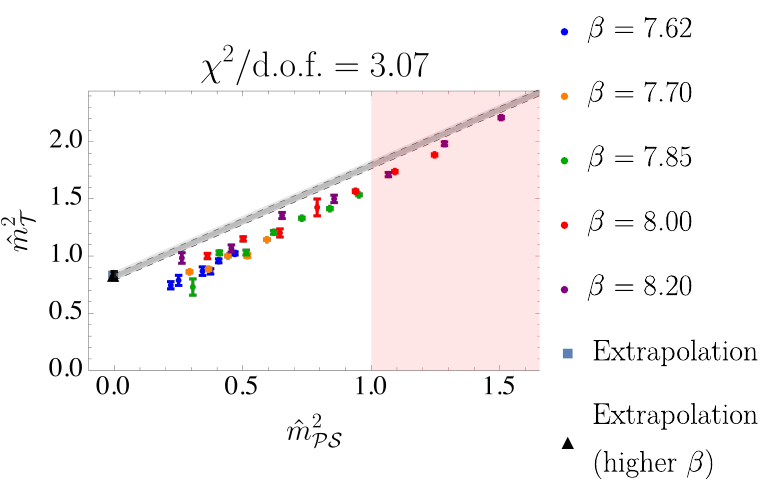}\\
    \includegraphics[width=0.4\textwidth]{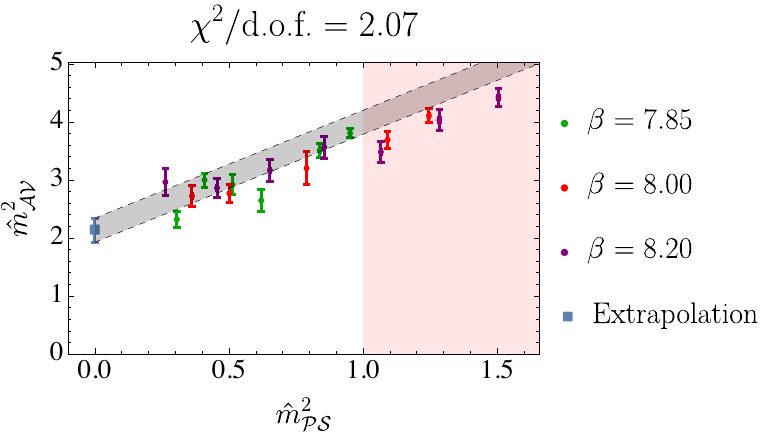} & \includegraphics[width=0.4\textwidth]{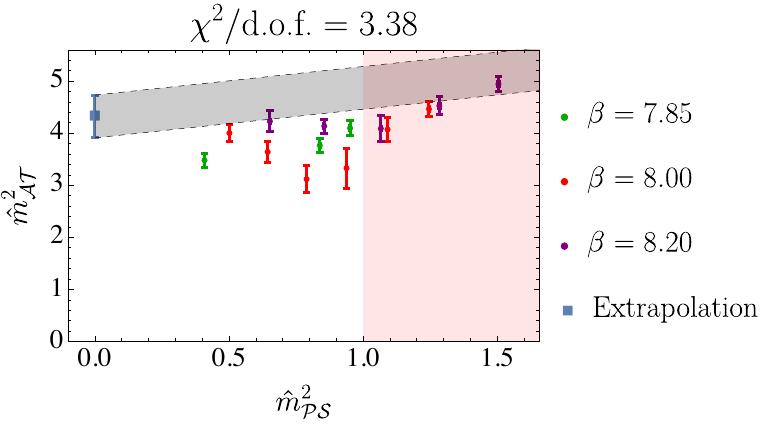}\\
    \multicolumn{2}{c}{\includegraphics[width=0.4\textwidth]{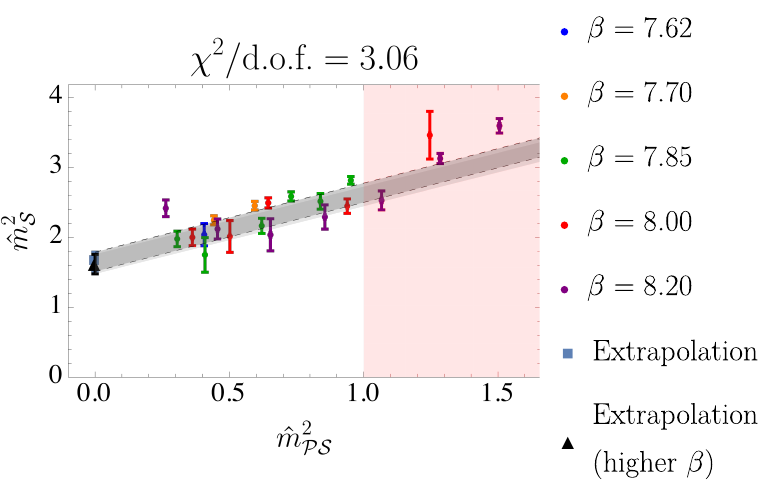}}
    \end{tabular}
    \caption[$Sp(4)$ chiral masses for symmetric fermions]{Masses squared in the $\cal V$, $\cal T$,  $\cal AV$,  $\cal AT$ and $ \cal S$  channels comprised of fermions in the symmetric representation of $Sp(4)$. The reduced chi-squared value is printed at the top of each plot. Data points in the pink shaded region are not included in the curve-fitting procedure. The grey band represents the continuum and massless extrapolation with the blue square being the observable  and the vertical width corresponding to the statistical error. In instances where a reliable extrapolation cannot be made, no grey band is shown. All quantities are expressed in units of the gradient flow scale, $w_0$. The extrapolation with the smallest $\beta$ value removed is shown as a lighter grey band and a black triangle in cases where data were available at the smallest $\beta$ value.
\label{fig:sp4Schiralmass}}
\end{figure*}

 \subsection{Sum rules}
  \label{Sec:Weinberg}

The Weinberg sum rules~\cite{Weinberg:1967kj} are  exact results, that can be formulated as follows:
 \beqs
  \sum_i \left( \hat{f}_{{\rm V},i}^2-  \hat{f}_{{\rm AV},i}^2\right)&=&\hat{f}_{\rm PS}^2\,,\\
 \sum_i \left(\hat{m}_{{\rm V},i}^2 \hat{f}_{{\rm V},i}^2- \hat{m}_{{\rm AV},i}^2 \hat{f}_{{\rm AV},i}^2\right)&=&0\,,
 \eeqs
where the summation is over the whole tower of states sourced by the $\rm V$ and  $\rm AV$ meson operators.
It is interesting to question whether these rules can be saturated by  restricting the sums to the lightest state in each channel.
Our numerical results do not support saturation, as we shall see.

An extension of the sum rules is given by the quantity:
\beqs
S&\equiv & 4\pi  \sum_i \left( \frac{\hat{f}_{{\rm V},i}^2}{\hat{m}_{{\rm V},i}^2}-\frac{ \hat{f}_{{\rm AV},i}^2}{ \hat{m}_{{\rm AV},i}^2}\right)\,,
\eeqs
where, again, the sum runs over all the  states in the $\rm V$ and $\rm AV$ channels.
 In the case of a 2-flavor QCD-like theory, this is one of the many, equivalent, definitions of the Peskin-Takeuchi precision parameter, $S$~\cite{Peskin:1991sw}, if we interpret the underlying dynamics in terms of a technicolor model of electroweak symmetry breaking.
Interestingly, this quantity is dimensionless, therefore does not depend on the scale-setting procedure adopted.
Extrapolating to small Higgs masses the combination of indirect tests of the electroweak theory,
following Ref.~\cite{Barbieri:2004qk}, yields a conservative bound $|S|\lsim 0.4$, at the  $3\sigma$ confidence level.
We can only provide a rough estimate for this quantity, obtained by saturating the defining sum with the first resonance,
as is the case for the Weinberg sum rules, reminding the reader that, since this has not proved to be a valid approximation in the latter case, the result should be taken with a grain of salt.

For ease of comparison, we define three dimensionless quantities,  involving only the lightest states:
\beqs
s_0&\equiv&  4\pi   \left( \frac{\hat{f}_{V}^2}{\hat{m}_{{\rm V}}^2}-\frac{ \hat{f}_{{\rm AV}}^2}{ \hat{m}_{{\rm AV}}^2}\right)\,,\\
s_1&\equiv&1-  \frac{\hat{f}_{{\rm AV}}^2+\hat{f}_{\rm PS}^2}{ \hat{f}_{{\rm V}}^2}\,,\\
s_2&\equiv&1-\frac{ \hat{m}_{{\rm AV}}^2 \hat{f}_{{\rm AV}}^2}{\hat{m}_{{\rm V}}^2 \hat{f}_{\rm V}^2}\,.
\eeqs
We compute them with massless and continuum limit extrapolations,
and report the results  in Table~\ref{Tab:s}.
The numerical evidence we collected indicates that neither $s_1$ nor $s_2$  vanish, which would discourage one from using the approximation of saturating the sum rules on the first resonance only. These results suggest to use caution, as in general $s_0$ will also differ from  $S$.
It would be  interesting to repeat this exercise with lattice calculations that
involve dynamical fermions, to see how the dynamics affects them.
\Edit{For completeness, and to facilitate comparison, we include in the table also 
the estimates of the same quantities for 2-flavor QCD, for which we borrow the input from Table~II of
Ref.~\cite{Erlich:2005qh}, based in turn on data from Ref.~\cite{ParticleDataGroup:2004fcd}, even if these numerical results are obtained with a non-zero mass for the quarks ($m_{\pi}=139.6$ MeV)}:
$f_{\pi}=92.4\pm 0.35$ MeV, 
$f_{\rho}=153.4\pm7.2$ MeV,
$f_{a_1}=152.4\pm10.4$ MeV,
$m_{\rho}=775.8\pm 0.5$ MeV,
$m_{a_1}=1230\pm 40$ MeV.

\section{Conclusions and Outlook}
\label{Sec:conclusions}

We reported the results of a first systematic study of the spectra of mesons in $Sp(2N)$ lattice gauge theories
with fermions in three different representations, in the quenched approximation, for $N=2, 3, 4$.
We applied next-to-leading order W$\chi$PT to extract the continuum and massless limits of the spectroscopy observables.
We also performed a first simplified extrapolation towards the large-$N$ limit.
Finally, we computed non-trivial quantities, related to the Weinberg sum rules, using the 
lattice numerical results, with the additional drastic approximation of including only the ground states.
For all these measurements, we also performed an extensive exploration of the lattice
parameter space, to assess the magnitude of finite-size effects.
Details about the intermediate steps of these calculations can be found in the public releases in Refs.~\cite{datarelease}.

In principle, our results are applicable to phenomenological studies of models of new physics that extend the standard model, 
particularly when the number of fermion species is small, and when the quenched approximation is
sufficient to provide useful estimates of masses and decay constants for the mesons.
This includes the context of composite Higgs models and models of dark matter with strong-coupling origin. However, the systematics highlighted in our discussion, which affect more severely some of the states we have analysed, would suggest to exercise judicious caution if using these results for phenomenological applications.

This study sets the stage for future, extensive and high precision measurements of spectroscopy observables
in the corresponding lattice gauge theories with dynamical fermions, by benchmarking the lattice parameter space.
A first 
study of the spectrum of fermion bound states (chimera baryons),
that have  model-building  relevance in the context of top partial compositeness, performed 
in the quenched approximation and for $Sp(4)$ gauge theories, can be found in Ref.~\cite{Bennett:2023mhh}.
An ongoing, extensive research programme of study of the  dynamical theories with fermions transforming in multiple representations will provide precision measurements and explore complementary regions of parameter space, relevant for some phenomenological applications, for which one does not expect the quenched approximation to hold.

\begin{acknowledgments}

The work of E.~B. has been funded by the Supercomputing Wales project, which is part-funded by the European Regional Development Fund (ERDF) via Welsh Government and by the UKRI Science and Technologies Facilities Council (STFC) Research Software Engineering Fellowship EP/V052489/1. 

J.~H. is partially supported by the Center for Frontier Nuclear Science at Stony Brook University.

The work of D.~K.~H. was supported by Basic Science Research Program through the National Research Foundation of Korea (NRF) funded by the Ministry of Education (NRF-2017R1D1A1B06033701) and also  by the Korea government (MSIT) (2021R1A4A5031460). 

The work of J.-W. L. was supported in part by the National Research Foundation of Korea (NRF) grant funded by the Korea government(MSIT) (NRF-2018R1C1B3001379) and by IBS under the project code, IBS-R018-D1.

The work of C.-J.~D.~L. and of H.~H. is supported by the Taiwanese NTSC Grant No. 112-2112-M-A49-021-MY3. 

The work of B.~L. and M.~P. has been supported in part by the STFC Consolidated Grants No. ST/P00055X/1, ST/T000813/1, and ST/X000648/1. B.~L. and M.~P. received funding from the European Research Council (ERC) under the European Union’s Horizon 2020 research and innovation program under Grant Agreement No.~813942. The work of B. L. is further supported in part by the Royal Society Wolfson Research Merit Award No. WM170010 and by the Leverhulme Trust Research Fellowship No. RF-2020-4619. 

The work of D.~V. is supported by
STFC under the consolidated grant N. ST/X000680/1.

Numerical simulations have been performed on the Swansea SUNBIRD cluster (part of the Supercomputing Wales project) and AccelerateAI A100 GPU system.
The Swansea SUNBIRD system and AccelerateAI are part funded by the European Regional Development Fund (ERDF) via Welsh Government. 

\vspace{1.0cm}

{\bf Open Access Statement}---For the purpose of open access, the authors have applied a Creative Commons 
Attribution (CC BY) licence  to any Author Accepted Manuscript version arising.

\vspace{1.0cm}

{\bf Research Data Access Statement}---\Edit{Full raw data for correlation functions and gradient flow histories, and all data presented in plots and tables in this work, can be downloaded in machine-readable format at Ref.~\cite{datarelease}. The analysis workflow used to generate the latter from the former, and to produce the plots and tables presented in this work, can be downloaded at Ref.~\cite{workflowrelease}.}

\end{acknowledgments}


\appendix


\section{Finite volume effects}
\label{Sec:FV}
Finite size effects arise from the limited extent of the lattice as well as its toroidal nature. These artefacts can contaminate our measurements of observable quantities.
It is possible to extrapolate to infinite volume (the analogue of the thermodynamic limit of statistical mechanics) 
from a lattice of finite extent, $L=N_s a$, by assuming that the mass of the lightest state at finite volume, which we denote generically
as  $m_{\pi}$ in this Appendix, is related to the infinite-volume  limit, $m_{\infty}$, via the relation
\begin{equation}
    m_{\pi}=m_{\infty}\left(1+A\frac{e^{-m_{\infty}L}}{(m_{\infty}L)^{3/2}}\right)\,
\end{equation}
\Edit{first established in Ref.~\cite{Luscher:1980fr}.}

As a preliminary study, propaedeutic to the one reported in the body of this paper, we examined the finite size effect 
by plotting $m_{\pi}/m_{\infty}$ as a function of $m_{\infty}L$ as emerges from different lattice volumes as well as with different bare masses, $m_0$, for the relevant fermion. We choose the value of $m_{\infty}L$ such that the finite volume result is within a few {\it per mille} of the infinite volume one, such that this source of systematics can be ignored in comparison with the statistical uncertainties.
Details about this study can be found in Refs.~\cite{datarelease,workflowrelease}, while here we only provide one example,
in Fig.~\ref{fig:sp6FFiniteSize}, for the $Sp(6)$ theory.

The finite size effects for $Sp(4)$ quenched mesons in the fundamental and antisymmetric representations were studied in 
Ref.~\cite{Bennett:2019cxd}, and it was found that one should restrict the analysis to cases in which 
 $m_{\pi}L \gtrsim 7.5$ for both fundamental and antisymmetric representations, 
 with the identification $m_{\pi}=m_{\rm PS}$.
 Interestingly, we find that this requirement is even more severe for the $Sp(6)$ and $Sp(8)$ theories.
 The ensembles and choice of fermions masses  used in the analysis the forms the body of this paper 
 are lead to satisfying these requirements.
To prevent non-physical processes, such as the analogous process to the $\rho\rightarrow\pi\pi$ decay,
we also demand that $0.5< m_{\rm PS}/m_{\rm V} <1$, for all fermion species,
 so that the quenched approximation can be justified.

\section{Continuum and massless extrapolations}\label{sec:CMExtrapolations}
\begin{figure*}[t]
    \centering
    \begin{tabular}{cc}
        \includegraphics[width=0.4\textwidth]{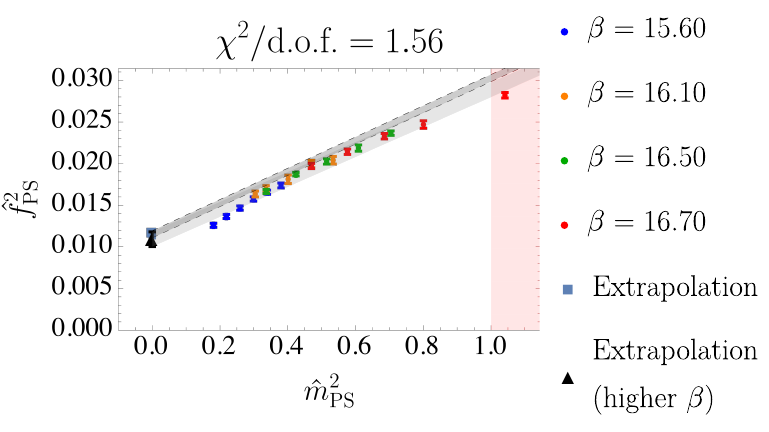} & \includegraphics[width=0.4\textwidth]{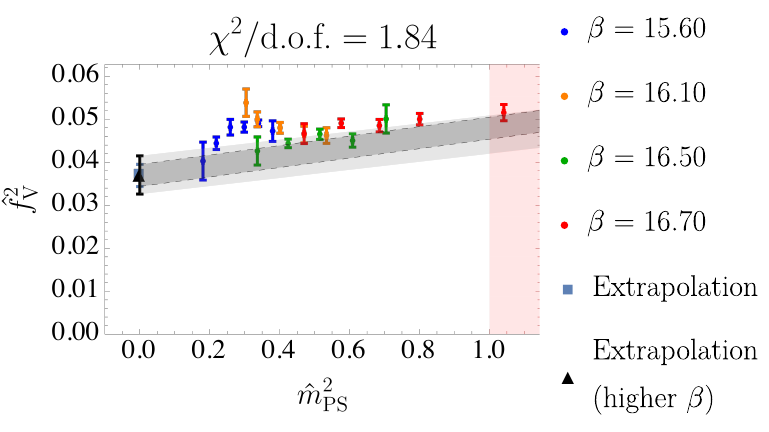}\\
        \multicolumn{2}{c}{\includegraphics[width=0.4\textwidth]{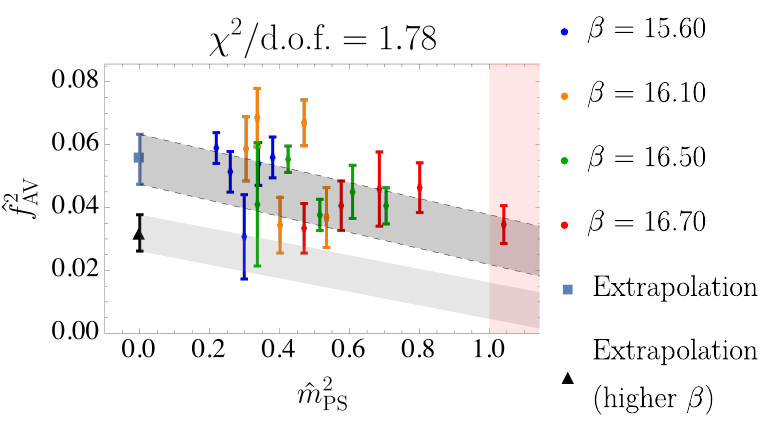}}
    \end{tabular}
    \caption[$Sp(6)$ chiral decay constants for fundamental fermions]{Decay constants squared in the $\rm PS$,  $\rm V$, and $\rm AV$ channels comprised of fermions in the fundamental representation of $Sp(6)$. The reduced chi-squared value is printed at the top of each plot. Data points in the pink shaded region are not included in the curve-fitting procedure. The grey band represents the continuum and massless extrapolation with the blue square being the observable  and the vertical width corresponding to the statistical error. In instances where a reliable extrapolation cannot be made, no grey band is shown. All quantities are expressed in units of the gradient flow scale, $w_0$. The extrapolation with the smallest $\beta$ value removed is shown as a lighter grey band and a black triangle in cases where data were available at the smallest $\beta$ value.}
\label{fig:sp6Fchiraldecay}
\end{figure*}

\begin{figure*}[t]
    \centering
    \begin{tabular}{cc}
        \includegraphics[width=0.4\textwidth]{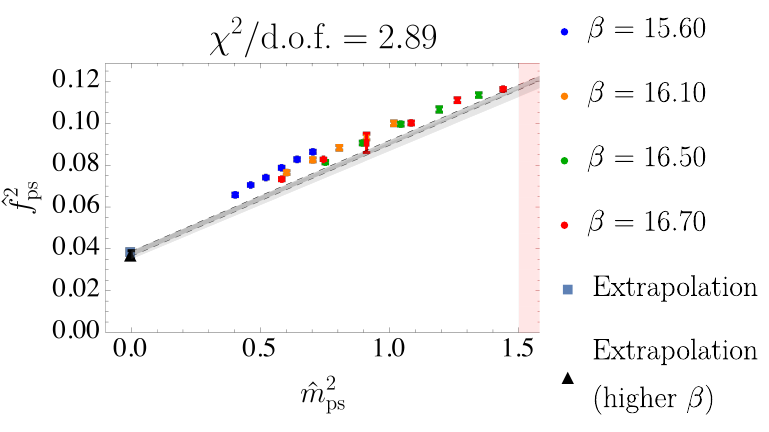} & \includegraphics[width=0.4\textwidth]{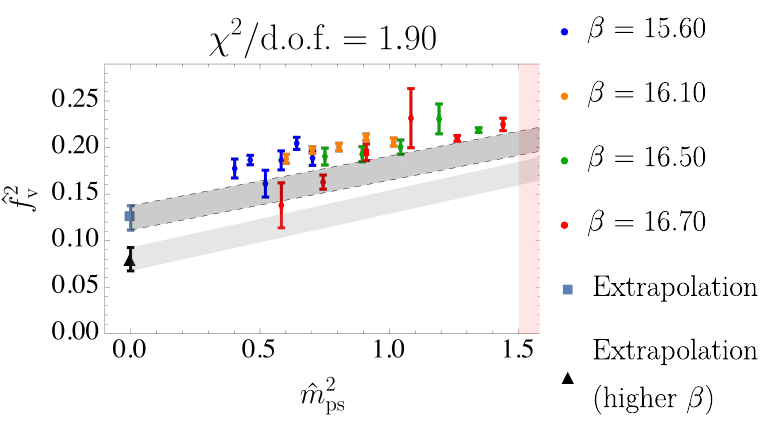}\\
        \multicolumn{2}{c}{\includegraphics[width=0.4\textwidth]{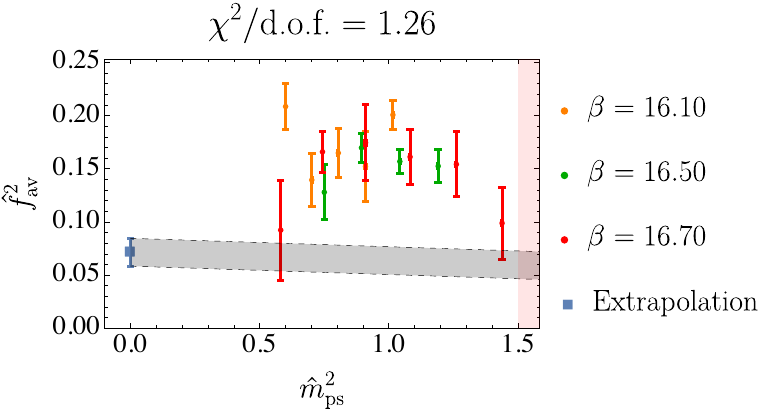}}
    \end{tabular}
    \caption[$Sp(6)$ chiral decay constants for antisymmetric fermions]{Decay constants squared for $\rm ps$, $\rm v$, and $\rm av$ channels comprised of fermions in the antisymmetric representation of $Sp(6)$. The reduced chi-squared value is printed at the top of each plot. Data points in the pink shaded region are not included in the curve-fitting procedure. The grey band represents the continuum and massless extrapolation with the blue square being the observable in the massless limit and the vertical width corresponding to the statistical error. In instances where a reliable extrapolation cannot be made, no grey band is shown. All quantities are expressed in units of the gradient flow scale, $w_0$. The extrapolation with the smallest $\beta$ value removed is shown as a lighter grey band and a black triangle in cases where data were available at the smallest $\beta$ value.}
\label{fig:sp6ASchiraldecay}
\end{figure*}

\begin{figure*}[t]
    \centering
    \begin{tabular}{cc}
        \includegraphics[width=0.4\textwidth]{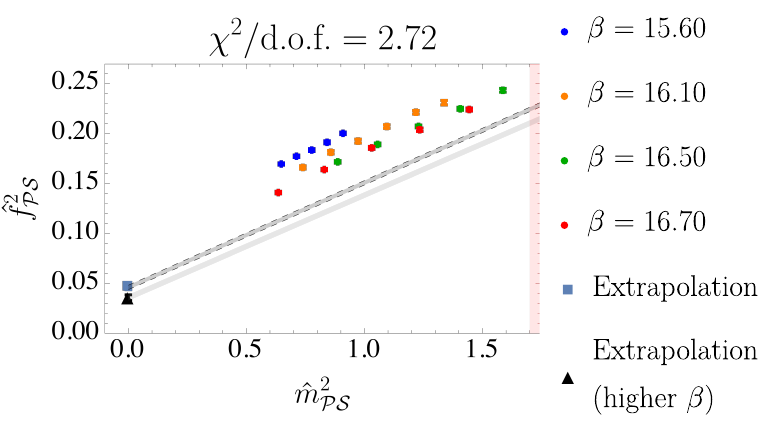} & \includegraphics[width=0.4\textwidth]{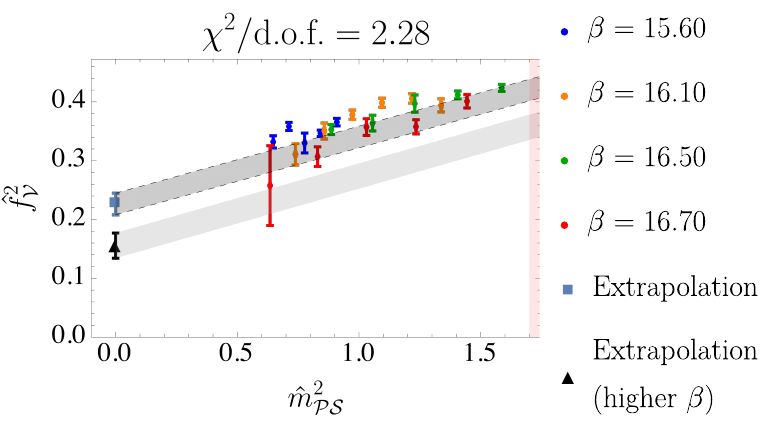}\\
        \multicolumn{2}{c}{\includegraphics[width=0.4\textwidth]{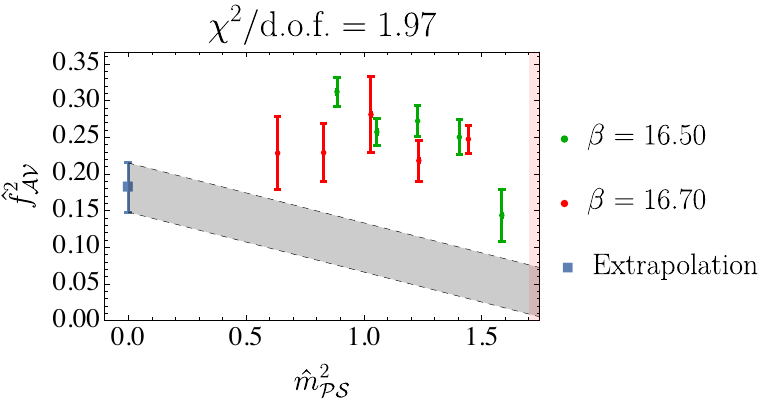}}
    \end{tabular}
    \caption[$Sp(6)$ chiral decay constants for symmetric fermions]{Decay constants squared in the $\cal PS$, $\cal V$,  and $\cal AV$ channels comprised of fermions in the symmetric representation of $Sp(6)$. The reduced chi-squared value is printed at the top of each plot. Data points in the pink shaded region are not included in the curve-fitting procedure. The grey band represents the continuum and massless extrapolation with the blue square being the observable and the vertical width corresponding to the statistical error. In instances where a reliable extrapolation cannot be made, no grey band is shown. All quantities are expressed in units of the gradient flow scale, $w_0$. The extrapolation with the smallest $\beta$ value removed is shown as a lighter grey band and a black triangle in cases where data were available at the smallest $\beta$ value.}
\label{fig:sp6Schiraldecay}
\end{figure*}

\begin{figure*}[t]
    \centering
    \begin{tabular}{cc}
    \includegraphics[width=0.4\textwidth]{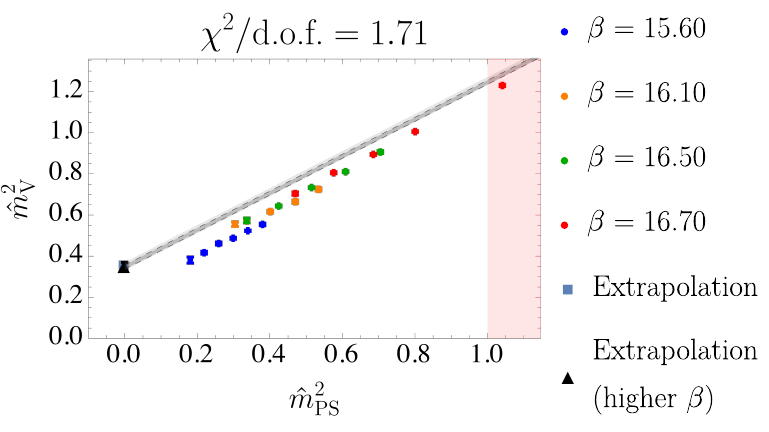} & \includegraphics[width=0.4\textwidth]{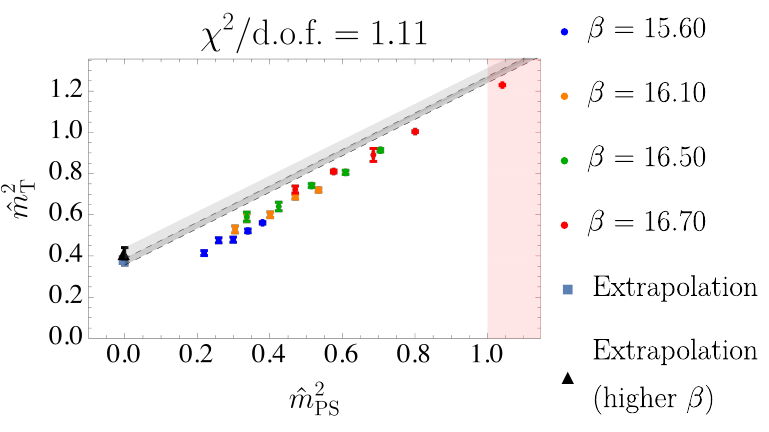}\\
    \includegraphics[width=0.4\textwidth]{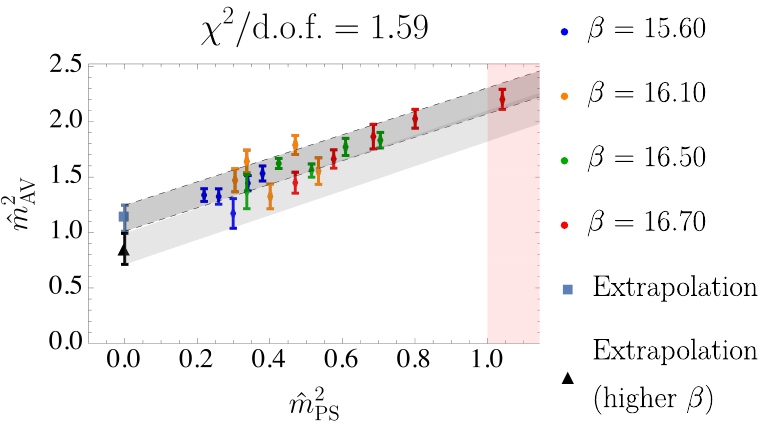} & \includegraphics[width=0.4\textwidth]{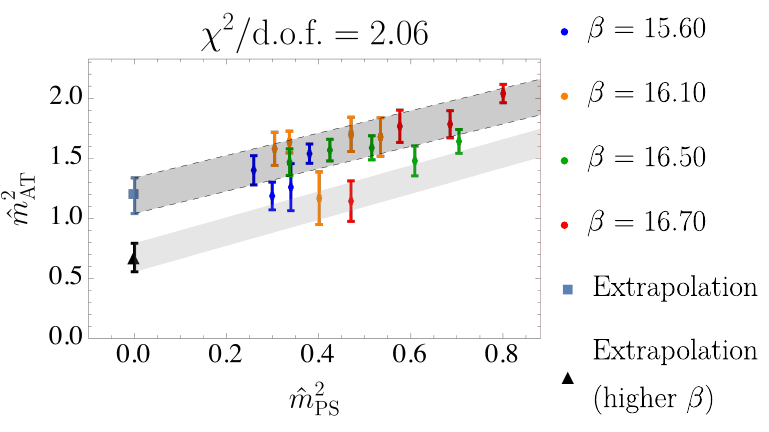}\\
    \multicolumn{2}{c}{\includegraphics[width=0.4\textwidth]{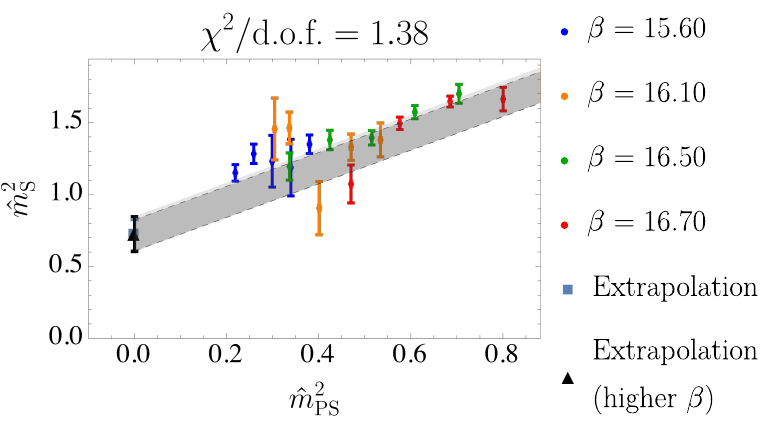}}
    \end{tabular}
    \caption[$Sp(6)$ chiral masses for fundamental fermions]{Masses squared in the $\rm V$, $\rm T$,  $\rm AV$,  $\rm AT$, and $ \rm S$ channels comprised of fermions in the fundamental representation of $Sp(6)$. The reduced chi-squared value is printed at the top of each plot. Data points in the pink shaded region are not included in the curve-fitting procedure. The grey band represents the continuum and massless extrapolation with the blue square being the observable and the vertical width corresponding to the statistical error. In instances where a reliable extrapolation cannot be made, no grey band is shown. All quantities are expressed in units of the gradient flow scale, $w_0$. The extrapolation with the smallest $\beta$ value removed is shown as a lighter grey band and a black triangle in cases where data were available at the smallest $\beta$ value.}
\label{fig:sp6Fchiralmass}
\end{figure*}

\begin{figure*}[t]
    \centering
    \begin{tabular}{cc}
    \includegraphics[width=0.4\textwidth]{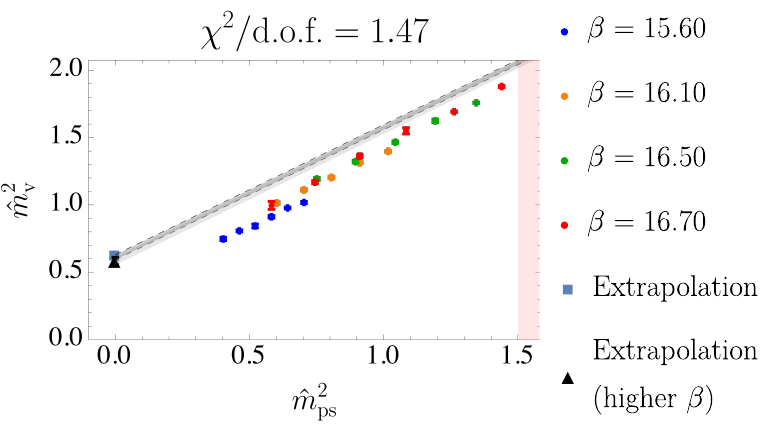} & \includegraphics[width=0.4\textwidth]{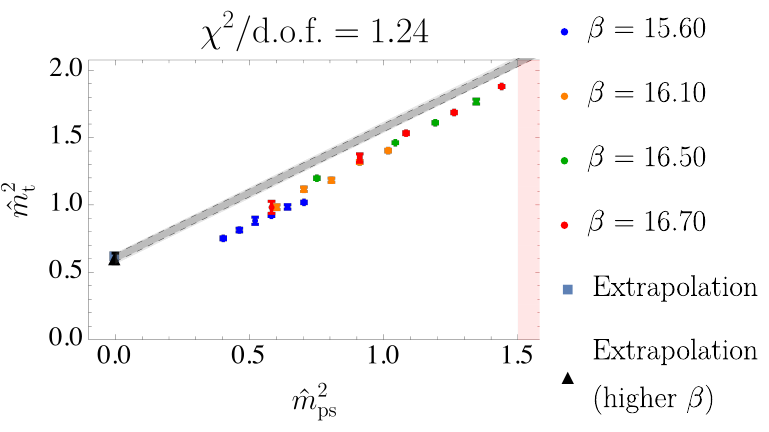}\\
    \includegraphics[width=0.4\textwidth]{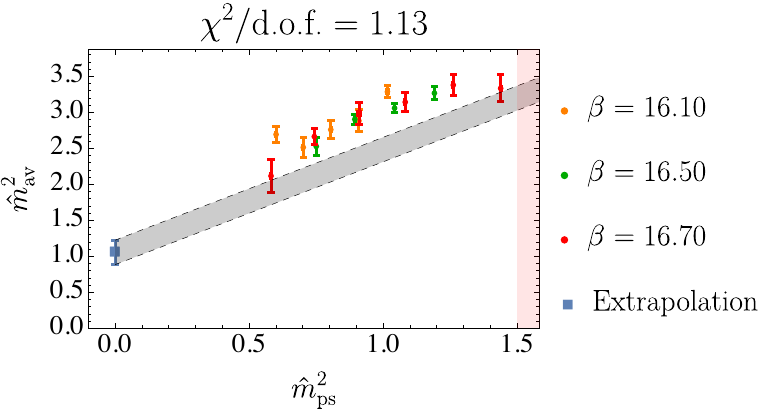} & \includegraphics[width=0.4\textwidth]{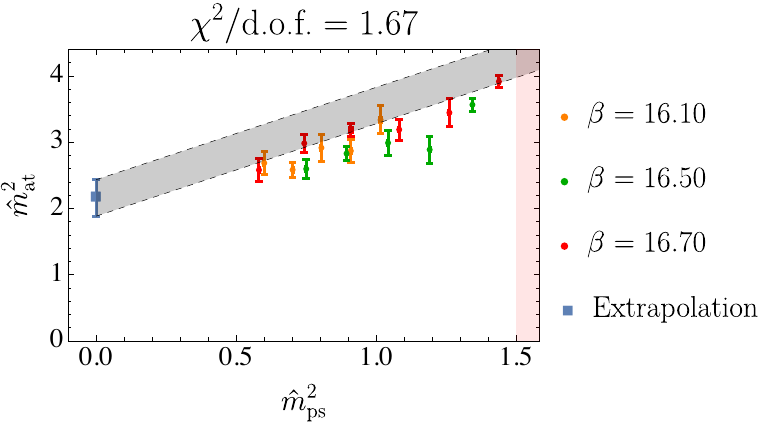}\\
    \multicolumn{2}{c}{\includegraphics[width=0.4\textwidth]{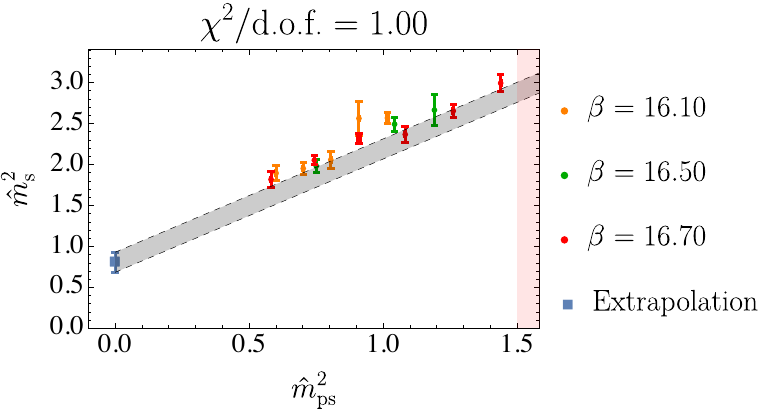}}
    \end{tabular}
    \caption[$Sp(6)$ chiral masses for antisymmetric fermions]{Masses squared in the $\rm v$, $\rm t$,  $\rm av$,  $\rm at$, and $ \rm s$  channels comprised of fermions in the antisymmetric representation of $Sp(6)$. The reduced chi-squared value is printed at the top of each plot. Data points in the pink shaded region are not included in the curve-fitting procedure. The grey band represents the continuum and massless extrapolation with the blue square being the observable and the vertical width corresponding to the statistical error. In instances where a reliable extrapolation cannot be made, no grey band is shown. All quantities are expressed in units of the gradient flow scale, $w_0$. The extrapolation with the smallest $\beta$ value removed is shown as a lighter grey band and a black triangle in cases where data were available at the smallest $\beta$ value.}
\label{fig:sp6ASchiralmass}
\end{figure*}

\begin{figure*}[t]
    \centering
    \begin{tabular}{cc}
    \includegraphics[width=0.4\textwidth]{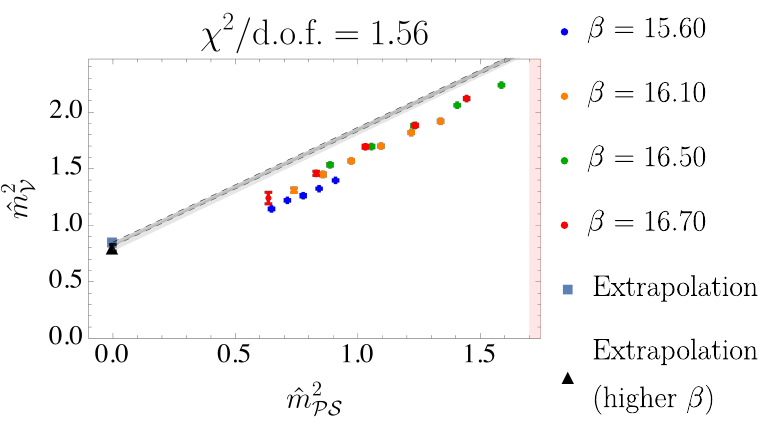} & \includegraphics[width=0.4\textwidth]{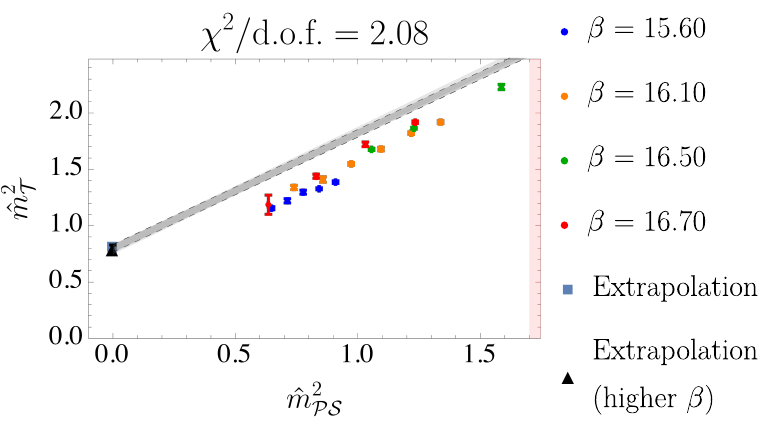}\\
    \includegraphics[width=0.4\textwidth]{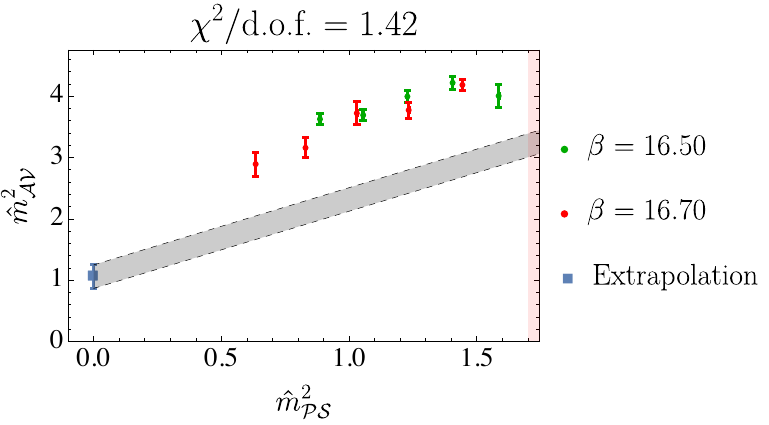} & \includegraphics[width=0.4\textwidth]{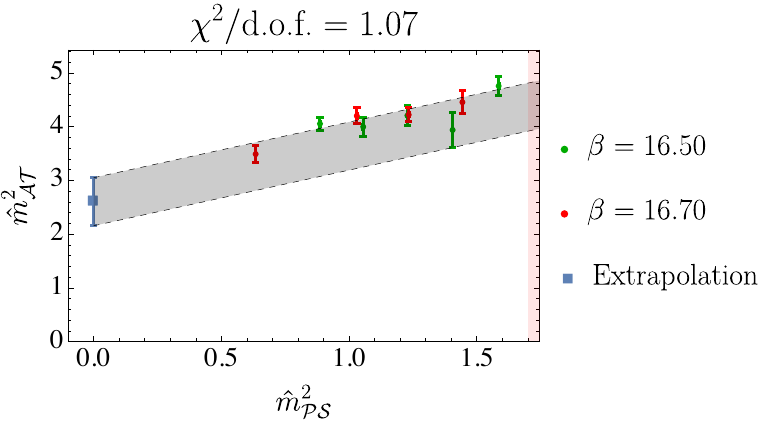}\\
    \multicolumn{2}{c}{\includegraphics[width=0.4\textwidth]{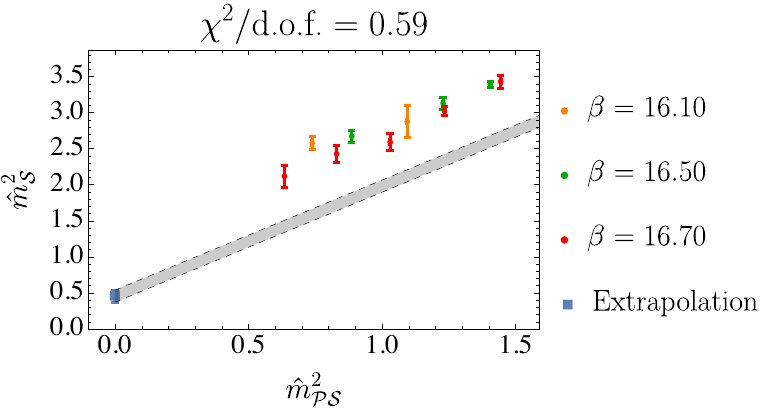}}
    \end{tabular}
    \caption[$Sp(6)$ chiral masses for symmetric fermions]{Masses squared in the $\cal V$, $\cal T$,  $\cal AV$,  $\cal AT$ and $ \cal S$  channels comprised of fermions in the symmetric representation of $Sp(6)$. The reduced chi-squared value is printed at the top of each plot. Data points in the pink shaded region are not included in the curve-fitting procedure. The grey band represents the continuum and massless extrapolation with the blue square being the observable  and the vertical width corresponding to the statistical error. In instances where a reliable extrapolation cannot be made, no grey band is shown. All quantities are expressed in units of the gradient flow scale, $w_0$. The extrapolation with the smallest $\beta$ value removed is shown as a lighter grey band and a black triangle in cases where data were available at the smallest $\beta$ value.}
\label{fig:sp6Schiralmass}
\end{figure*}

\begin{figure*}[t]
    \centering
    \begin{tabular}{cc}
        \includegraphics[width=0.4\textwidth]{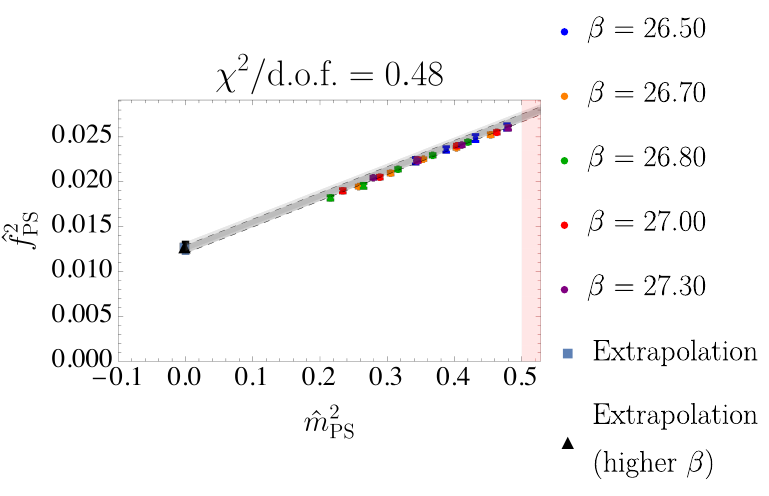} & \includegraphics[width=0.4\textwidth]{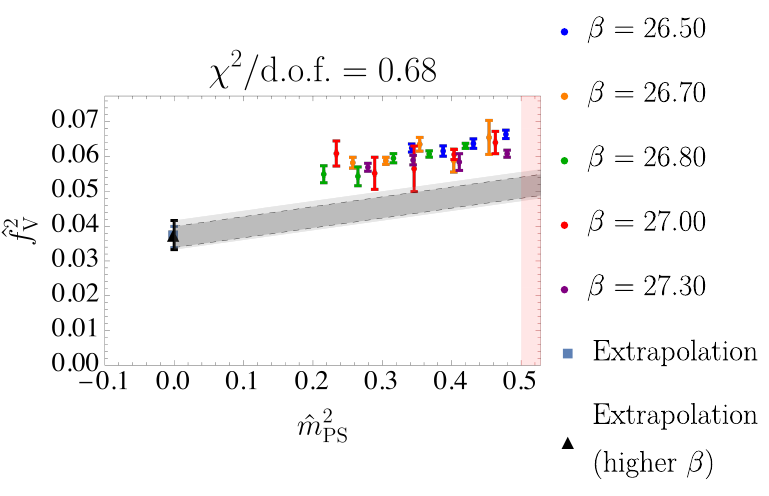}\\
        \multicolumn{2}{c}{\includegraphics[width=0.4\textwidth]{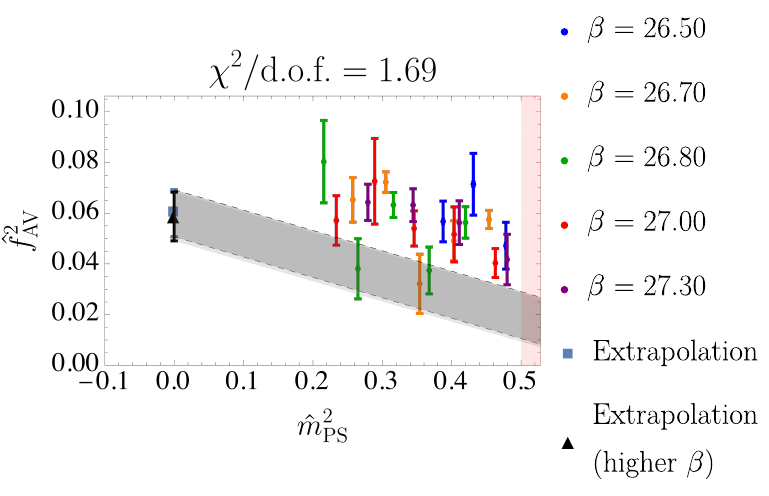}}
    \end{tabular}
    \caption[$Sp(8)$ chiral decay constants for fundamental fermions]{Decay constants squared in the $\rm PS$, $\rm V$ and $\rm AV$  channels comprised of fermions in the fundamental representation of $Sp(8)$. The reduced chi-squared value is printed at the top of each plot. Data points in the pink shaded region are not included in the curve-fitting procedure. The grey band represents the continuum and massless extrapolation with the blue square being the observable and the vertical width corresponding to the statistical error. In instances where a reliable extrapolation cannot be made, no grey band is shown. All quantities are expressed in units of the gradient flow scale, $w_0$. The extrapolation with the smallest $\beta$ value removed is shown as a lighter grey band and a black triangle in cases where data were available at the smallest $\beta$ value.}
\label{fig:sp8Fchiraldecay}
\end{figure*}

\begin{figure*}[t]
    \centering
    \begin{tabular}{cc}
        \includegraphics[width=0.4\textwidth]{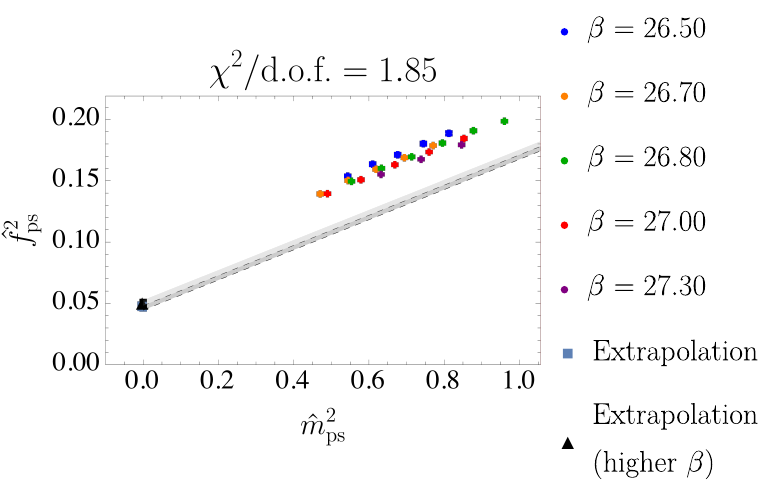} & \includegraphics[width=0.4\textwidth]{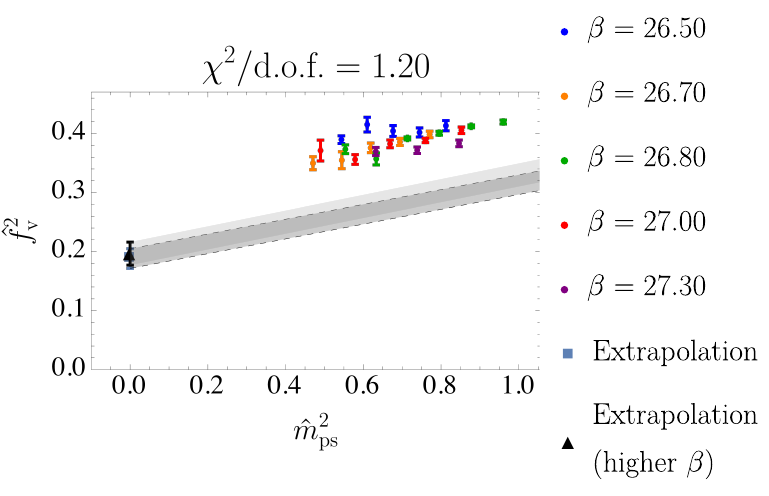}\\
        \multicolumn{2}{c}{\includegraphics[width=0.4\textwidth]{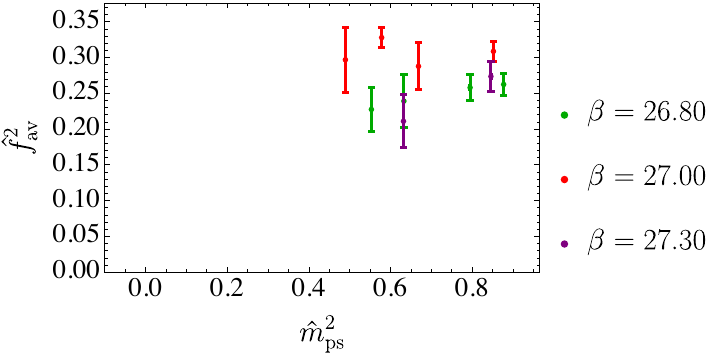}}
    \end{tabular}
    \caption[$Sp(8)$ chiral decay constants for antisymmetric fermions]{Decay constants squared in the $\rm ps$, $\rm v$, and $\rm av$ channels comprised of fermions in the antisymmetric representation of $Sp(8)$. The reduced chi-squared value is printed at the top of each plot. Data points in the pink shaded region are not included in the curve-fitting procedure. The grey band represents the continuum and massless extrapolation with the blue square being the observable  and the vertical width corresponding to the statistical error. In instances where a reliable extrapolation cannot be made, no grey band is shown. All quantities are expressed in units of the gradient flow scale, $w_0$. The extrapolation with the smallest $\beta$ value removed is shown as a lighter grey band and a black triangle in cases where data were available at the smallest $\beta$ value.}
\label{fig:sp8ASchiraldecay}
\end{figure*}

\begin{figure*}[t]
    \centering
    \begin{tabular}{cc}
        \includegraphics[width=0.4\textwidth]{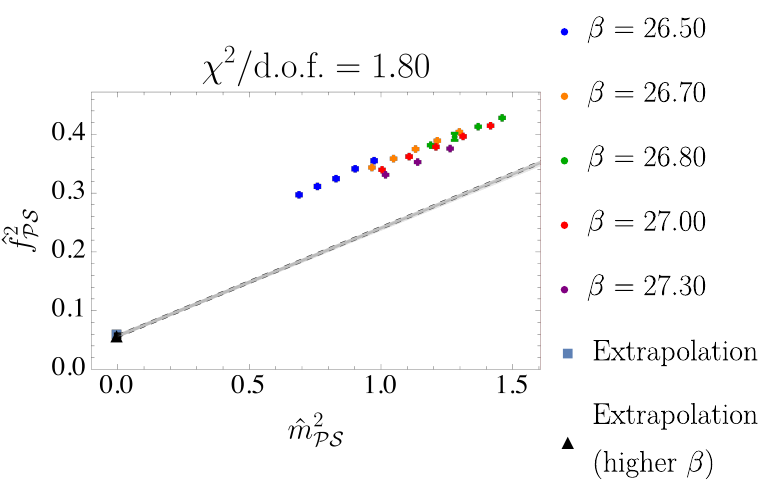} & \includegraphics[width=0.4\textwidth]{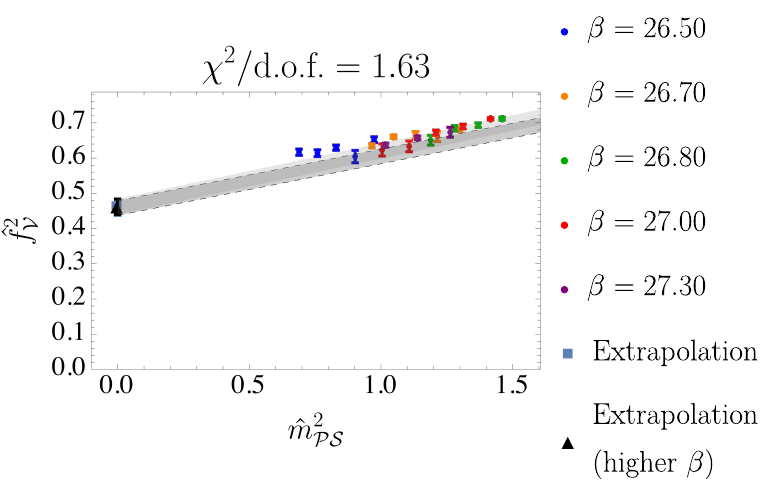}\\
    \end{tabular}
    \caption[$Sp(8)$ chiral decay constants for symmetric fermions]{Decay constants squared in the $\cal PS$ and $\cal V$ channels comprised of fermions in the symmetric representation of $Sp(8)$. The reduced chi-squared value is printed at the top of each plot. Data points in the pink shaded region are not included in the curve-fitting procedure. The grey band represents the continuum and massless extrapolation with the blue square being the observable and the vertical width corresponding to the statistical error. In instances where a reliable extrapolation cannot be made, no grey band is shown. All quantities are expressed in units of the gradient flow scale, $w_0$. The extrapolation with the smallest $\beta$ value removed is shown as a lighter grey band and a black triangle in cases where data were available at the smallest $\beta$ value.}
\label{fig:sp8Schiraldecay}
\end{figure*}

\begin{figure*}[t]
    \centering
    \begin{tabular}{cc}
    \includegraphics[width=0.4\textwidth]{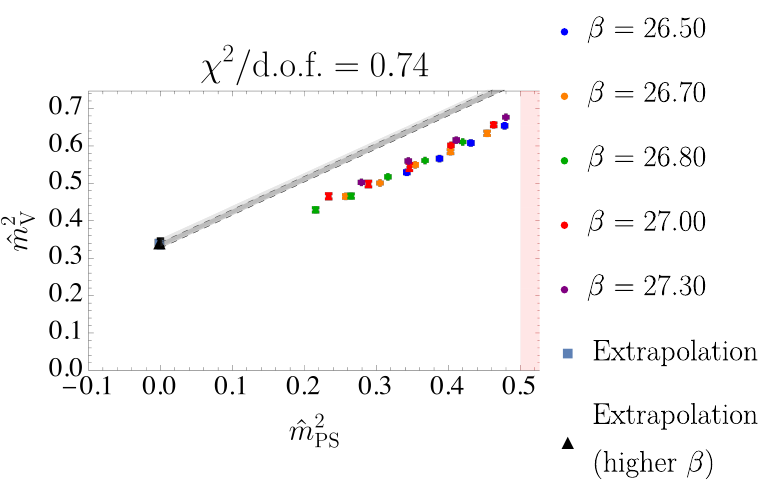} & \includegraphics[width=0.4\textwidth]{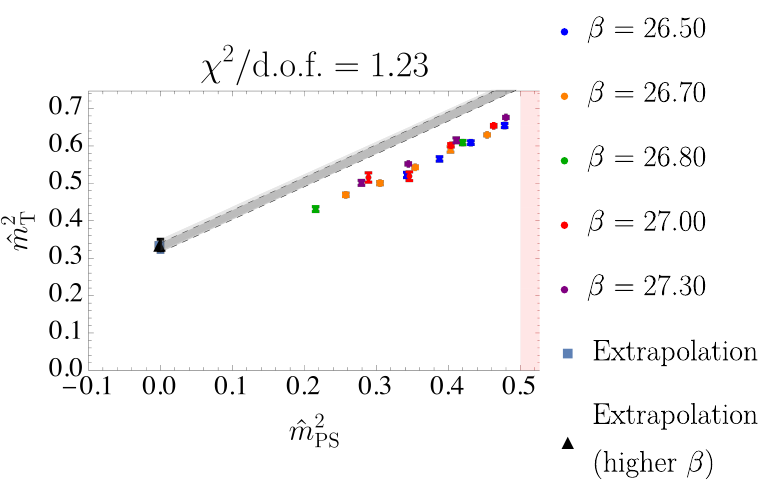}\\
    \includegraphics[width=0.4\textwidth]{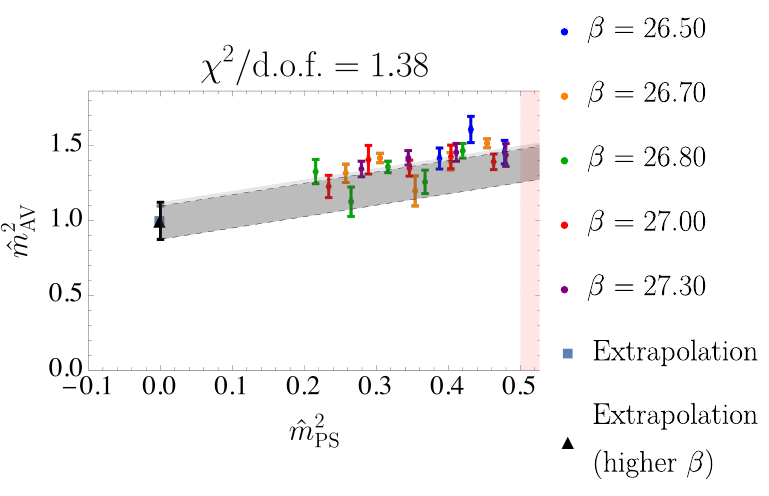} & \includegraphics[width=0.4\textwidth]{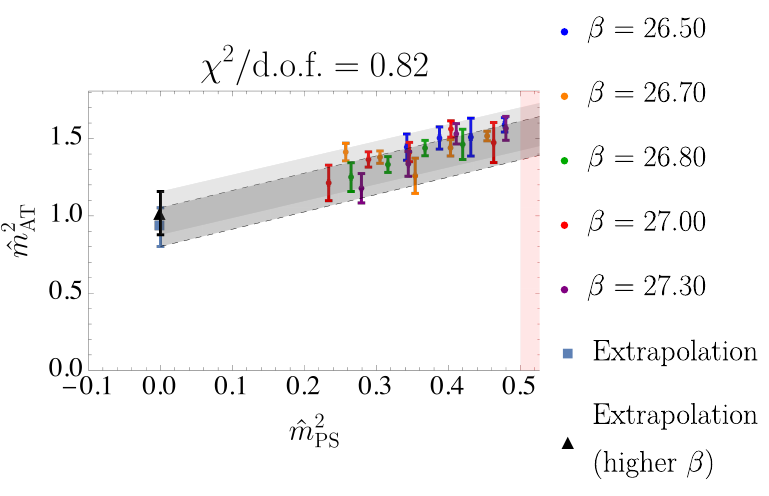}\\
    \multicolumn{2}{c}{\includegraphics[width=0.4\textwidth]{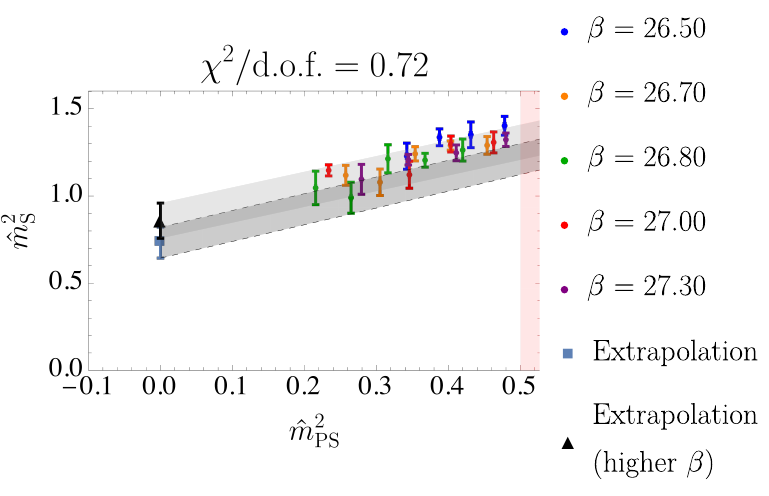}}
    \end{tabular}
    \caption[$Sp(8)$ chiral masses for fundamental fermions]{Masses squared in the $\rm V$, $\rm T$,  $\rm AV$,  $\rm AT$, and $ \rm S$ channels comprised of fermions in the fundamental representation of $Sp(8)$. The reduced chi-squared value is printed at the top of each plot. Data points in the pink shaded region are not included in the curve-fitting procedure. The grey band represents the continuum and massless extrapolation with the blue square being the observable  and the vertical width corresponding to the statistical error. In instances where a reliable extrapolation cannot be made, no grey band is shown. All quantities are expressed in units of the gradient flow scale, $w_0$. The extrapolation with the smallest $\beta$ value removed is shown as a lighter grey band and a black triangle in cases where data were available at the smallest $\beta$ value.}
\label{fig:sp8Fchiralmass}
\end{figure*}

\begin{figure*}[t]
    \centering
    \begin{tabular}{cc}
    \includegraphics[width=0.4\textwidth]{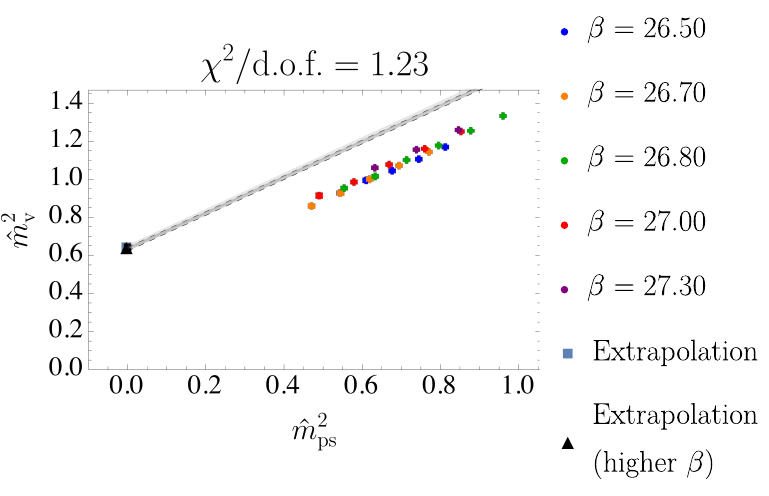} & \includegraphics[width=0.4\textwidth]{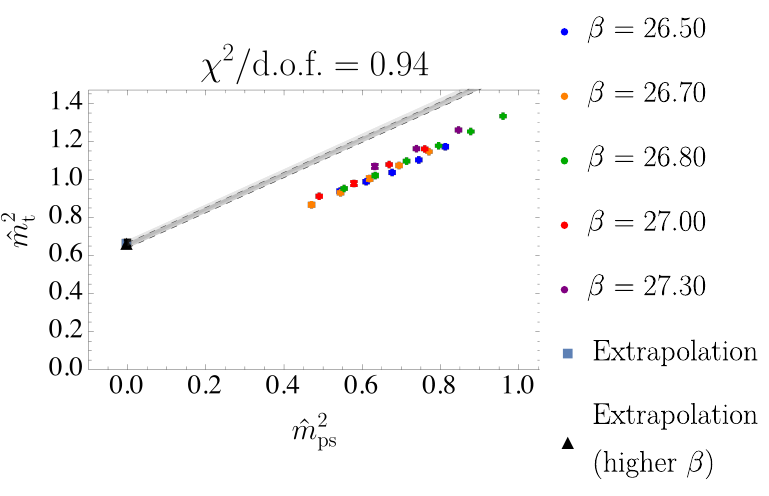}\\
    \includegraphics[width=0.4\textwidth]{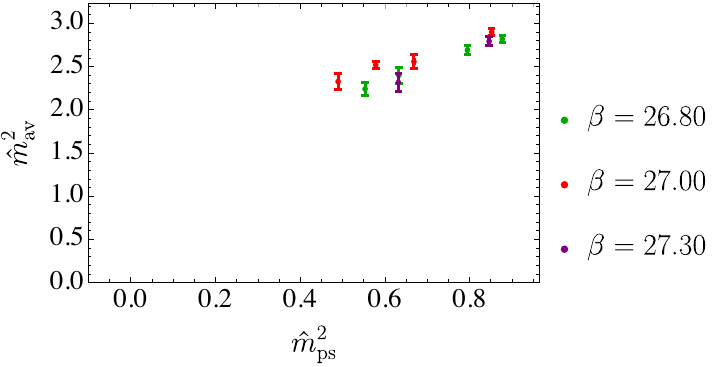} & \includegraphics[width=0.4\textwidth]{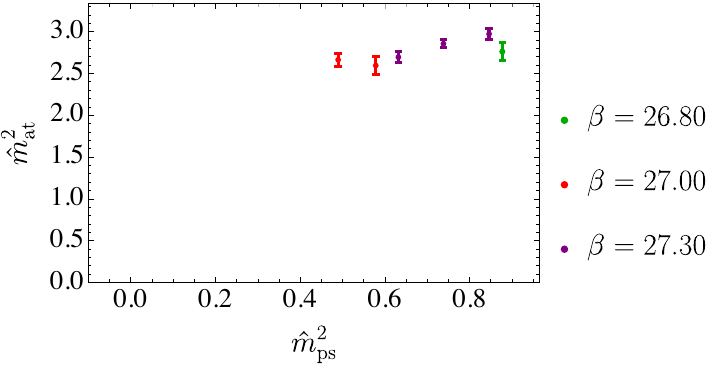}\\
    \multicolumn{2}{c}{\includegraphics[width=0.4\textwidth]{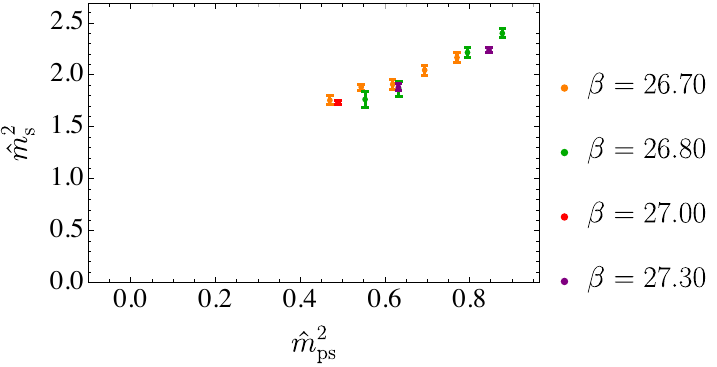}}
    \end{tabular}
    \caption[$Sp(8)$ chiral masses for antisymmetric fermions]{Masses squared in the $\rm v$, $\rm t$,  $\rm av$,  $\rm at$, and $ \rm s$  channels comprised of fermions in the antisymmetric representation of $Sp(8)$. The reduced chi-squared value is printed at the top of each plot. Data points in the pink shaded region are not included in the curve-fitting procedure. The grey band represents the continuum and massless extrapolation with the blue square being the observable  and the vertical width corresponding to the statistical error. In instances where a reliable extrapolation cannot be made, no grey band is shown. All quantities are expressed in units of the gradient flow scale, $w_0$. The extrapolation with the smallest $\beta$ value removed is shown as a lighter grey band and a black triangle in cases where data were available at the smallest $\beta$ value.}
\label{fig:sp8ASchiralmass}
\end{figure*}

\begin{figure*}[t]
    \centering
    \begin{tabular}{cc}
    \includegraphics[width=0.4\textwidth]{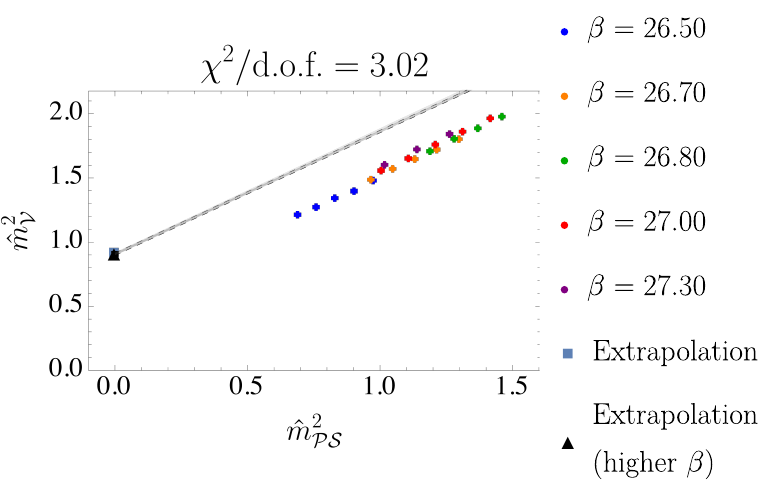} &
     \includegraphics[width=0.4\textwidth]{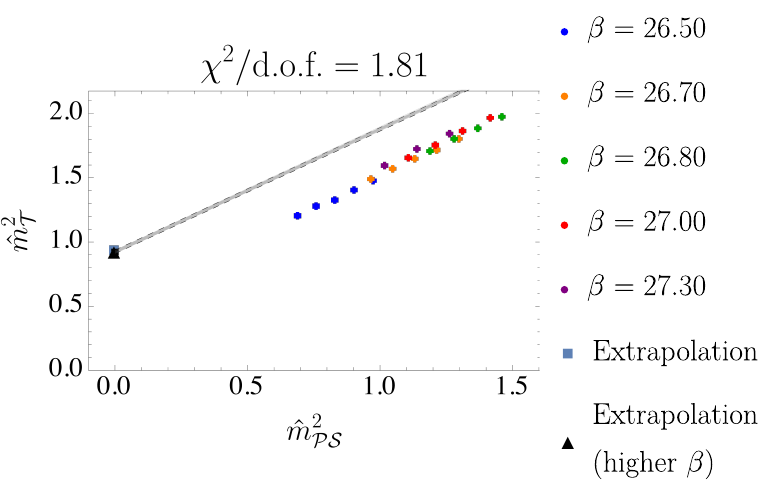}\\
    \end{tabular}
    \caption[$Sp(8)$ chiral masses for symmetric fermions]{Masses squared in the $\cal V$ and $\cal T$ channels comprised of fermions in the symmetric representation of $Sp(8)$. The reduced chi-squared value is printed at the top of each plot. Data points in the pink shaded region are not included in the curve-fitting procedure. The grey band represents the continuum and massless extrapolation with the blue square being the observable and the vertical width corresponding to the statistical error. In instances where a reliable extrapolation cannot be made, no grey band is shown. All quantities are expressed in units of the gradient flow scale, $w_0$. The extrapolation with the smallest $\beta$ value removed is shown as a lighter grey band and a black triangle in cases where data were available at the smallest $\beta$ value.}
\label{fig:sp8Schiralmass}
\end{figure*}

Having computed (quenched) meson masses and decay constants in a discrete spacetime lattice at finite bare mass, 
we then extrapolate to the continuum and massless limits simultaneously, by means of next-to-leading order W$\chi$PT. 
We plot  our measurements, and the extrapolations, for $Sp(4)$, $Sp(6)$ and $Sp(8)$ in 
Figs.~\ref{fig:sp4Fchiraldecay} to~\ref{fig:sp8Schiralmass}, while the numerical details can be found in Refs.~\cite{datarelease,workflowrelease}. For the extrapolations to the massless limit, we have excluded the points for which $\hat{f}_{\mathrm{PS}}$ does not exhibit a linear behaviour in $\hat{m}_{\mathrm{PS}}$. We exclude this set of points from all massless extrapolations at fixed $N$ and representation.

\begin{figure*}[t]
    \centering
    \begin{tabular}{cc}
        \includegraphics[width=85mm]{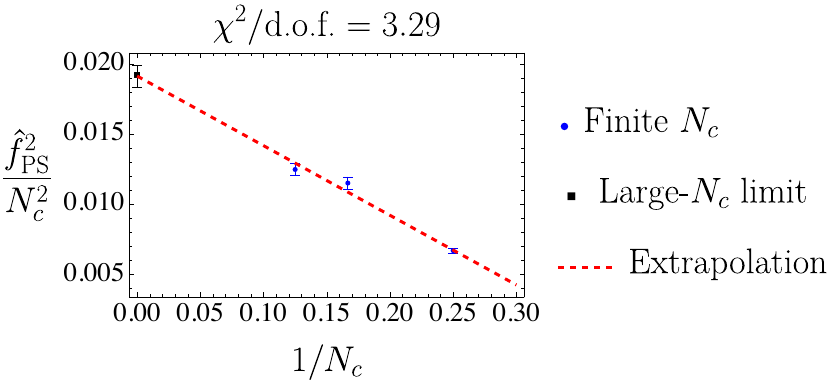} & \includegraphics[width=85mm]{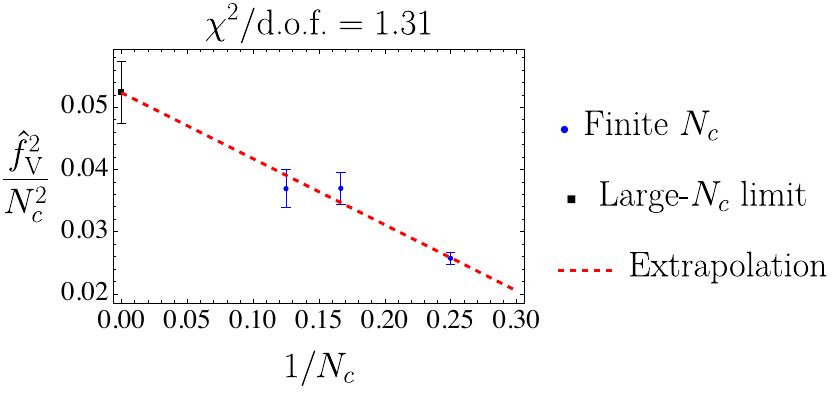}\\
        \multicolumn{2}{c}{\includegraphics[width=85mm]{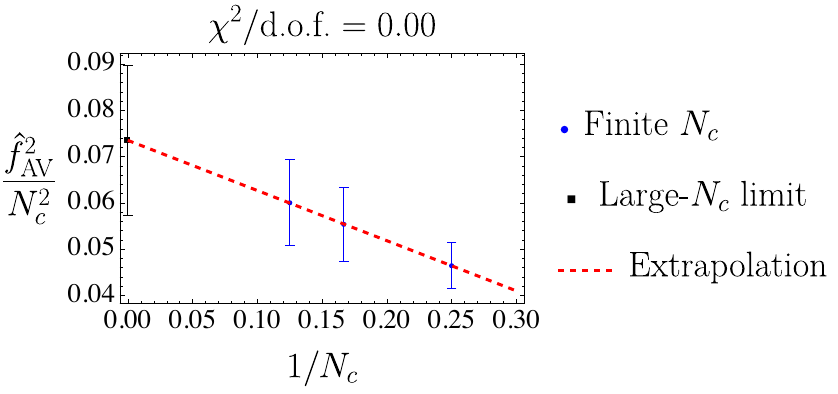}}
    \end{tabular}
    \caption[$Sp(\infty)$ chiral decay constants for fundamental fermions]{Decay constants in $\rm PS$, $\rm V$ and $\rm AV$  channels, with fermions in the fundamental representation extrapolated to $N_c\to\infty$. Reduced chi-squared values are printed at the top of each plot. All quantities are expressed in units of the gradient flow scale, $w_0$.}
\label{fig:largeNFchiraldecay}
\end{figure*}

\begin{figure*}[t]
    \centering
    \begin{tabular}{cc}
        \includegraphics[width=85mm]{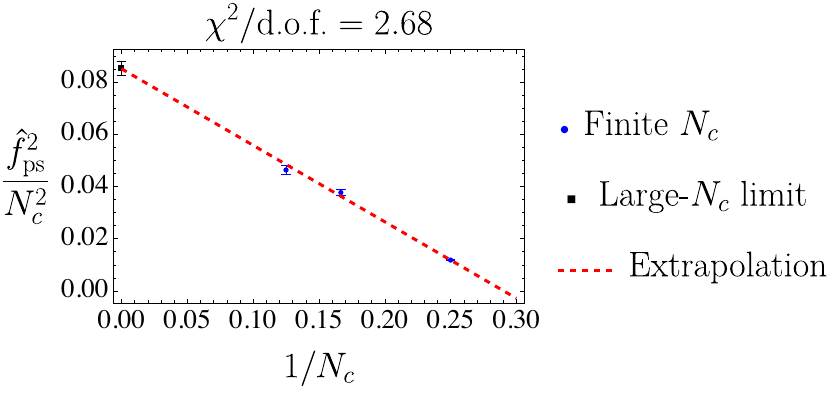} & \includegraphics[width=85mm]{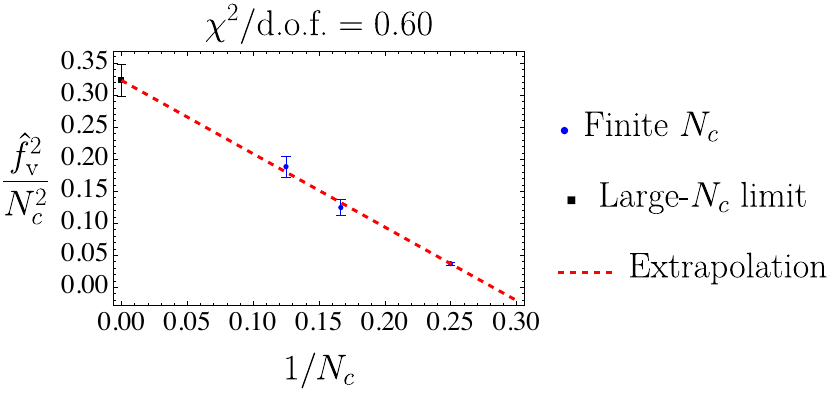}\\
        \multicolumn{2}{c}{\includegraphics[width=85mm]{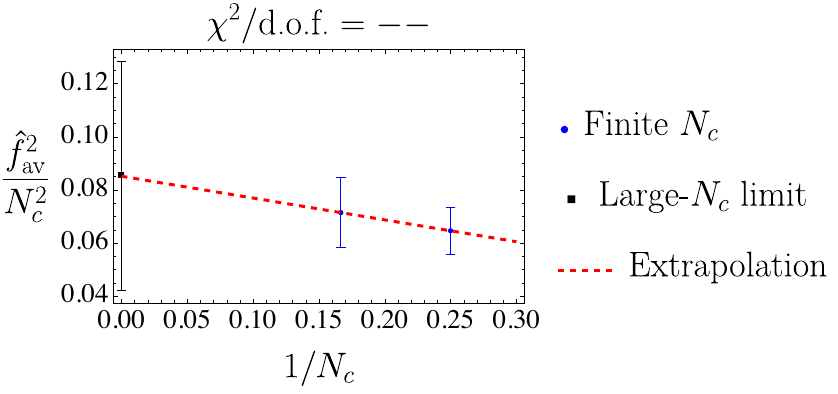}}
    \end{tabular}
    \caption[$Sp(\infty)$ chiral decay constants for antisymmetric fermions]{Decay constants in $\rm ps$, $\rm v$, and $\rm av$  channels, with fermions in the antisymmetric representation extrapolated to $N_c\to\infty$. Reduced chi-squared values are printed at the top of each plot. All quantities are expressed in units of the gradient flow scale, $w_0$.}
\label{fig:largeNASchiraldecay}
\end{figure*}

\begin{figure*}[t]
    \centering
    \begin{tabular}{cc}
        \includegraphics[width=85mm]{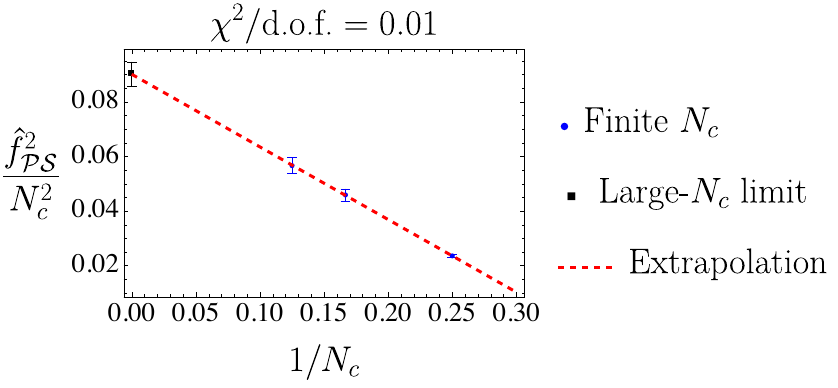} & \includegraphics[width=85mm]{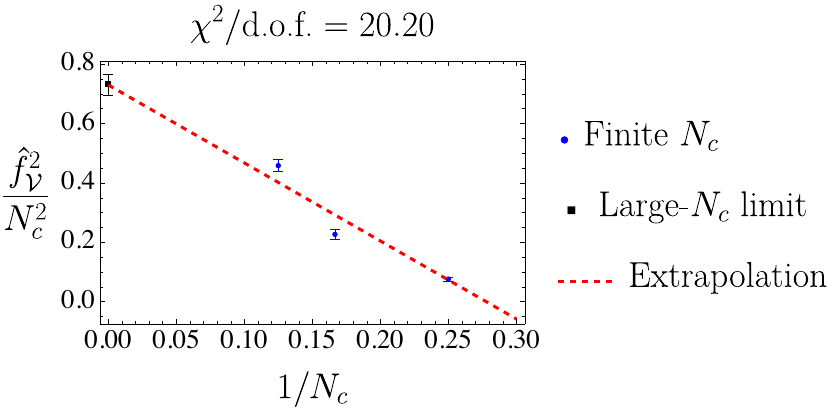}\\
        \multicolumn{2}{c}{\includegraphics[width=85mm]{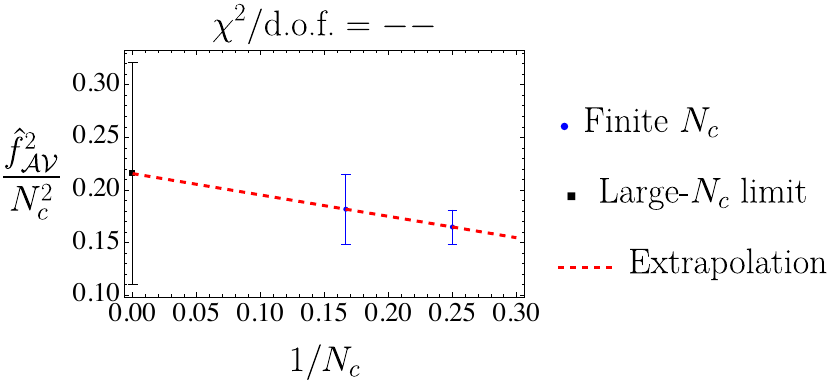}}
    \end{tabular}
    \caption[$Sp(\infty)$ chiral decay constants for symmetric fermions]{Decay constants in  ${\cal PS}$, $\cal V$, and ${\cal AV}$  channels, with fermions in the symmetric representation extrapolated to $N_c\to\infty$. Reduced chi-squared values are printed at the top of each plot. All quantities are expressed in units of the gradient flow scale, $w_0$.}
\label{fig:largeNSchiraldecay}
\end{figure*}

\begin{figure*}[t]
    \centering
    \begin{tabular}{cc}
        \includegraphics[width=85mm]{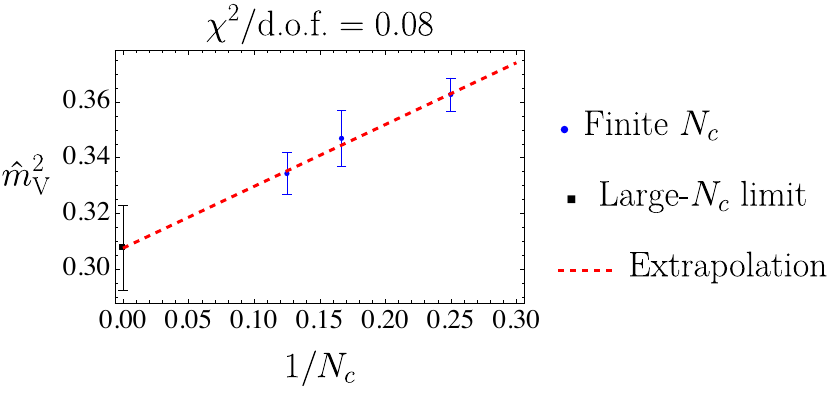} & \includegraphics[width=85mm]{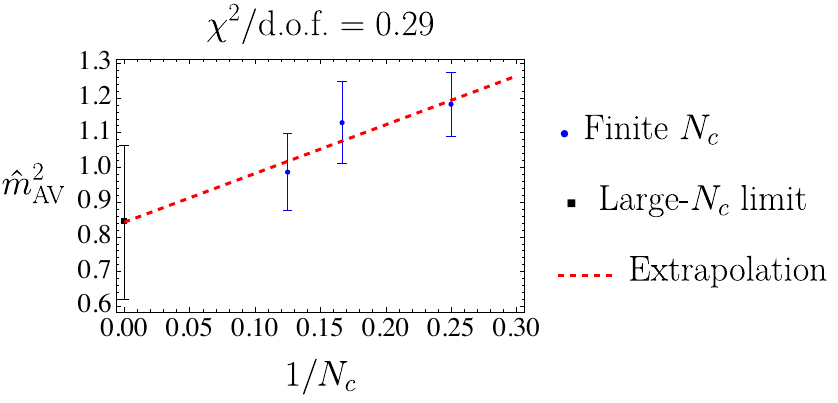}\\
        \includegraphics[width=85mm]{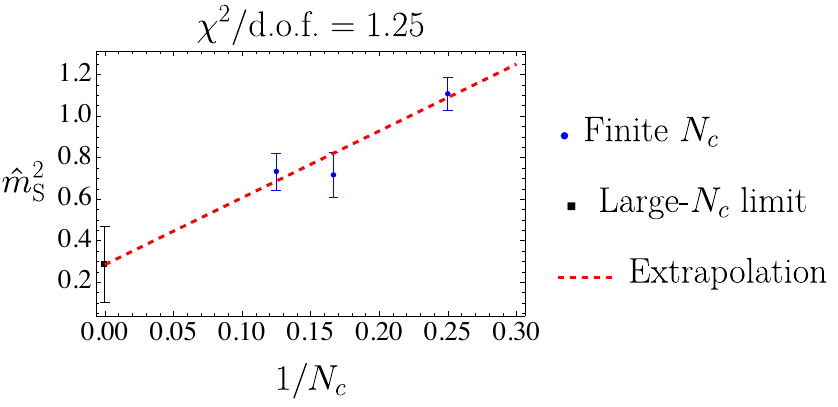} & \includegraphics[width=85mm]{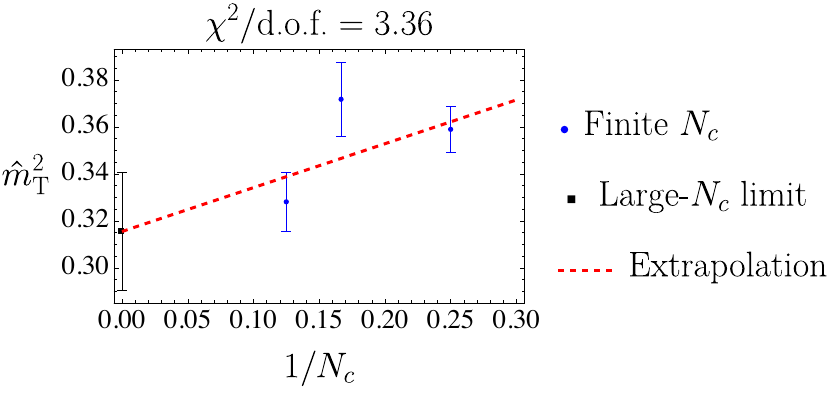}\\
        \multicolumn{2}{c}{\includegraphics[width=85mm]{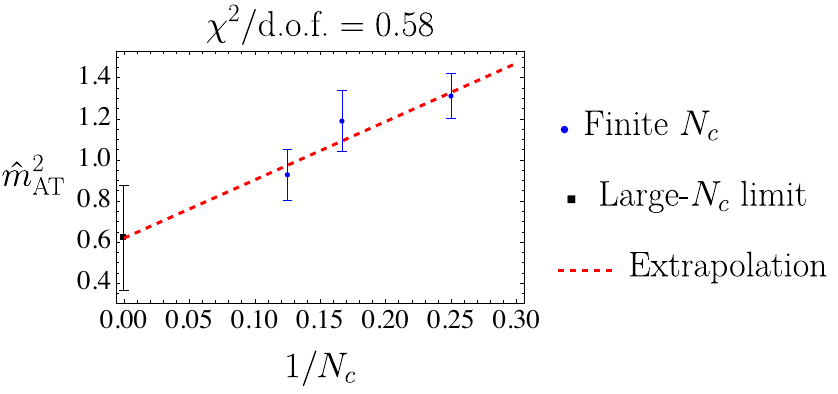}}
    \end{tabular}
    \caption[$Sp(\infty)$ chiral masses for fundamental fermions]{Masses in $\rm V$, $\rm T$,  $\rm AV$,  $\rm AT$, and $ \rm S$ channels, with fermions in the fundamental representation extrapolated to $N_c\to\infty$. Reduced chi-squared values are printed at the top of each plot. All quantities are expressed in units of the gradient flow scale, $w_0$.}
\label{fig:largeNFchiralmass}
\end{figure*}

\begin{figure*}[t]
    \centering
    \begin{tabular}{cc}
        \includegraphics[width=85mm]{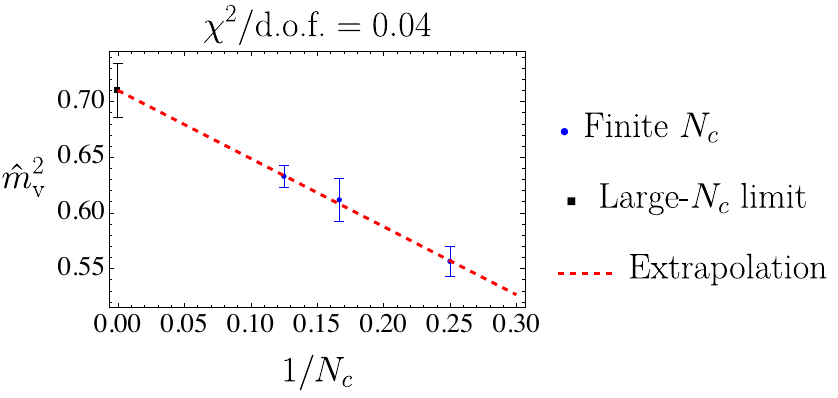} & \includegraphics[width=85mm]{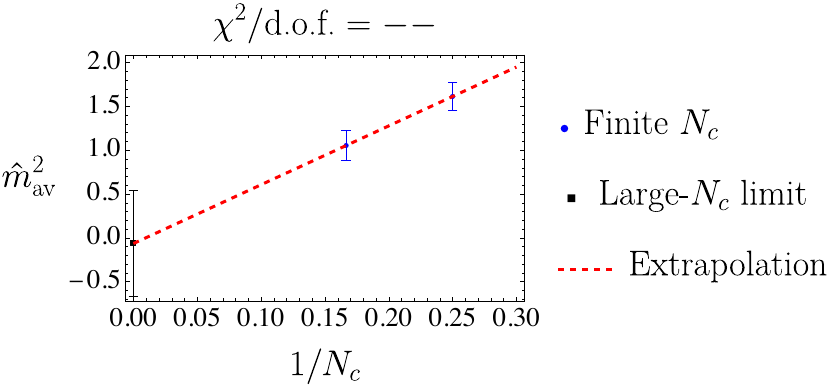}\\
        \includegraphics[width=85mm]{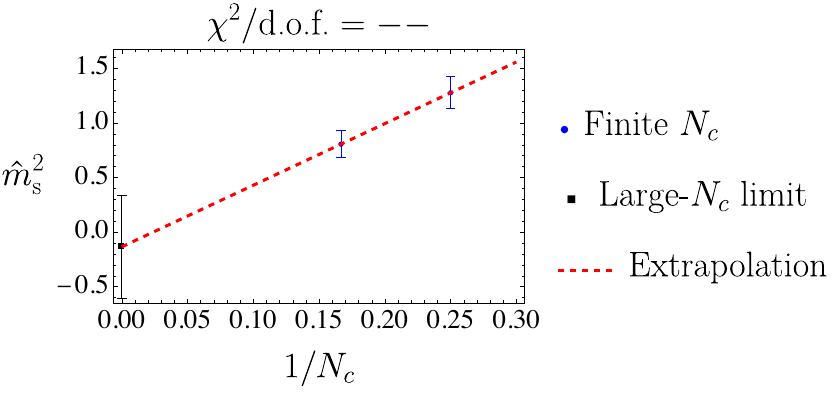} & \includegraphics[width=85mm]{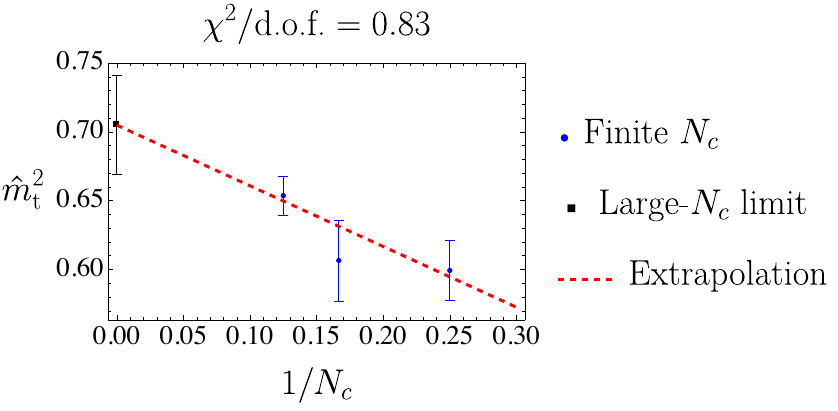}\\
        \multicolumn{2}{c}{\includegraphics[width=85mm]{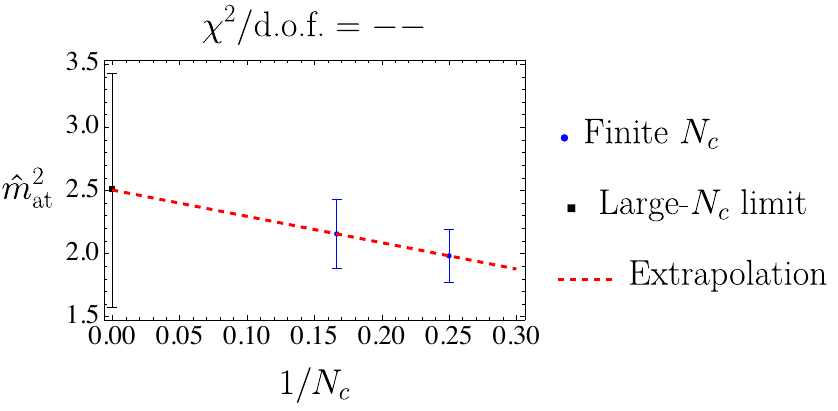}}
    \end{tabular}
    \caption[$Sp(\infty)$ chiral masses for antisymmetric fermions]{Masses in  $\rm v$, $\rm t$,  $\rm av$,  $\rm at$, and $ \rm s$  channels, with fermions in the antisymmetric representation extrapolated to $N_c\to\infty$. Reduced chi-squared values are printed at the top of each plot. All quantities are expressed in units of the gradient flow scale, $w_0$.}
\label{fig:largeNASchiralmass}
\end{figure*}

\begin{figure*}[t]
    \centering
    \begin{tabular}{cc}
        \includegraphics[width=85mm]{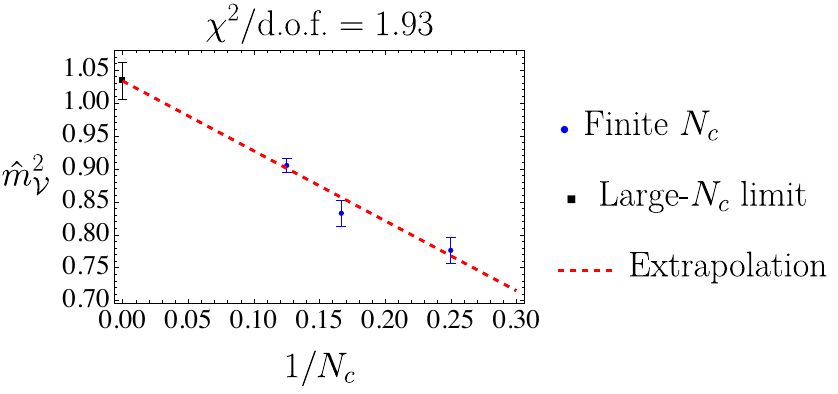} & \includegraphics[width=85mm]{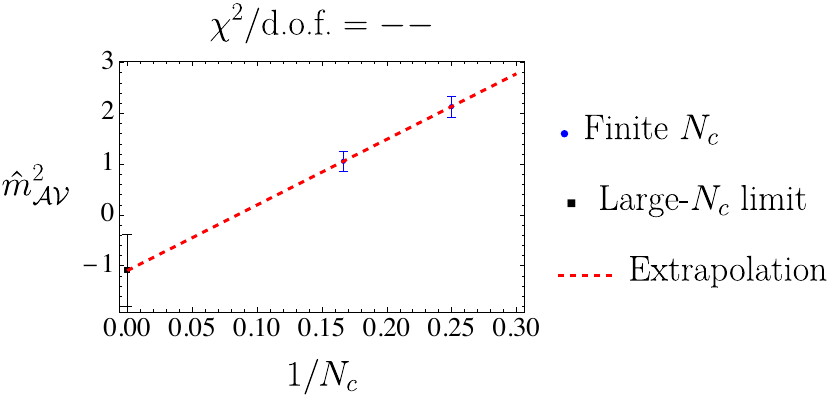}\\
        \includegraphics[width=85mm]{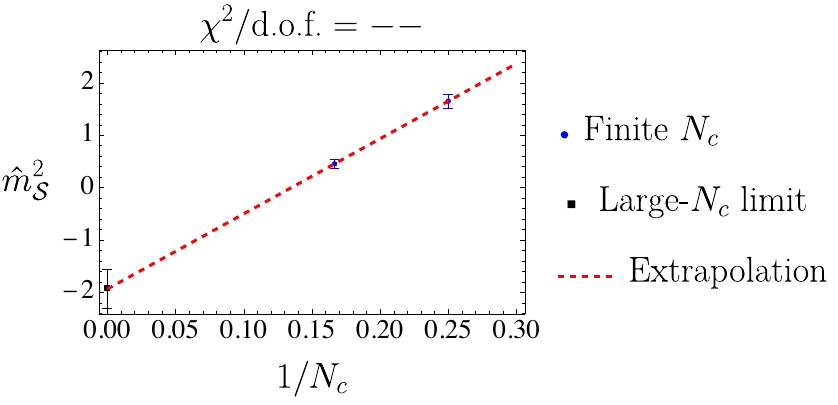} & \includegraphics[width=85mm]{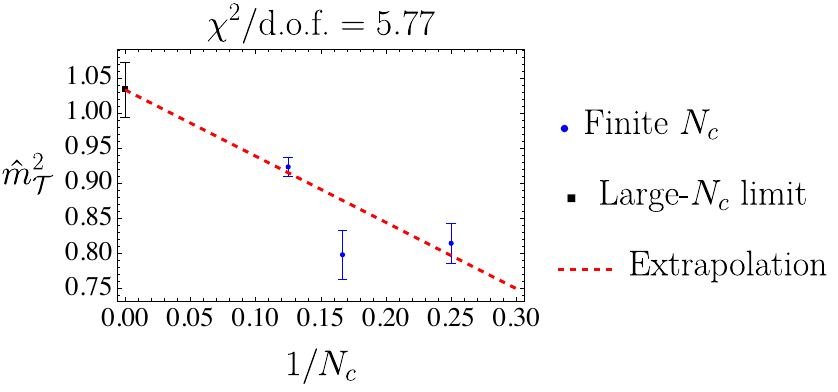}\\
        \multicolumn{2}{c}{\includegraphics[width=85mm]{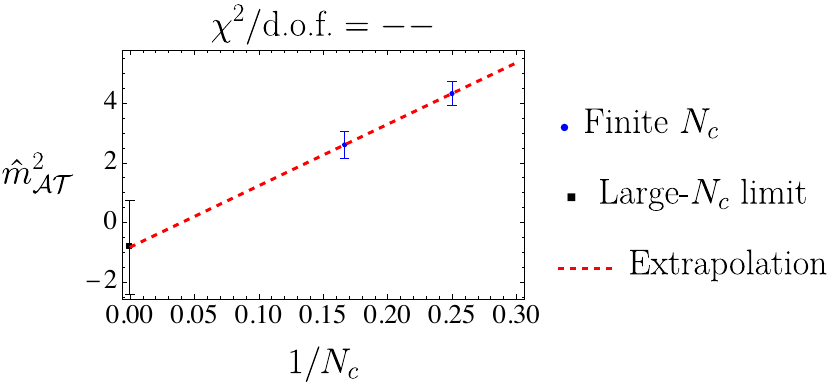}}
    \end{tabular}
    \caption[$Sp(\infty)$ chiral masses for symmetric fermions]{Masses in   $\cal V$, $\cal T$,  $\cal AV$,  $\cal AT$ and $ \cal S$   channels, with fermions in the symmetric representation extrapolated to $N_c\to\infty$. Reduced chi-squared values are printed at the top of each plot. All quantities are expressed in units of the gradient flow scale, $w_0$.}
\label{fig:largeNSchiralmass}
\end{figure*}

In a similar spirit, we display our large-$N$ extrapolations of the massless and continuum extrapolations,
for all three fermion representations, in Figs.~\ref{fig:largeNFchiraldecay} to~\ref{fig:largeNSchiralmass},
while intermediate results and the complete set of fitted parameters can be downloaded from Ref.~\cite{datarelease}.

\bibliography{ref}

\end{document}